\documentclass[aps,pre,reprint,letterpaper,superscriptaddress,twocolumn,10pt,amsfonts,amssymb,amsmath]{revtex4-1}
\usepackage[hidelinks]{hyperref}
\usepackage{graphicx,graphics}
\usepackage{color}

\usepackage[makeroom]{cancel}

\usepackage{subfigure}
\renewcommand{\thesubfigure}{\alph{subfigure}}
\makeatletter
  \renewcommand{\@thesubfigure}{(\thesubfigure)\hskip\subfiglabelskip}
\makeatother

\usepackage{subfloat}
\usepackage{float}
\usepackage[format=plain,justification=centerlast]{caption}

\usepackage[scale=1.0]{OldStandard}
\usepackage{libertinust1math}

\usepackage{cyrdg}
\def\lt{<}

\newcount\fontnumber
\newif\ifoneaccent\oneaccenttrue
\def\ifnextchar#1#2#3{\let\tempe #1\def\tempa{#2}\def\tempb{#3}\futurelet
  \tempc\ifnch}
\def\ifnch{\ifx\tempc\tempe\let\tempd\tempa\else\let\tempd\tempb\fi\tempd}
\def\gobble#1{}
\def\greekmode{%
\catcode`\<=13%
\catcode`\>=13%
\catcode`\'=11%
\catcode`\`=13%
\catcode`\~=11%
\catcode`\"=13%
\lccode`\<=`\<%
\lccode`\>=`\>%
\lccode`\'=`\'%
\lccode`\~=`\~%
\lccode`\"=`\"%
\def\rg{\fontnumber=1\tengr}%
\def\sl{\fontnumber=2\tengrsl}%
\def\it{\fontnumber=3\tengrit}%
\def\bf{\fontnumber=4\tengrbf}%
\def\smc{\fontnumber=5\tengrsmc}%
\def\I##1{\setbox0\hbox{##1}\ifdim\ht0=1ex\accent'174 ##1%
  \else{\ooalign{\hidewidth\char'174\hidewidth\crcr\unhbox0}}\fi}}%
\newcount\vwl
\newcount\acct
{
  \greekmode
  \gdef>{\ifnextchar ~{\expandafter\smoothcircumflex\gobble}{\char\lq\>}}
  \gdef<{\ifnextchar ~{\expandafter\roughcircumflex\gobble}{\char\lq\<}}
  \gdef\smoothcircumflex#1{\acct=\rq134 \vwl=\lq#1 \dobreathingcircumflex}
  \gdef\roughcircumflex#1{\acct=\rq100 \vwl=\lq#1 \dobreathingcircumflex}
  \gdef\dobreathingcircumflex{\ifnum\vwl\lt\rq140 %if uppercase
    \char\the\acct\kern -.2em\char\the\vwl\else
    \doaccent\fi}
  \gdef\doaccent{\accent\the\acct \char\the\vwl\relax}
  \gdef"{\ifnextchar '{\expandafter\diaeresisacute\gobble}{\accent\lq\"}}
  \gdef\diaeresisacute#1{\accent\rq043 #1}
  \gdef`{'}
}
 
\newif\ifgreek\greekfalse
 
\def\begingreek{\bgroup\font\tengr=rgrrg10\font\tengrsl=rgrsl10%
\font\tengrbf=rgrbf10\font\tengrit=rgrti10\font\tengrsmc=rgrsc10%
\greektrue\greekmode\rg}
\def\beginmgreek{\bgroup\font\tengr=mrgrrg10\font\tengrsl=mrgrsl10%
\font\tengrbf=mrgrbf10\font\tengrit=mrgrti10\font\tengrsmc=rgrsc10%
\greektrue\greekmode\rg}
\def\endgreek{\egroup}
\def\endmgreek{\egroup}
\def\monotoniko{%
\font\tengr=mrgrrg10\font\tengrsl=mrgrsl10%
\font\tengrbf=mrgrbf10\font\tengrit=mrgrti10%
\ifnum\fontnumber=5\smc%
  \else\ifnum\fontnumber=4\bf%
    \else\ifnum\fontnumber=3\it%
      \else\ifnum\fontnumber=2\sl%
        \else\rg%
      \fi%
    \fi%
  \fi%
\fi%
}
\def\polutoniko{%
\font\tengr=rgrrg10\font\tengrsl=rgrsl10%
\font\tengrbf=rgrbf10\font\tengrit=rgrti10%
\ifnum\fontnumber=5\smc%
  \else\ifnum\fontnumber=4\bf%
    \else\ifnum\fontnumber=3\it%
      \else\ifnum\fontnumber=2\sl%
        \else\rg%
      \fi%
    \fi%
  \fi%
\fi%
}
\let\math=$%
{\catcode`\$=13%
}
\def\grave#1{{\edef\next{\the\font}\smc\accent\rq001\next#1}}
\def\roughgrave#1{{\edef\next{\the\font}\smc\accent\rq002\next#1}}
\def\smoothgrave#1{{\edef\next{\the\font}\smc\accent\rq003\next#1}}
\def\diaeresisgrave#1{{\edef\next{\the\font}\smc\accent\rq004\next#1}}
\def\diaeresiscircumflex#1{{\edef\next{\the\font}\smc\accent\rq005\next#1}}
\def\breve#1{{\edef\next{\the\font}\smc\accent\rq006\next#1}}
\def\macron#1{{\edef\next{\the\font}\smc\accent\rq007\next#1}}
\def\rhorough{{\tengrsmc
\ifnum\fontnumber=5\char\rq162
  \else\ifnum\fontnumber=4\char\rq016
    \else\ifnum\fontnumber=3\char\rq014
      \else\ifnum\fontnumber=2\char\rq012
        \else\char\rq010
      \fi
    \fi
  \fi
\fi
}}
\def\rhosmooth{{\tengrsmc
\ifnum\fontnumber=5\char\rq162
  \else\ifnum\fontnumber=4\char\rq017
    \else\ifnum\fontnumber=3\char\rq015
      \else\ifnum\fontnumber=2\char\rq013
        \else\char\rq011
      \fi
    \fi
  \fi
\fi
}}
\def\digamma{{\smc\char\rq135}}

\def\Digamma{{\tengrsmc
\ifnum\fontnumber=5\char\rq021
  \else\ifnum\fontnumber=4\char\rq027
    \else\ifnum\fontnumber=3\char\rq025
      \else\ifnum\fontnumber=2\char\rq023
        \else\char\rq021
      \fi
    \fi
  \fi
\fi
}}
\def\vardigamma{{\tengrsmc
\ifnum\fontnumber=5\char\rq020
  \else\ifnum\fontnumber=4\char\rq026
    \else\ifnum\fontnumber=3\char\rq024
      \else\ifnum\fontnumber=2\char\rq022
        \else\char\rq020
      \fi
    \fi
  \fi
\fi
}}

\newcommand{\greeky} [1]{\mbox{\begingreek#1\endgreek}}    % polytonic; it works also in math environment with mbox
\newcommand{\sgreeky}[2]{\scalebox{#1}{#2}} % scaled

\newcommand{\greeksizeinfootnote}{0.75}  % 0.80
\newcommand{\betasizeinfootnote}{0.70}
\newcommand{\greeksizeinsubscript}{0.66} % 0.70
\newcommand{\greeksizeinfootnotesubscript}{0.5725} % 0.55

\newcommand{\alphay}{\greeky{a}}
  
\newcommand{\betay}{\greeky{b}}
  
  \newcommand{\betayf}{\sgreeky{\betasizeinfootnote}{\betay}}              % footnote

\newcommand{\gammay}{\greeky{g}}

\newcommand{\deltay}{\greeky{d}}

\newcommand{\epsilony}{\greeky{e}}

\newcommand{\zetay}{\greeky{z}}

\newcommand{\etay}{\greeky{h}}
  \newcommand{\etayf}{\sgreeky{\greeksizeinfootnote}{\etay}} % footnote
  \newcommand{\etays}{\sgreeky{\greeksizeinsubscript}{\etay}} % sub/superscript
\newcommand{\thetay}{\greeky{j}}
  \newcommand{\thetays}{\sgreeky{\greeksizeinsubscript}{\thetay}} % sub/superscript
  \newcommand{\thetaysf}{\sgreeky{\greeksizeinfootnotesubscript}{\thetay}} % sub/superscript in footnote

\newcommand{\muy}{\greeky{m}}

\newcommand{\xiy}{\greeky{x}}
  \newcommand{\xiyf}{\sgreeky{\greeksizeinfootnote}{\xiy}}

\newcommand{\piy}{\greeky{p}}
\newcommand{\rhoy}{\greeky{r}}
  
  \newcommand{\brhoy}{\overline{\rhoy}}
  \newcommand{\trhoy}{\tilde{\rhoy}}
\newcommand{\phiy}{\greeky{f}}

\newcommand{\psiy}{\greeky{y}}

\usepackage{dg-latexcmds}
\newcommand{\figdir}{.}

  \newcommand{\ma}{\cite{dg2019ejmb}}
  \newcommand{\subma}{$_{\ssub{0.65}{\mbox{\ma}}}$}
  \newcommand{\Reqma}[1]{\mbox{Eq. (#1)}\subma}
  
  \newcommand{\Rma}[1]{\mbox{#1}\subma}

  \newcommand{\oT}{\overline{T}}
  \newcommand {\pnd} {\mathfrak{p}}
  \newcommand {\pc}  {\mathfrak{p}_{c}}
  \newcommand {\tc}  {\thetay_{c}}
  \newcommand {\svc} {\zetay_{c}}
  \newcommand {\dc}  {\xiy_{c}}
  \newcommand {\Zc}  {Z_{c}}
  \newcommand {\phic}{\phi_{c}}
  \newcommand {\Tl}  {T_{l}}
  \newcommand {\Tg}  {T_{g}}
  \newcommand {\vl}  {v_{l}}
  \newcommand {\vg}  {v_{g}}
  \newcommand {\xil} {\xiy_{l}}
  \newcommand {\xig} {\xiy_{g}}
  \newcommand {\phii}{\phi^{\ast}}
  \newcommand {\etai}{\etay^{\ast}}
  \newcommand {\etaim}{\etay^{\ast}_{\ssub{0.65}{\mbox{m}}}}
  \newcommand {\Ni}  {N_{\ssub{0.5}{\mbox{G}}}}     %{N^{\ast}}
  \newcommand {\NM}  {N_{\ssub{0.5}{\mbox{M}}}}

\begin{document}

\title{Fluid statics of a self-gravitating isothermal sphere of van der Waals' gas}
\date{\today}
\author{Domenico Giordano} \email{dg.esa.retired@gmail.com} \affiliation{European Space Agency - ESTEC (retired), The Netherlands}
\author{Pierluigi Amodio} \author{Felice Iavernarto}        \affiliation{Dipartimento di Matematica, Universit\`a di Bari Aldo Moro, Italy}
\author{Francesca Mazzia}                                   \affiliation{Dipartimento di Informatica, Universit\`a di Bari Aldo Moro, Italy}
\author{P\'{e}ter V\'an}  \author{M\'{a}ty\'{a}s Sz\"{u}cs}
\affiliation{Department of Theoretical Physics, HUN-REN Wigner Research Center for Physics, Budapest, Hungary}
\affiliation{Department of Energy Engineering, Faculty of Mechanical Engineering, Budapest University of Technology and Economics, Hungary}

\begin{abstract}
We subject to scrutiny the physical consistency of adopting the perfect-gas thermodynamic model within self-gravitation circumstances by studying the fluid statics of a self-gravitating isothermal sphere with the van der Waals' thermodynamic model, whose equation of state features well-known terms that account for molecular attraction and size.
The governing equations are formulated for any thermodynamic model with two intensive degrees of freedom, applied with the van der Waals' model and solved numerically in nondimensional form by finite-difference algorithms.
After a brief summary of thermodynamic characteristics possessed by the van der Waals' model, and relevant to the present study, we proceed to the description of the results in terms of comparative graphs illustrating radial profiles of density, pressure and gravitational field.
We complement them with graphs that compare the dependence of central and wall densities on gravitational number for both perfect-gas and van der Waals' models and that attest dramatically and unequivocally how the presence of molecular-attraction and -size terms removes questionable fluid-statics results systematically found accompanying the perfect-gas model in standard treatments.
We also describe, within a very brief and preliminary digression, how the sanitising action of the mentioned terms affects the thermodynamics of the isothermal sphere by providing evidence of how the gravitational correction to entropy corresponding to the van der Waals' model makes sure that there is no risk of gravothermal catastrophes, negative specific heats, and thermal instabilities.
Furthermore, we investigate the phenomenology related to self-gravitationally induced both liquid-gas phase equilibria and metastable-gas states and we describe how they arise naturally and self-consistently from the governing equations.
We conclude with a summary of the main results and with a challenging proposal of future work meant to attempt a revalorisation the perfect-gas model.
\end{abstract}

\maketitle

\section{Introduction}\label{intro}
\flushbottom
The present article supplements the contribution \cite{dg2019ejmb} published by some of us in 2019 on the theme ``self-gravitating fluids''.
We target the physical consistency of the supposition that a self-gravitating gas behaves according to the perfect-gas thermodynamic model, an assumption systematically made by the overwhelming majority of authors, including us in \cite{dg2019ejmb}, who have dealt with self-gravitating isothermal spheres since the times of the pioneers \cite{hl1870ajs,eb1880inc,ar1882adp,wt1887pma,gh1888aom,gd1889ptrs,re1907}.
We were driven into this investigation by our dissatisfaction and lack of confidence in, somewhat uncomfortable, results related to the perfect-gas thermodynamic model that have escaped explanation and challenged physical acceptability for about 150 years, namely the upper boundedness of the gravitational number \itm{N\leq N_{m}\simeq 2.5175}, the spiralling behaviour of the peripheral density (\mbox{Fig. 5} in \cite{dg2019ejmb}), the oscillating behaviour of the central density (\mbox{Fig. 6} in \cite{dg2019ejmb}), and the non-uniqueness of density, pressure and gravitational-field radial profiles (\mbox{Figs. 7 and 8} in \cite{dg2019ejmb}) in the gravitational-number range \itm{1.84\leq N \leq N_{m}}.
In \cite{dg2019ejmb}, our hopes of clarification were put on thermodynamics but to no avail.
%in \cite{dg2019ejmb}, we had put our hopes of clarification in thermodynamics but to no avail.
Already in the Joint European Thermodynamics Conference of 2019, we presented a communication in the concluding slide of which we proposed %(see \Rfi{sphere}, item 5)
to investigate better the applicability of the perfect-gas assumption in the presence of a self-imposed gravitational field, in our opinion, the only weak
%and, to the best of our knowledge, unexplored
corner left.
That event marked the beginning of a fruitful collaboration among the present authors that led to the study described in this article. %; it is basically the accomplishment of the task defined in item 5.
In May 2022, we were revising the contents of \cite{dg2019ejmb} and the idea \footnote{The credit for the idea goes to {M. Sz\"{u}cs}.} popped up to consider the van der Waals' (vdW) thermodynamic model \cite{jvdw1873phd}.
The idea met some initial hesitation due to the limitations, well known in the engineering community tasked with equation-of-state development, affecting the quantitative accuracy of the vdW model.
On second thoughts, however, we realised the idea's worthiness because our main interest was more focused on physical consistency, even if only qualitative, rather than on quantitative accuracy.
As a matter of fact, the vdW equation of state has built in the capability to account for molecular attraction, specifically gravitational in our case, so that, conceptually speaking, the vdW model should certainly be nearer to a real self-gravitating gas than the perfect-gas model is.
We felt gratified to discover, later while exploring the literature, that our decision to opt for the vdW model finds imprimatur in the gravitational-thermodynamics analysis of Saslaw \cite{ws1987} and, in particular, in the statistical-mechanics analysis of van Kampen \cite{nvk1964pr}, who retraced, with a more modern formalism, the inquiry pattern laid by Ornstein  \cite{lso1908phd} in his doctoral thesis \footnote{Ornstein's thesis, produced under the supervision of Lorentz, is really a valuable piece of scientific work. In §23 of chapter III, Ornstein re-derived vdW equation of state as first-order solution [Eq. (58)] of his method based, according to the thesis title, on the application of Gibbs' statistical mechanics to molecular-theoretical problems. In §24, he achieved an improvement by obtaining the second-order solution [Eq. (59)], with due acknowledgment to the same formula presented by Boltzmann, together with alternative equations of state, in his famous lectures on gas theory \cite[Eq. (156) and §54]{lb1986}. In §25, another interesting generalisation [Eq. (69)] of vdW equation of state, that boils down to the introduction of a temperature dependence of its coefficients, was obtained by Ornstein with the help of microcanonical-ensemble theory. Regrettably, the thesis was never published; fortunately, it is publicly available online at
{\footnotesize https://web.archive.org/web/20090729212141/http://igitur-archive.library.uu.nl/phys/2006-0117-200056/UUindex.html} but, obviously, the contents are accessible only to Dutch-speaking readers. } defended 35 years after van der Waals'.
An accessory feature associated to the vdW equation of state is the capability, also built-in, to predict liquid-gas phase equilibria; and that opens a window on the interesting possibility to study their gravitationally prompted formation.

The awareness of the interest in the astrophysical community regarding self-gravitating fluids brought us to familiarise with the pertinent literature.
We detected two studies that probed the ground beyond the perfect-gas equation of state.
Aronson and Hansen \cite{ea1972aj} \footnote{We thank our colleague R. Trasarti-Battistoni for bringing to our attention Aronson and Hansen's paper.} also selected the vdW equation of state but, inexplicably, in the form deprived of the molecular-attraction term [see their Eq. (8)]; their calculations were also replicated by Padmanabhan \cite[Sec. 4.5]{tp1990pr}.
Stahl et al. \cite{bs1995pss} founded their investigation on the Carnahan and Starling's \cite{nc1969tjocp} equation of state, another seemingly peculiar choice in view of the fact that such an equation of state was developed for a gas composed by \textit{non-attracting} rigid spheres, as unambiguously stated in the title of \cite{nc1969tjocp}.
The lack of molecular attraction in the mentioned equations of state apart, we were pleased to find out that both teams considered the Gaussian gravitational-field boundary condition at the wall of the container \{compare Eq. (14) of \cite{ea1972aj} and Eq. (7) of \cite{bs1995pss} with Eq. (16) of \cite{dg2019ejmb}\}, although, in the end, Stahl et al.'s path to calculations was rerouted towards the over-popular central-density boundary condition \{Eq. (20) of \cite{bs1995pss}\}.
We will explore the findings of \cite{ea1972aj,bs1995pss} more detailedly in \Rse{nsr}.

In the sequel, we assume the reader to be familiar with \cite{dg2019ejmb} because we will make frequent reference to material contained therein and consistently use the same notation, with exceptions duly indicated.
Also, from now on we adopt the convention to subscript numerical labels cross-referencing to \cite{dg2019ejmb}; for example, \Rma{Eq. (3)} refers to \mbox{Eq. (3)} in \ma.
%\begin{figure}[h]
%%  \includegraphics[keepaspectratio=true, trim= 5ex 8ex 4ex 10ex , clip , width=\columnwidth]{\figdir/jetc2019.png}
%  \centering\includegraphics[keepaspectratio=true, clip , width=0.95\columnwidth]{\figdir/jetc2019.png}
%  \caption{Concluding slide of our communication in the Joint European Thermodynamics Conference of 2019.\hfill\ }\label{sphere}
%\end{figure}    It is regrettable that

\section{Gravitofluid-static fields} \label{fsgsf}

\subsection{Governing equations and boundary conditions} \label{ge+bc}
The microscopic picture we have in mind is a collection of structureless particles, each couple interacting according to the Newtonian gravitational potential,  contained in the spherical solid shell of internal radius $a$ and thickness $\deltay$ illustrated in \Rfi{sphere}. %\Rma{Fig. 1}.
\begin{figure}[t]
	\includegraphics[keepaspectratio=true, trim= 5ex 8ex 4ex 10ex , clip , width=\columnwidth]{\figdir/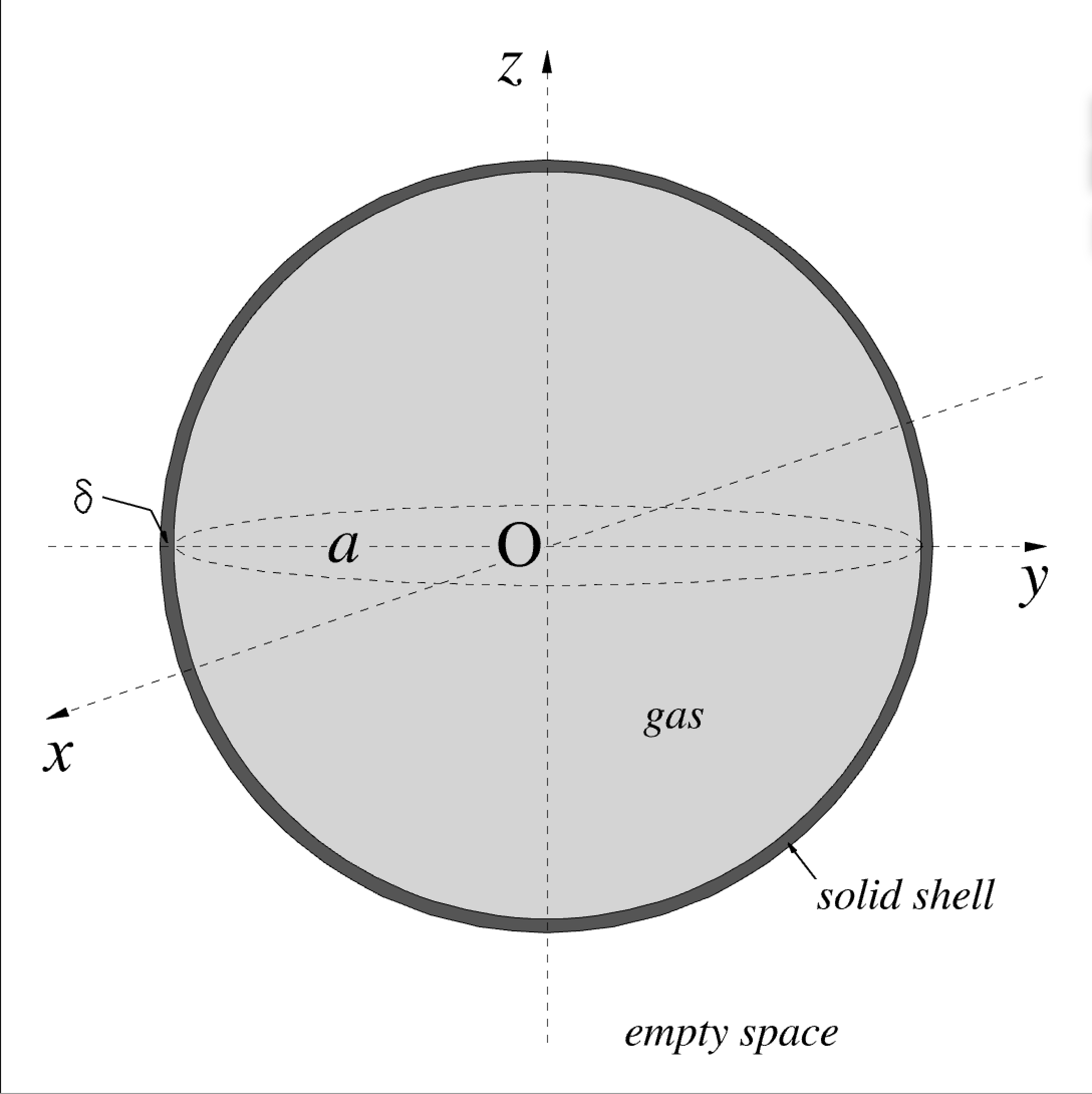}
	\caption{Study test case: fluid statics of a self-gravitating vdW gas inside a spherical solid shell.\hfill\ }\label{sphere}
\end{figure}
Macroscopically, we look at it as a prescribed mass $m_{f}$ of fluid in mechanical equilibrium, sum of the masses of gas and liquid
\begin{equation}\label{masses}
  m_{f} = m_{g} + m_{l}
\end{equation}
if a phase equilibrium is present; if the phase equilibrium is absent then the mass of the liquid vanishes \itm{(m_{l}=0)}.
Accordingly, the average density is defined as
\begin{equation}\label{ad}
  \brhoy = \dfrac{m_{f}}{V} = \dfrac{m_{f}}{\tfrac{4}{3} \piy a^{3}}
\end{equation}
and its reciprocal defines the average specific volume
\begin{equation}\label{asv}
  \overline{v} = \dfrac{1}{\brhoy}
\end{equation}

The governing equations comprise the Newtonian gravity's formulation in gas, shell and empty space, and the static versions of the traditional balance equations of mass, momentum and energy in the gas, duly accompanied by the equations of state of the adopted thermodynamic model; they are described in \Rma{Sec. 2.1} whose contents apply basically unchanged from its beginning up to \Rma{Eq. (21)} except for the suspension of the perfect-gas equation of state [\Rma{Eq.~(3)}] and the average-pressure definition [\Rma{Eq.~(14b)}]. %given, respectively, by \Rma{Eqs. (3)} and \Rma{(14b)}.
We replace the former with the vdW equation of state
\begin{equation}\label{vdwse}
%  p = \frac{RT}{v - B} - \frac{A}{v^{2}} = \frac{\rhoy RT}{1 - B\rhoy} - A\rhoy^{2}
  p = \frac{\rhoy RT}{1 - B\rhoy} - A\rhoy^{2}
\end{equation}
and leave the latter on hold for the time being.
%In \REq{vdwse}, $A$ and $B$ are the notorious coefficients that bring into account, respectively, the existence of intermolecular forces and the finiteness of molecular size; in our case, we entrust to $A$ the responsibility of representing, even if at least qualitatively, the gravitational attraction among the molecules of the gas.\footnote{The list of intermolecular forces is rather long (https://en.wikipedia.org/wiki/Van\_der\_Waals\_force); among them, there are the electrostatic interactions among permanent electric charges.
%Our conceptual approximation consists in replacing ``electrostatic'' with ``gravitational'' and ``electric charges'' with ``masses''. }
In \REq{vdwse}, $A$ and $B$ are the notorious coefficients that bring into account, respectively, the existence of intermolecular forces and the finiteness of molecular size; in our case, we entrust to $A$ the responsibility of representing, even if qualitatively at least, the gravitational attraction among the gas particles.
%\footnote{It is well known that electrostatic interactions among electric charges belong to the list of intermolecular forces; our conceptual approximation consists in replacing ``electrostatic'' with ``gravitational'' and ``electric charges'' with ``masses''.}

The temperature field is again uniform %[from \Rma{Eq. (21)} with the change of notation $T_{s} \rightarrow \oT$]
\begin{equation} \label{Tf}
	T(r) = \oT
\end{equation}
and, in principle, we could try to follow the same procedure that, for a perfect gas, leads to the isothermal Lane-Emden [\Rma{Eq. (27a)}].
However, the vdW equation of state introduces a complication brought forth by the squared isothermal characteristic velocity's dependence on density
\begin{equation}\label{dpdr}
  \pds{}{p}{\rhoy}{T}  = \frac{RT}{(1 - B\,\rhoy)^{2}} - 2A\,\rhoy
\end{equation}
Indeed, with the availability of \REqd{vdwse}{dpdr}, the gravitational field follows from the momentum equation [\Rma{Eq. (2b)}]
\begin{equation} \label{gf}
    \g = \pds{}{p}{\rhoy}{T=\oT} \Grad\ln\rhoy
\end{equation}
and the comparison of \REq{gf} with \Rma{Eq. (1b)} yields
\begin{equation}   \label{gp}
     \Grad \psiy + \pds{}{p}{\rhoy}{T=\oT} \Grad\ln\rhoy = 0
\end{equation}
For a perfect gas \itm{(A=B=0)} under isothermal conditions, \REq{gp} reduces to a global gradient [\Rma{Eq. (25)}] because the squared isothermal characteristic velocity \itm{(R\oT)} is constant.
Thus, the logically consequent question that arises by looking at \REq{gp} is whether or not it turns out to be possible to reduce it to the form of a global gradient also in the case of the vdW model to which \REq{dpdr} applies.
But the prospect of investigating such a possibility  induces some justified hesitation because, provided that the reduction has been accomplished and the gravitational potential's expression as a function of density has been obtained, the substitution of such an expression into the Poisson equation [\Rma{Eq. (1a)} (gas)] that governs gravity in the gas should be expected to lead to a second-order differential equation for the density whose suspected mathematical cumbersomeness may turn out to be not as easily and smoothly handled from a numerical-calculation point of view as one would wish.
Therefore, we decided to hold on \REq{gp} and considered another possible strategy that consists in the simultaneous solution of the density-adapted momentum equation
\begin{equation}\label{meq.s}
%  \pds{}{p}{\rhoy}{T=\oT} \Grad\rhoy - \rhoy \g = 0
  \pds{}{p}{\rhoy}{T=\oT} \pd{}{\rhoy}{r} - \rhoy g = 0
\end{equation}
and the gravitational-field divergence equation
\begin{equation} \label{gfd.s}
%   \Dive{\g} = - 4\piy G \rhoy
   \pd{}{g}{r} + \frac{2}{r} g + 4\piy G \rhoy = 0
\end{equation}
as they look like after simplification due to spherical symmetry. % simplifies \REqd{meq}{gfd} to the scalar forms
%\begin{equation}\label{meq.s}
%  \pds{}{p}{\rhoy}{T=\oT} \pd{}{\rhoy}{r} - \rhoy g = 0
%\end{equation}
%\begin{equation} \label{gfd.s}
%   \pd{}{g}{r} + \frac{2}{r} g + 4\piy G \rhoy = 0
%\end{equation}
We refer to the system composed by \REqd{meq.s}{gfd.s} as the P scheme; it is a first-order differential-equation system for density and gravitational field that requires \REq{dpdr} as auxiliary equation.
%{\itshape The P scheme works swiftly in the absence of liquid-gas phase equilibrium, we found out, but, in the presence of phase equilibrium, it is affected by an undesirable weakness that requires careful handling from a numerical point of view: it is \textit{strongly} affected by the discontinuity of the density across the phase-equilibrium spherical interface whose radial location, by the way, is an ulterior unknown of the problem.}
%
The P scheme works swiftly in the absence of phase equilibrium but, in the case of presence, we found out it to be \textit{strongly} affected by the discontinuity of the density across the phase-equilibrium spherical interface, whose radial location, by the way, is an ulterior unknown of the problem.
This undesirable weakness adds complexity and requires careful handling from a numerical point of view.
A possible valid alternative, promoted by the continuity of pressure across the interface, consists in the replacement of \REq{meq.s} with the nominal momentum equation
\begin{equation}\label{nmeq.s}
  \pd{}{p}{r} - \rhoy g = 0
\end{equation}
The introduction of the pressure, a third unknown, requires the incorporation in the system of the vdW equation of state [\REq{vdwse}] and presupposes the latter's inversion to extract the density, but that is a less serious concern from a numerical point of view.
Of course, \REq{nmeq.s} is not immune to the effect of the density discontinuity, if any, but the latter's impact on the numerical machinery is appreciably attenuated.
We refer to \REq{vdwse}, \REq{gfd.s} and \REq{nmeq.s} as the M$_1$ scheme.

In the absence of phase equilibrium, the solution of both P and M$_1$ schemes must be accomplished with the boundary conditions enforcing the vanishing of the gravitational field at the sphere center [\Rma{Eq. (11a)}] and the Gaussian value of the gravitational field at the shell's internal wall [\Rma{Eq. (16)} with the change of notation \itm{m_{g}\rightarrow m_{f}}] respectively.
In the presence of phase equilibrium, the mentioned boundary conditions are not sufficient anymore because the interface plays a fundamental role as boundary between liquid and gas and, in so doing, its physics deeply affects the mathematical structure of the computational schemes.
However, we believe it would be premature to engage now on the discussion of the details related to that situation because such a discussion presupposes a clear understanding of the phase-equilibrium thermodynamics which we will deal with in \Rse{lgpe}; for this reason, we put on hold this matter for the time being and postpone its consideration to \Rse{agepei}.

The P and M$_1$ schemes constitute a solid base from which we could have proceeded to numerical-solution operations and forgotten the uncompleted path left at \REq{gp}; and indeed, so we did at first.
Yet, we were somehow irked by the apparent barrier met there and, in the \textit{background}, we kept looking from different angles at ways of breaking through it until, finally, our patience and tenacity were rewarded.
The inspiring spark came from thermodynamics.
The specific Helmholtz potential of a fluid with two thermodynamic intensive degrees of freedom reads formally
\begin{equation}\label{hp}
  f = f(T,v)
\end{equation}
Its natural state parameters are temperature and specific volume \itm{v=1/\rhoy}; the corresponding first derivatives constitute, by definition, the state equations and represent entropy and pressure
\begin{subequations}\label{hpse}
  \begin{align}
    \pds{}{f}{T}{v} & = - \, s\,(T,v) \label{hpse.s} \\[.5\baselineskip] \pds{}{f}{v}{T} & = - \, p\,(T,v) \label{hpse.p}
  \end{align}
\end{subequations}
The application of the gradient operator to \REq{hp} gives
\begin{equation}\label{ghp}
  \Grad f = -\,\cancel{\ s\,\Grad T \ } -p\,\Grad v = - \Grad (pv) + v \Grad p
\end{equation}
The crossed term vanishes because the gas sphere is isothermal.
From \REq{ghp}, we obtain
\begin{equation}\label{cp}
  v \Grad p = \Grad (f+pv) = \Grad\muy
\end{equation}
in which \itm{\muy=\muy(T,v)} is the chemical potential of the fluid \footnote{\label{gibbs}The sum \itm{f+pv} is also the Legendre transform of the Helmholtz potential with respect to the specific volume and defines the Gibbs potential \itm{g=g(T,p)} whose natural state parameters are temperature and pressure; in our case, the chemical-potential interpretation is preferable because we find more convenient to work with temperature and specific volume.
%\REqb{cp} will not be looked at as a surprising novelty by well read fluid dynamicists because convenient rearrangements of the term \itm{v\Grad p} in the momentum balance equation occur from time to time in the literature.
\REqb{cp} does not strike the eye of well-read fluid dynamicists as a surprising novelty because convenient rearrangements of the term \itm{v\Grad p} in the momentum balance equation occur from time to time in the literature.
To mention a few examples, it appeared in Fridman's (or Friedmann's) doctoral thesis, defended in 1922, to reformulate the right-hand side of the momentum equation for an inviscid ideal gas in terms of gradients of temperature and entropy; see Eq.~(94) in the Russian publication \cite{af1934} appeared in 1934 and available online at http://books.e-heritage.ru/Book/10087382. It played a role in Crocco's derivation \cite{lc1937zamm} of his famous theorem in the case of  homenthalpic flow of ideal gas as well as in Vaszonyi's generalisation \cite{av1945qam} of Crocco's theorem to the form, taught nowadays, in terms of gradients of enthalpy and entropy applicable to any fluid with two intensive thermodynamic degrees of freedom}.
With \REq{cp} in hand, we turn to the momentum equation [\Rma{Eq. (2b)}], multiply it by the specific volume, solve it for the gravitational field
\begin{equation}\label{gf.meq}
  \g = v \Grad p
\end{equation}
%and compare \REq{gf.meq} with \REq{cp} to deduce the remarkable result\footnote{It appears worth to remark that this can be considered a particular example of classical-holography property of ideal fluids, a concept that, in its embryonic form, traces back to Fridman's doctoral thesis defended in 1922 and officially published 12 years after \cite{af1934r}; a digitised copy is available online at http://books.e-heritage.ru/book/10073889 but unfortunately only in Russian. A more recent and thorough elaboration regarding holographic fluids has been published by V\'{a}n \cite{pv2023pof}.}
and compare \REq{gf.meq} with \REq{cp} to deduce the remarkable result \footnote{It appears worth to remark that, according to \REq{gf.g}, the chemical potential acts as a mechanical potential and that can be considered a particular example of the classical-holography property of ideal fluids. A thorough elaboration regarding holographic fluids has been recently published by V\'{a}n \cite{pv2023pof}.}
\begin{equation}\label{gf.g}
  \g = \Grad\muy
\end{equation}
Further comparison between \REq{gf.g} and the standard definition of the gravitational field as gradient of gravitational potential [\Rma{Eq. (1b)}] leads to
\begin{equation}\label{gp.g}
  \Grad(\psiy + \muy) = 0
\end{equation}
from which we deduce the gravitational potential
\begin{equation}\label{gp.g.e}
  \psiy = C - \muy
\end{equation}
in terms of the chemical potential, save for an arbitrary inessential constant $C$.
With the path illuminated by \REq{cp}, we discovered rather easily how \REqd{gf}{gp} are contained, respectively, in \REqd{gf.g}{gp.g} \footnote{Indeed, once we independently understood how to obtain \REqd{gf.g}{gp.g}, the way to break through the apparent barrier we saw in \REqd{gf}{gp} became immediately clear and we realised that we could not see it because we were somehow biased by the mathematical complexity of \REq{dpdr}. We leave the mathematical passages as exercise for the interested reader.}.
%With the path illuminated by \REq{cp}, we found out rather easily how \REqd{gf}{gp} are contained, respectively, in \REqd{gf.g}{gp.g}; we could not see through the barrier because we were biased by \REq{dpdr}.\footnote{Indeed, the way to surmount the apparent barrier we saw in \REqd{gf}{gp} became immediately clear once we independently understood how to obtain \REqd{gf.g}{gp.g}. We could not see through the barrier because we were somehow biased by \REq{dpdr}. We leave the mathematical passages as an exercise for the interested reader.}
The substitution of the gravitational potential [\REq{gp.g.e}] into the Poisson equation [\Rma{Eq. (1a)} (gas)] that governs gravity in the gas leads to the second-order differential equation we were looking for
\begin{equation}\label{sode}
  	    \Lap\muy  = -  \frac{4\piy G}{v}
\end{equation}
and that expands into the spherically-symmetric form
\begin{equation}\label{sode.ss}
    \frac{1}{r^{2}}\pd{}{}{r}\left( r^{2} \pd{}{\muy}{r} \right) +\frac{4\piy G}{v} = 0
\end{equation}
whose integration requires the gravitational boundary conditions [\Rma{Eq. (11a)} and \Rma{Eq. (16)}] reformulated, by account of \REq{gf.g}, in terms of the chemical potential
\begin{subequations} \label{bc.cp}
  \begin{align}
    g(0) = {\left. \pd{}{\muy}{r} \right|}_{r=0} & =  0                      \label{bc.cp.r=0} \\[.5\baselineskip]
    g(a) = {\left. \pd{}{\muy}{r} \right|}_{r=a} & =  -\frac{G m_{f}}{a^{2}} \label{bc.cp.r=a}
  \end{align}
\end{subequations}
These boundary conditions are applicable in the circumstance of absence of phase equilibrium and, here again, we postpone the consideration of the case with presence of phase equilibrium to \Rse{agepei}, similarly to what we have planned for the P and M$_{1}$ schemes and for the same reason.
In \REqd{sode}{sode.ss}, we have switched from density to specific volume in the source term for consistency with the natural state parameters of the Helmholtz potential [\REq{hp}].
In our opinion, the second-order differential equation of the chemical potential and the corresponding boundary conditions represent an elegant and appealing formulation of the mathematical problem from a physical point of view mainly because of two important features associated with them: (a) they are applicable to \textit{any} fluid with two thermodynamic intensive degrees of freedom; (b) as we will see soon in \Rse{agepei}, chemical potential and its first (radial) derivative, that is the gravitational field [\REq{gf.g}], are continuous across a phase-equilibrium interface, if any.
Of course, \REqd{sode.ss}{bc.cp} are in open form; they become operative only upon assignment of the thermodynamic model, that is, the explicit knowledge of the Helmholtz potential [\REq{hp}] from which the chemical potential [\REq{cp}] can be straightforwardly deduced.
In general, and certainly for perfect-gas and vdW models, the chemical potential takes the form
\begin{equation}\label{cp.pgvdw}
   \muy(T,v) = \muy_{0}(T) + RT\cdot\phi(T,v)
\end{equation}
We do not need to expand the contribution \itm{\muy_{0}(T)} because it depends only on temperature and is, therefore, a constant under isothermal circumstances.
The gravitational field is obtained from \REq{gf.g} via \REq{cp.pgvdw} and, in scalar form, reads
\begin{equation}\label{gf.vdw}
  g = RT \pd{}{\phi}{r}
\end{equation}
In the case of the perfect-gas model, the reduced chemical potential does not depend on temperature and is simply
\begin{equation}\label{rcp.pg}
   \phi(T,v) = 1-\ln v
\end{equation}
It is a rather straightforward exercise to verify that the substitution of \REqd{cp.pgvdw}{rcp.pg} into \REqd{sode.ss}{bc.cp} leads to the Lane-Emden equation [\Rma{Eq. (27b)}] and its boundary conditions [\Rma{Eqs. (28)}].
In the case of the vdW model, the reduced chemical potential is a bit more complex
\begin{equation}\label{rcp.vdw}
   \phi(T,v) = - \ln ( v - B ) + \frac{v}{v - B} - \frac{2A}{RTv}
\end{equation}
and it is clearly not convenient to substitute \textit{both} \REqd{cp.pgvdw}{rcp.vdw} into \REq{sode.ss} because the cost of doing so is the loss of linearity of the second-order differential equation;
but we can certainly substitute \textit{only} \REq{cp.pgvdw} into \REqd{sode.ss}{bc.cp} to obtain the differential equation
\begin{equation}\label{sode.ss.vdw}
    \frac{1}{r^{2}}\pd{}{}{r}\left( r^{2} \pd{}{\phi}{r} \right) +\frac{4\piy G}{RTv} = 0
\end{equation}
with the boundary conditions
\begin{subequations} \label{bc.rcp}
  \begin{align}
    {\left. \pd{}{\phi}{r} \right|}_{r=0} & =  0                         \label{bc.rcp.r=0} \\[.5\baselineskip]
    {\left. \pd{}{\phi}{r} \right|}_{r=a} & =  -\frac{G m_{f}}{a^{2}RT}  \label{bc.rcp.r=a}
  \end{align}
\end{subequations}
and, then, associate to them the algebraic equation of the reduced chemical potential [\REq{rcp.vdw}] to proceed to their numerical solution.
We refer to this set of equations as the M$_{2}$ scheme.

\subsection{Nondimensional formulation\label{ndf}}
We have repeated the nondimensional analysis along the guidelines indicated in \Rma{Sec. 2.2} and, for obvious reasons of convenience, we have selected the scale factors
\begin{equation} \label{sf}
    \begin{bmatrix} \;
        \tilde{r} = a                      & %\negthickspace
        \tilde{T} = \oT                    & %\negthickspace
        \trhoy    = \brhoy                 & %\negthickspace
        \tilde{v} = \overline{v}           & %\negthickspace
        \tilde{p} = \brhoy R \oT           & %\negthickspace
        \tilde{g} = \dfrac{G m_{f}}{a^{2}} \;
    \end{bmatrix}
\end{equation}
in a manner consistent with the choice made therein so that the nondimensional variables turn out to be defined as
\begin{equation} \label{ndv}
    \begin{bmatrix} \,
        \etay = \dfrac{r}{a}               & %\negthickspace
        \thetay = \dfrac{T}{\oT}           & %\negthickspace
        \xiy    = \dfrac{\rhoy}{\brhoy}    & %\negthickspace
        \zetay  = \dfrac{v}{\overline{v}}  & %\negthickspace
        \pnd    = \dfrac{p}{\brhoy R \oT}  & %\negthickspace
        \gammay = \dfrac{a^{2}g}{G m_{f}} \,
    \end{bmatrix}
\end{equation}
\begin{subequations}\label{cn}%
We discovered three physical characteristic numbers this time.
The gravitational number
    \begin{equation}\label{gn}
       N = \frac{G m_{f}}{aR\oT}
    \end{equation}
    that we introduced and described in the vicinity of \Rma{Eqs. (37)}, comes accompanied by two other numbers
      \begin{align}
        \alphay & =  \dfrac{A\,\brhoy}{R \oT}   \label{cn.a} \\[.5\baselineskip]
        \betay  & =  B\,\brhoy                  \label{cn.b}
      \end{align}
    whose origin can be clearly traced back to the vdW equation of state [\REq{vdwse}]; they vanish (\itm{\alphay=\betay=0}) for a perfect gas.
\end{subequations}

The vdW equation of state [\REq{vdwse}] reads
\begin{equation}\label{vdwse.nd}
  \pnd = \frac{\xiy\,\thetay}{1 - \betay\,\xiy} - \alphay\xiy^{2} % = \frac{\xiy}{1 - \betay\xiy} - \alphay\xiy^{2}
\end{equation}
and the incorporation of the isothermal condition [\REq{Tf}]
\begin{equation} \label{Tf.nd}
	\thetay(\etay) = \frac{T(r)}{\oT} = \frac{\oT}{\oT} = 1
\end{equation}
allows the slight simplification
\begin{equation}\label{vdwse.nd.s}
  \pnd =  \frac{\xiy}{1 - \betay\,\xiy} - \alphay\xiy^{2}
\end{equation}
whose derivation with respect to $\xiy$ provides the isothermal characteristic velocity
\begin{equation}\label{dpdr.nd}
  \pds{}{\pnd}{\xiy}{\!\!\thetays=1}  = \frac{1}{(1 - \betay\,\xiy)^{2}} - 2\alphay\,\xiy
\end{equation}
We ought to point out that, on the contrary of what happens in the perfect-gas case, the pressure scale factor $\tilde{p}$ [\REq{sf}] does not coincide with the average pressure $\overline{p}$ for the vdW model.
The latter quantity follows from the vdW equation of state [\REq{vdwse}] after setting \itm{\rhoy=\brhoy} or, in nondimensional language, from \REq{vdwse.nd.s} after setting \itm{\xiy=1}
\begin{equation}\label{ndap}
  \overline{\pnd} =  \frac{1}{1 - \betay} - \alphay  % = \pnd_{\,\xi=1}
\end{equation}
Then, the simple multiplication
\begin{equation}\label{popa}
  \frac{p}{\overline{p}} = \frac{p}{\tilde{p}} \cdot \frac{1}{\dfrac{\overline{p}}{\tilde{p}}} = \pnd\cdot\frac{1}{\overline{\pnd}}    %{\pnd_{\,\xi=1}}
\end{equation}
provides the adequate rescaling, if wished.

The computational schemes P, M$_{1}$ and M$_{2}$ are easily reformulated in nondimensional form by taking into account the definitions indicated in \REqd{sf}{ndv}.
The P scheme includes the differential equations
%\begin{subequations}\label{Pscheme}
\begin{equation}\label{meq.s.nd}
  \pds{}{\pnd}{\xiy}{\!\!\thetays=1} \pd{}{\xiy}{\etay} - N\,\xiy\,\gammay = 0
\end{equation}
\begin{equation} \label{gfd.s.nd}
   \pd{}{\gammay}{\etay} + \frac{2}{\etay} \gammay  + 3 \xiy = 0
\end{equation}
obtained from \REqd{meq.s}{gfd.s} and require the auxiliary \REq{dpdr.nd}.
%; in \REq{meq.s.nd}, the nondimensional form of the isothermal characteristic velocity [\REq{dpdr}] reads
%\begin{equation}\label{dpdr.nd.1}
%  \pds{}{\pnd}{\xiy}{\!\!\thetays=1}  = \frac{1}{(1 - \betay\,\xiy)^{2}} - 2\alphay\,\xiy
%\end{equation}
%\end{subequations}
In the M$_{1}$ scheme, \REq{meq.s.nd} is replaced with the nondimensional form
\begin{equation}\label{nmeq.s.nd}
  \pd{}{\pnd}{\etay} - N\,\xiy\,\gammay = 0
\end{equation}
of the momentum equation [\REq{nmeq.s}]
and \REq{dpdr.nd} is superseded by the vdW equation of state [\REq{vdwse.nd.s}] which is bound to require algebraic inversion because, expectedly, it has to be solved for $\xiy$ for a specified $\pnd$.
Both schemes require the gravitational boundary conditions
\begin{subequations} \label{bc.nd}
  \begin{align}
    \gammay(0) =  &   0  \label{bc.r=0.nd} \\[.5\baselineskip]
    \gammay(1) =  &  -1  \label{bc.r=a.nd}
  \end{align}
\end{subequations}

The M$_{2}$ scheme includes the reduced chemical potential
\begin{equation}\label{rcp.vdw.nd}
   \phi(\thetay=1,\zetay) = - \ln ( \zetay - \betay ) + \frac{\zetay}{\zetay - \betay} - \frac{2\alphay}{\zetay}
\end{equation}
the second-order differential equation for it
\begin{equation}\label{sode.ss.vdw.nd}
    \frac{1}{\etay^{2}}\pd{}{}{\etay}\left( \etay^{2} \pd{}{\phi}{\etay} \right) +\frac{3N}{\zetay} = 0
\end{equation}
and the boundary conditions
\begin{subequations} \label{bc.rcp.nd}
  \begin{align}
    {\left. \pd{}{\phi}{\etay} \right|}_{\etays=0} & =  0   \label{bc.rcp.r=0.nd} \\[.5\baselineskip]
    {\left. \pd{}{\phi}{\etay} \right|}_{\etays=1} & =  -N  \label{bc.rcp.r=a.nd}
  \end{align}
\end{subequations}
obtained, respectively, from \REq{rcp.vdw}, \REq{sode.ss.vdw} and \REqq{bc.rcp}.
The gravitational field
\begin{equation}\label{gf.vdw.nd}
  \gammay = \frac{1}{N} \pd{}{\phi}{\etay}
\end{equation}
follows from \REq{gf.vdw}.

\section{Features of the vdW equation of state and reduced chemical potential\label{fvdw}}
In compliance with thermodynamics practice, we switch from density $\xiy$ to specific volume \itm{\zetay=1/\xiy} and rewrite \REq{vdwse.nd} accordingly
\begin{equation}\label{vdwse.nd.sv}
  \pnd = \frac{\thetay}{\zetay - \betay} - \dfrac{\alphay}{\zetay^{2}}
\end{equation}
Typical isotherms are illustrated in \Rfi{vdwisot} (solid lines).
\begin{figure}[b!]
  \includegraphics[keepaspectratio=true, trim = 9ex 6ex 3ex 6ex , clip , width=\columnwidth]{\figdir/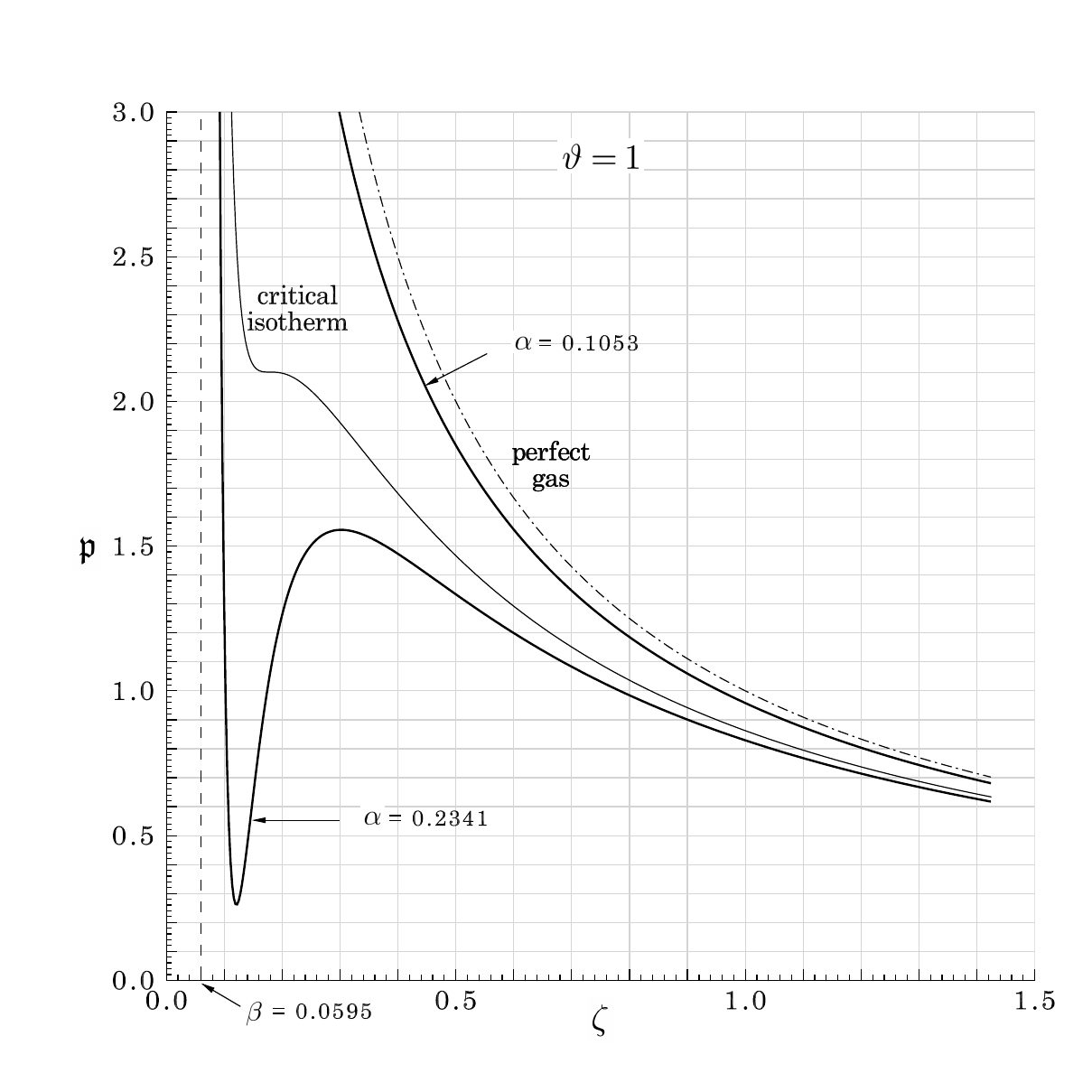} % 9ex 6ex 3ex 6ex
  \caption{Typical isotherms of the vdW model. The perfect-gas isotherm (dash-dot curve) is shown for comparison.\hfill\ }\label{vdwisot}
\end{figure}
A first concern to take care of in connection with \REq{vdwse.nd.sv} is that it can produce specific-volume intervals of unphysical negative pressures.
Their detection follows from imposing the mathematical vanishing of \REq{vdwse.nd.sv} and by solving the corresponding second-order algebraic equation for the specific volume; that sequence of steps determines the specific-volume interval
\begin{equation}\label{zeta.np}
   \zetay_{1,2} = \frac{\alphay}{2\thetay} \left( 1 \pm \sqrt{1-\frac{4\betay\thetay}{\alphay}} \right)
\end{equation}
\begin{subequations}\label{tic}
    The condition
    \begin{equation}\label{tic.eq1}
      \frac{4\betay\thetay}{\alphay} = 1
    \end{equation}
    identifies the tangent isotherm, the one whose minimum is tangent to the specific-volume axis.
    Isotherms above the tangent one have
    \begin{equation}\label{tic.gt1}
      \frac{4\betay\thetay}{\alphay} > 1
    \end{equation}
    and are not affected by any negative-pressure concern; on the other hand, isotherms below imply
    \begin{equation}\label{tic.lt1}
      \frac{4\betay\thetay}{\alphay} < 1
    \end{equation}
    and require some attention when operated with because the solutions provided by \REq{zeta.np} are real and distinct.
\end{subequations}

Another important facet regards the analysis of the critical point, the locus at which minimum, inflexion point and maximum coalesce and that constitutes the terminus of the liquid-gas phase-equilibrium line.
The corresponding critical values of pressure, temperature and specific volume are obtained from the standard conditions
\begin{subequations}\label{cc}
    \begin{align}
       \pc                      & =  \frac{\tc} {\svc - \betay}       - \dfrac{ \alphay}{\svc^{2}}     \label{cc.p}  \\[.5\baselineskip]
       \pds{} {\pnd}{\zetay}{c} & = -\frac{\tc} {(\svc - \betay)^{2}} + \dfrac{2\alphay}{\svc^{3}} = 0 \label{cc.p1} \\[.5\baselineskip]
       \pds{2}{\pnd}{\zetay}{c} & =  \frac{2\tc}{(\svc - \betay)^{3}} - \dfrac{6\alphay}{\svc^{4}} = 0 \label{cc.p2}
    \end{align}
\end{subequations}
that yield the solutions
\begin{subequations}\label{ctp}
    \begin{align}
       \pc  & = \frac{\alphay} {27\,\betay^{2}}  \label{ctp.p}  \\[.5\baselineskip]
       \tc  & = \frac{8\alphay}{27\,\betay}      \label{ctp.t}  \\[.5\baselineskip]
       \svc & = 3\,\betay                        \label{ctp.sv}
    \end{align}
The critical density follows from \REq{ctp.sv}
\begin{equation}\label{ctp.d}
  \dc = \frac{1}{\svc} = \frac{1}{3\betay}
\end{equation}
and the compressibility ratio takes the well known value
\begin{equation}\label{ctp.z}
  \Zc = \frac{\pc\,\svc}{\tc} = \frac{3}{8}
\end{equation}
independent from the characteristic numbers \itm{\alphay,\betay}.
\end{subequations}
\REqb{ctp.t} is the selector between absence or presence of phase equilibrium
\begin{equation}\label{pes}
  \tc = \frac{T_{c}}{\oT} = \frac{8\alphay}{27\betay} \rightarrow
  \begin{cases}
    < 1     \qquad \text{certain absence}   \\[1ex]
    \ge 1   \qquad \text{possible presence}
  \end{cases}
\end{equation}
The top condition in \REq{pes} is necessary and sufficient (upper solid-line isotherm in \Rfi{vdwisot}).
The bottom condition, instead, is necessary but not sufficient (lower solid-line isotherm in \Rfi{vdwisot}); it merely indicates possibility of phase-equilibrium occurrence but the responsibility for its happening falls on the gravitational number, as we shall soon see.
When the equality prevails, the gas sphere is thermodynamically critical.

We have considered helium as candidate gas for the purpose of selecting physically meaningful values of the characteristic numbers \itm{\alphay, \betay}.
The relevant data are listed in \Rta{cnc}.
\begin{table}[htb]
  \caption{Gas data for the selection of $\protect\alphay,\protect\betay$. The vdW \textit{molar} constants $a,b$ reflect notation and units in \cite{rw1978}.\hfill\ \label{cnc}} \vspace*{-2ex}
  \includegraphics[keepaspectratio=true, trim= 0ex 0ex 0ex 0ex , clip , width=\columnwidth]{\figdir/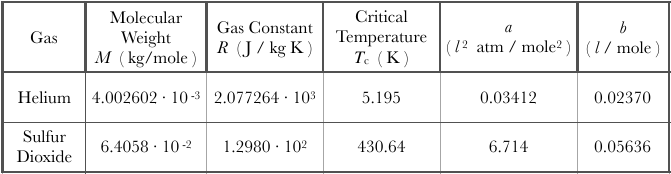}
\end{table}
We took molecular-weight data from NIST's periodic table of elements and calculated the gas constants from the standard definition
\begin{equation}\label{gc}
  R = \frac{\Rugc}{M}
\end{equation}
based on the universal gas constant (\itm{\Rugc=8.31446261815324} J/K mole).
The critical temperatures are tabulated at pages 4-117 (helium) and 6-93 (sulfur dioxide) of \cite{wh2016}.
The vdW \textit{molar} constants for gases are extracted from the table at page D-178 of \cite{rw1978}; the constants $a,b$, in the original notation of the source, are not given in SI units and, therefore, require conversion.
After that, finally, the constants $A,B$ appearing in \REq{vdwse} are obtained as
\begin{subequations}\label{vdwgc}
  \begin{align}
     A = & \frac{a}{M^{2}} = 2.158\cdot\,10^{2}  \quad \text{J\,m$^{3}$/kg$^{2}$} \label{vdwgc.A} \\
     B = & \frac{b}{M}     = 5.921\cdot\,10^{-3} \quad \text{m$^{3}$/kg}          \label{vdwgc.B}
  \end{align}
\end{subequations}
Then, we have assumed an average density \itm{\brhoy\simeq 10} kg/m$^{3}$ and obtained (a rounded off)
\begin{equation}\label{cn.b.he}
    \betay =  0.0595
\end{equation}
from \REq{cn.b}; this value of $\betay$ has been systematically used in all numerical calculations.
Of course, we are interested in considering both the conditions in \REq{pes}; thus,
a temperature \itm{\oT \simeq 10\,}K, a few degrees above the critical temperature \itm{\Tc = 5.195\,}K, makes true the top condition and determines
\begin{equation}\label{cn.a.he.ca}
    \alphay =  0.1053
\end{equation}
while a temperature $\oT \simeq 4.5\,$K slightly lower than the critical temperature makes true the bottom condition and provides
\begin{equation}\label{cn.a.he.pp}
    \alphay =  0.2341
\end{equation}
In the latter case, we had to go not too low below the critical temperature in order to stay above the tangent isotherm [\REq{tic.eq1} with \itm{\thetay=1}].
The corresponding isotherms are shown in \Rfi{vdwisot} (solid bold lines).
We wish to emphasise \textit{strongly} that we have used helium's data only for the purpose of obtaining physically meaningful values for \itm{\alphay, \betay}; the findings of the forthcoming sections are not restricted to that gas but are applicable to any other gas for which \itm{\alphay, \betay} assume the same values.
Take, for example, sulfur dioxide, a gas whose thermodynamic data are substantially different from those of helium;
it is very easy to verify with the aid of the data provided in \Rta{cnc} that, with average density \itm{\brhoy\simeq 67.6} kg/m$^{3}$ and temperatures \itm{\oT \simeq 820\,}K and \itm{\oT \simeq 369\,}K, its \itm{\alphay, \betay} attain the same values indicated in \REqs{cn.b.he}{cn.a.he.pp}.
In other words, the nondimensional behaviour of two so different gases is exactly the same.

%The reduced chemical potential [\REq{rcp.vdw.nd}] is the fundamental unknown in the M$_{2}$ scheme.
The curves of the reduced chemical potential [\REq{rcp.vdw.nd}], the fundamental unknown in the M$_{2}$ scheme, are illustrated in \Rfi{vdwrcp} and are very similar to the isotherms of \Rfi{vdwisot}.
\begin{figure}[b!]
  \includegraphics[keepaspectratio=true, trim= 9ex 6ex 3ex 6ex , clip , width=\columnwidth]{\figdir/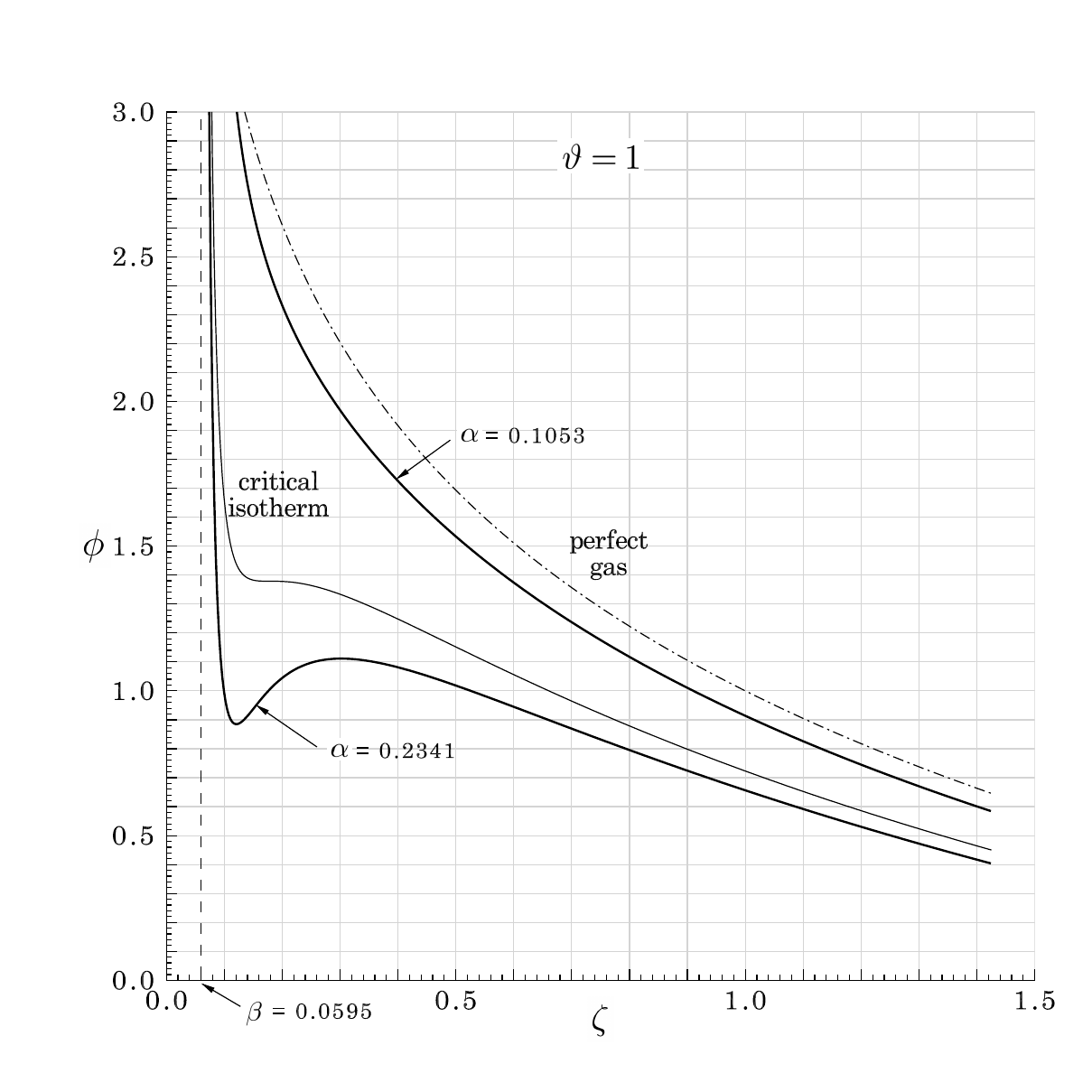}  % 9ex 6ex 3ex 6ex
  \caption{Reduced chemical potential of the vdW model. The perfect-gas curve (dash-dot) is shown for comparison.\hfill\ }\label{vdwrcp}
\end{figure}
Obviously, they possess same characteristics; for example, under critical conditions [\REqd{ctp.t}{ctp.sv}], minimum, inflexion point and maximum of the reduced chemical potential also coalesce
\begin{equation}\label{cc.rcp12}
  \pds{} {\phic}{\zetay}{c} = \pds{2}{\phic}{\zetay}{c} = 0
\end{equation}
into one single point at which the reduced chemical potential attains the critical value
\begin{equation}\label{cc.rcp0}
  \phic = - \ln \left( 2\,\betay \right) - \dfrac{3}{4}
\end{equation}

A final remark regarding nondimensional variables.
The attentive reader will not have missed that our choice of how making nondimensional the thermodynamic properties [\REq{ndv}] does not comply with the conventional one based on referring them to critical values for the purpose of obtaining and dealing with a nondimensional equation of state involving \textit{only} numerical constants.
As a matter of fact, we could have complied but the conventional manner is not convenient in our case.
Indeed, we probed the possibility but found two good reasons why the manner we opted for is preferable.
First, results' comparison between \ma\ and this study are direct and straightforward; there is no need for variable conversions and related risks of mistakes.
Second and more importantly, this problem is controlled by three physical characteristic numbers and there is no amount of algebraic funambulism that can make disappear two of them.
By following the conventional manner, \itm{\alphay, \betay} would certainly evanesce from the corresponding nondimensional equation of state but they would reemerge in the corresponding nondimensional differential equations analogous to, for example, \REqd{meq.s.nd}{gfd.s.nd} of the P scheme.
Unfortunately, their reemergence takes place in a form that makes the differential equations singular in the case of perfect gas (\itm{\alphay, \betay \rightarrow 0}) to which the \textit{critical-point} concept is completely foreign.
Our choice of nondimensional variables avoids all that cumbersomeness and presupposes a convenient share of responsibilities:
the gravitational number $N$ predominates the gravito-fluid statics while thermodynamics is competence of \itm{\alphay, \betay} \footnote{The presence of the thermodynamic derivative $\pdst{}{\pnd}{\protect\xiyf}{\protect\thetaysf=1}$ in \REq{meq.s.nd} slightly compromises this share in the P scheme.}.
%\footnote{The presence of the thermodynamic derivative $\pdst{}{\pnd}{\xiyf}{\thetaysf=1}$ in \REq{meq.s.nd} slightly compromises this share in the P scheme.}

\section{Liquid-gas phase equilibrium\label{lgpe}}
The thermodynamic analysis of liquid-gas phase equilibrium is disseminated throughout a vast majority of thermodynamics textbooks.
%We have followed the guidelines of the didactically remarkable exposition of Planck \cite{mp1903} based on entropy maximisation.
The material utilised in this section has been adapted from the contents of \cite{dg2001aiaa, dg2002jtht} in which the guidelines of the didactically remarkable exposition of Planck \cite{mp1903} based on entropy maximisation were followed; Callen's and Tisza's textbooks \cite{hc1963,lt1966,hc1985} have also been of great aid.

The equations that govern the phase equilibrium express the continuity of temperature, pressure and chemical potential
\begin{subequations}\label{c.tpmu}
    \begin{align}
       \Tl           & = \Tg           \label{c.tpmu.t}  \\[.5\baselineskip]
       p(\Tl,\vl)    & = p(\Tg,\vg)    \label{c.tpmu.p}  \\[.5\baselineskip]
       \muy(\Tl,\vl) & = \muy(\Tg,\vg) \label{c.tpmu.cp}
    \end{align}
\end{subequations}
across interfaces separating the two phases.
The subscripts \itm{l,g} in \REqq{c.tpmu} refer to liquid and gas, respectively.
The temperature continuity [\REq{c.tpmu.t}] fits nicely into the reasoning about the solution of the energy equation exposed in \ma, near \Reqma{5}, and with \REq{Tf}; in other words, the presence of the interface does not perturb the uniformity of the temperature field and we are certainly authorised to complete \REq{c.tpmu.t} with
\begin{equation}\label{c.tpmu.t.c}
  \Tl = \Tg = \oT
\end{equation}
The continuity of pressure and chemical potential [\REqd{c.tpmu.p}{c.tpmu.cp}] constitute an algebraic system for the unknown specific volumes \itm{\vl,\vg} of saturated liquid and gas, respectively.
With the specific-volume solution in hand, the saturation values of pressure and chemical potential follow from
\begin{subequations}\label{sat.pcp}
    \begin{align}
       p^{\ast}    & = p(\oT,\vl)    = p(\oT,\vg)    \label{sat.p}   \\[.5\baselineskip]
       \muy^{\ast} & = \muy(\oT,\vl) = \muy(\oT,\vg) \label{sat.cp}
    \end{align}
\end{subequations}
Of course, both the algebraic system formed by \REqd{c.tpmu.p}{c.tpmu.cp} and the saturation-value equations [\REqq{sat.pcp}] become operational only upon explicit assignment of the thermodynamic model, van der Waals' in our case [\REq{vdwse} with \itm{\rhoy=1/v}; \REqd{cp.pgvdw}{rcp.vdw}].
Thus, and naturally after switching to nondimensional language, \REqs{c.tpmu.p}{c.tpmu.t.c} turn into the explicit nonlinear system
\begin{subequations}\label{c.tpmu.vdw}
    \begin{align}
       \thetay_{l}                                                      & = \thetay_{g} = 1                                                    \label{c.tpmu.vdw.t}  \\[.5\baselineskip]
       \frac{1}{\zetay_{l} - \betay} - \dfrac{\alphay}{\zetay_{l}^{2}}  & = \frac{1}{\zetay_{g} - \betay} - \dfrac{\alphay}{\zetay_{g}^{2}}    \label{c.tpmu.vdw.p}  \\[.5\baselineskip]
      - \ln ( \zetay_{l} - \betay ) + \frac{\zetay_{l}}{\zetay_{l} - \betay} - \frac{2\alphay}{\zetay_{l}}
                                                                        & =
                         - \ln ( \zetay_{g} - \betay ) + \frac{\zetay_{g}}{\zetay_{g} - \betay} - \frac{2\alphay}{\zetay_{g}}                  \label{c.tpmu.vdw.cp}
    \end{align}
\end{subequations}
whose solution can be visualised in the neat graphical way proposed by Callen \cite{hc1963,hc1985}. % and illustrated in \Rfi{phaseeq}.
It consists in superposing the graph of the reduced chemical potential versus pressure \footnote{A convenient conceptual switch to the interpretation of the reduced chemical potential as Gibbs potential; see \cite{Note4}.} on the graph of the isotherm, as in the rightmost side of \Rfi{phaseeq}. % Note4 corresponds to \ref{gibbs}
\begin{figure}[t]
  \includegraphics[keepaspectratio=true, trim= 9ex 6ex 3ex 6ex , clip , width=\columnwidth]{\figdir/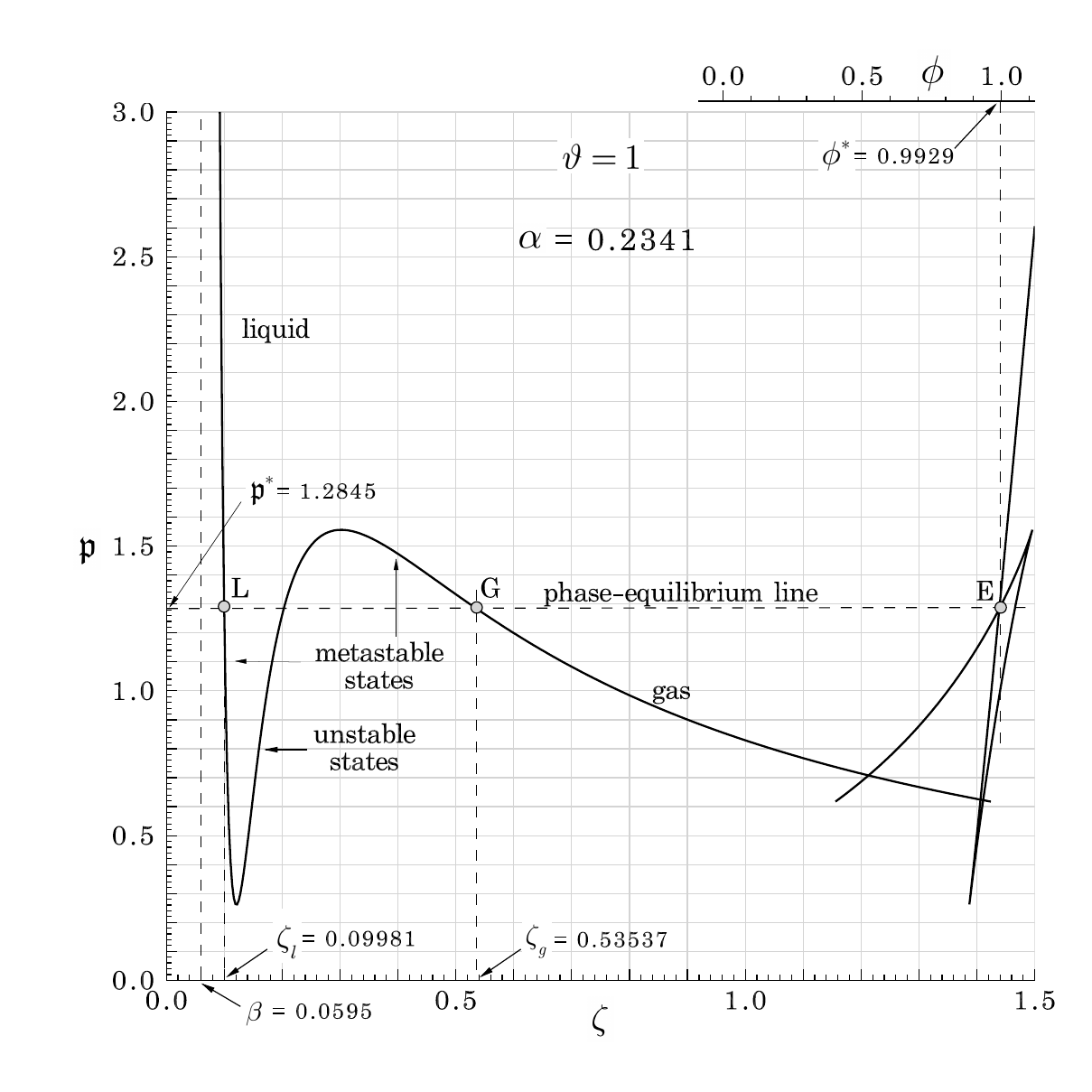}  % 9ex 6ex 3ex 6ex
  \caption{Graphical visualisation of the solution to \REqd{c.tpmu.vdw.p}{c.tpmu.vdw.cp} governing the liquid-gas phase equilibrium.\hfill\ }\label{phaseeq}
\end{figure}
The intersection E determined by the closed loop visualises the equality in \REq{c.tpmu.vdw.cp}.
The vertical line through E fixes the saturation value of the reduced chemical potential; the horizontal line through E pinpoints the saturation pressure and its intersections with the isotherm identify the phase-equilibrium points L,G that share that pressure [equality in \REq{c.tpmu.vdw.p}] together with the corresponding specific-volume saturation values.
The solution of \REqd{c.tpmu.vdw.p}{c.tpmu.vdw.cp} requires, obviously, a numerical calculation whose results are displayed in \Rfi{phaseeq} and collected in \Rta{phaseeqv} with the inclusion of the saturation densities of liquid and gas.
{The algorithm we use is based on a Netwon-Raphson iterative scheme whose choice of the proper initial-guess values of \itm{\zetay_{l},\zetay_{g}} turns out to be a critical issue since, in addition to the \textit{physical} solution of our interest (points E, L, G in \Rfi{phaseeq}), \REqd{c.tpmu.vdw.p}{c.tpmu.vdw.cp} admit an infinitude of \textit{mathematical} solutions, say \itm{\zetay_{l}=\zetay_{g}=n} with $n$ being any arbitrary numerical constant.
In this regard, we found out that the intersection of the two straight lines tangent to the curve in the cuspidal points is an accurate approximation to the point $E$ in \Rfi{phaseeq}.}
\begin{table}[htb]
  \caption{Saturation values of specific volume, density, pressure and reduced chemical potential corresponding to $\protect\alphay=0.2341, \protect\betay=0.0595$.\hfill\ \label{phaseeqv}} \vspace*{-2ex}
  \includegraphics[keepaspectratio=true, trim= 0ex 0ex 0ex 0ex , clip , width=\columnwidth]{\figdir/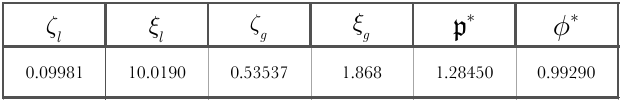}
\end{table}
Finally, we wish to point out a remarkable property that emerges from \REqq{c.tpmu.vdw}: they clearly indicate that the phase equilibrium is controlled exclusively by characteristic numbers \itm{\alphay, \betay} and turns out to be completely independent from the gravitational number.
\begin{figure}[t]
  \includegraphics[keepaspectratio=true, trim = 9ex 5ex 3ex 6ex , clip , width=\columnwidth]{\figdir/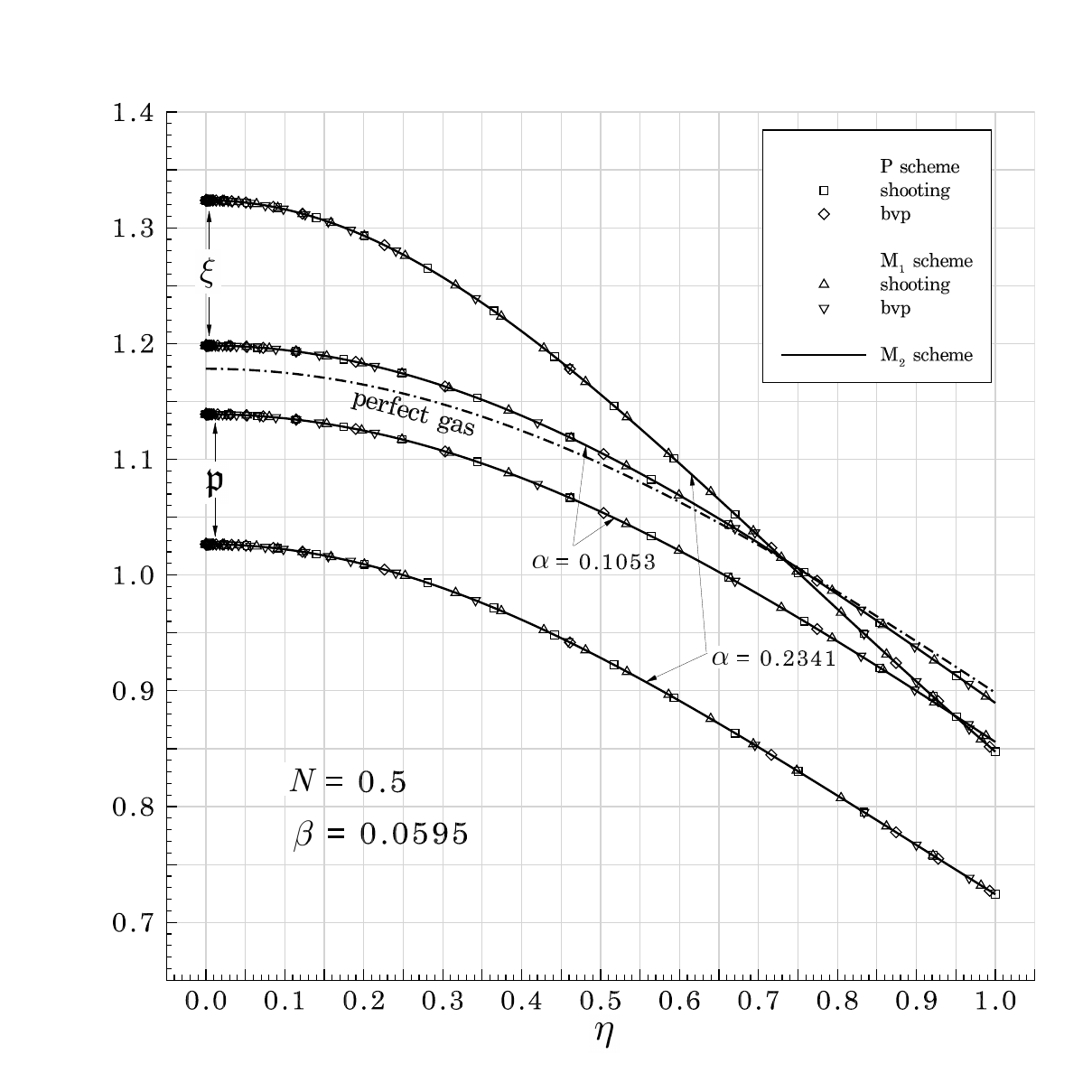}   % 9ex 6ex 3ex 6ex
  \caption{Radial profiles of density and pressure at \itm{N=0.5} for the vdW model; the perfect-gas profile (dash-dot curve) is included for comparison.\hfill\ }\label{N=0.5rpd}
\end{figure}
\begin{figure}[b!]
  \includegraphics[keepaspectratio=true, trim= 8ex 5ex 3ex 6ex , clip , width=\columnwidth]{\figdir/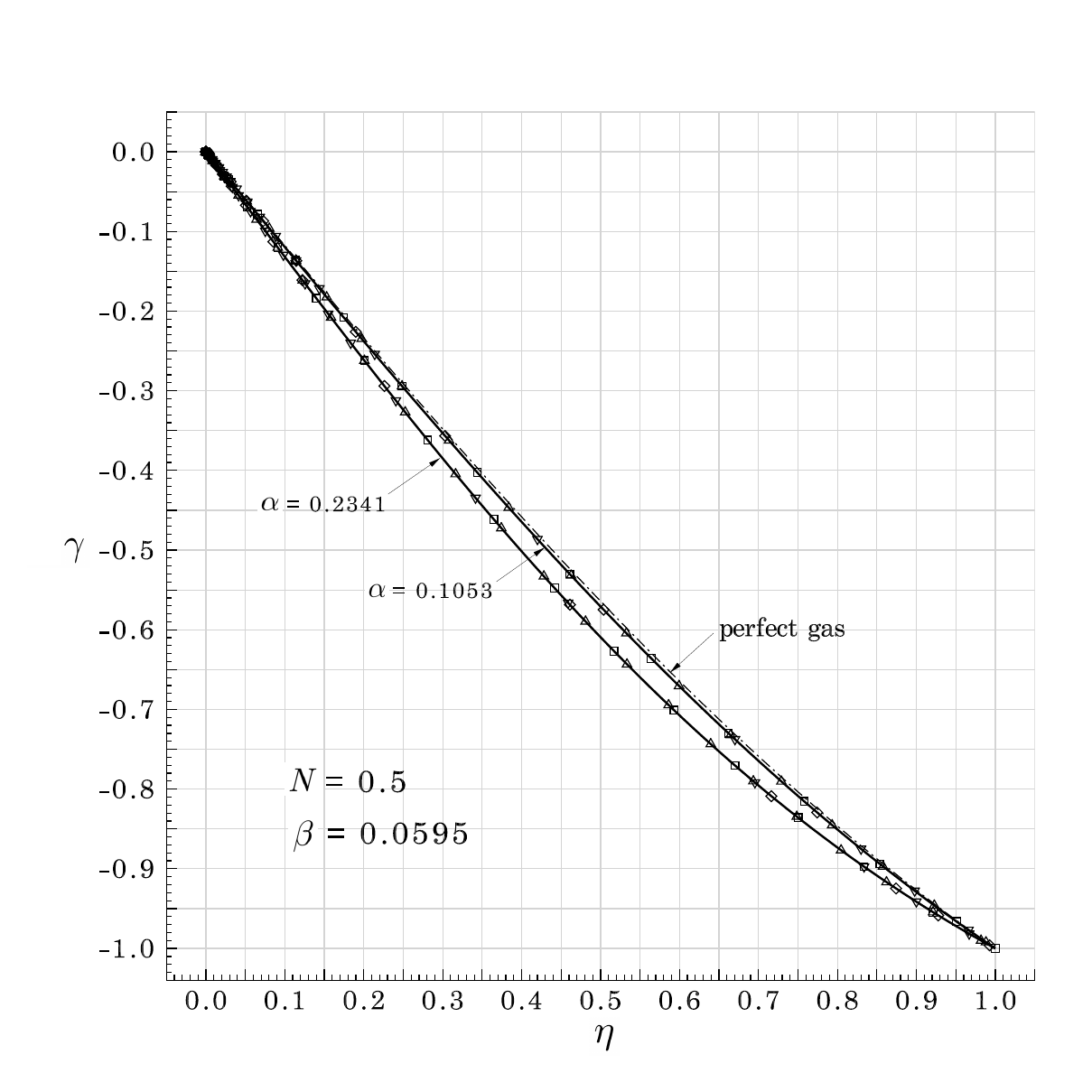}
  \caption{Radial profiles of the gravitational field at \itm{N=0.5} for the vdW model; the perfect-gas profile (dash-dot curve) is included for comparison.\hfill\ }\label{N=0.5rgf}
\end{figure}

\section{Numerical solution and results\label{nsr}}
\subsection{Preliminary considerations}
The nondimensional computational schemes P, M$_{1}$ and M$_{2}$ formulated at the end of \Rse{ndf} require a numerical calculation.
For that purpose, we have defined a few test cases corresponding to selected significative values of the gravitational number, more or less in accordance with the choice we made in \ma. %\ma; they are listed in \Rta{tct}.
%\begin{table}[h]
%  \centering\includegraphics[keepaspectratio=true, trim= 0ex 0ex 0ex 0ex , clip , height=.8\columnwidth]{\figdir/tct.pdf}
%  \caption{Computational test cases.}\label{tct}
%\end{table}
We have applied all schemes to calculate each case; the different and independent algorithms, five in total taking into account the variants ``shooting'' and ``boundary-value problem'' (bvp) for the first-order schemes P and M$_{1}$, produce extremely concordant results, as shown in the illustrative example of \Rfid{N=0.5rpd}{N=0.5rgf} in which the comparative graphs of radial profiles of density, pressure and gravitational field are displayed for $N=0.5$; we are, therefore, very confident about the capability of our codes.
By convention, the legend of \Rfi{N=0.5rpd} applies by default to subsequent figures; exceptions are explicitly indicated. % for \Rfid{xi1}{xi0}.
%\begin{figure}[ht]
%  \includegraphics[keepaspectratio=true, trim= 8ex 5ex 3ex 6ex , clip , width=\columnwidth]{\figdir/N=0.5rgf.pdf}
%  \caption{Radial profiles of the gravitational field at $N=0.5$ for the vdW model; the perfect-gas profile (dash-dot curve) is included for comparison.\hfill\ }\label{N=0.5rgf}
%\end{figure}

%\input{nm-pa}
{\color{black}%% pierluigi
The first-order boundary-value problems of the P and M$_{1}$ schemes [\REqd{meq.s.nd}{gfd.s.nd} or \REqd{gfd.s.nd}{nmeq.s.nd}, with \REqq{bc.nd}] are dealt with differently according to two variants.

In the ``shooting'' approach, they are converted to initial value problems, solved specifically by the MATLAB function \textsc{ode45}, in which an initial condition for either $\xiy(0)$ or $\pnd(0)$, according to the scheme, is introduced to start the iterative process, based on the MATLAB's root-finding function {\textsc{fzero}}, that proceeds until the peripheral boundary condition [\REq{bc.r=a.nd}], kept as target, is met.

In the ``bvp'' approach, they are solved directly by the three independent codes \textsc{bvp5c} \cite{bvp5c}, \textsc{twpbvp} \cite{Cash2013} and \textsc{tom}  \cite{Mazzia20061954,Mazzia2022555}, available in MATLAB, to check that all codes converge to same solutions.

The second-order boundary value problem of the M$_{2}$ scheme [\REq{sode.ss.vdw.nd} with \REqq{bc.rcp.nd}] is solved by the MATLAB code \textsc{HOFiD\_bvp} \cite{amse1,amse2}, currently still in a prototype state.
The mathematical problem is transformed from continuous to discrete by discretizing the derivatives separately through high-order finite-difference schemes that provides the solution in $[0,1]$ with the desired accuracy when solved by the Newton-Raphson method.
}

We discuss separately the cases above [\REq{pes} top; \itm{\alphay=0.1053}] and below [\REq{pes} bottom; \itm{\alphay=0.2341}] the critical isotherm in \Rsed{ape}{ppe}, respectively.

\subsection{Fluid-sphere temperature above critical temperature\label{ape}}
%\textit{sphere \textnormal{$\thetay_{c}$} above critical temperature}
%\subsubsection{Analytical solution for $N\rightarrow0$\label{as.N=0}}
\subsubsection{Vanishingly small gravitational number\label{as.N=0}}
A fluid-sphere temperature above the critical temperature makes sure that all the fluid is in gas phase.
In this situation, the first logical step consists in studying the limiting case of vanishingly small gravitational number, for which we expect to find a uniform-density isothermal sphere with linear profile of the gravitational field.
Indeed, there exists an analytical solution corresponding to such a situation.
If \itm{N \rightarrow 0} then \REq{sode.ss.vdw.nd} assumes the simplified form
\begin{equation}\label{sode.ss.vdw.nd.N=0}
    \frac{1}{\etay^{2}}\pd{}{}{\etay}\left( \etay^{2} \pd{}{\phi}{\etay} \right) \simeq 0
\end{equation}
that, together with the simplified peripheral boundary condition [\REq{bc.rcp.r=a.nd}]
\begin{equation}\label{bc.rcp.r=a.nd.N=0}
  {\left. \pd{}{\phi}{\etay} \right|}_{\etays=1}  \simeq  0
\end{equation}
enforces the radial uniformity of the reduced chemical potential
\begin{equation}\label{rcp.N=0}
   \phi \simeq K
\end{equation}
In turn, \REq{rcp.vdw.nd} transmits the same uniformity to specific volume and density
\begin{equation}\label{svxi.N=0}
   \zetay = \frac{1}{\xiy} \simeq K'
\end{equation}
%and density
%\begin{equation}\label{xi.N=0}
%   \xiy \simeq \frac{1}{K'}
%\end{equation}
Now, the mass constraint expressed by \Rma{Eq. (13a)} (with the change of notation \itm{m_{g}\rightarrow m_{f}}) and its nondimensional form given by the normalisation condition indicated in \Rma{Eq. (43)} are still applicable.
The substitution of \REq{svxi.N=0} into \Rma{Eq. (43)} and the execution of the integral fix the value of the constant to \itm{K'=1}. %to be
%\begin{equation}\label{K.N=0}
%  K' \simeq 1
%\end{equation}
Thus, we obtain the expected uniform solution
\begin{equation}\label{xisv.N=0}
  \xiy = \frac{1}{\zetay} \simeq 1
\end{equation}
obviously independent from $\alphay$ and $\betay$, with uniform reduced chemical potential
\begin{equation}\label{rcp.vdw.nd.N=0}
   \phi \simeq - \ln ( 1 - \betay ) + \frac{1}{1 - \betay} - 2\alphay
\end{equation}
and pressure
\begin{equation}\label{ndap.N=0}
  \pnd \simeq \frac{1}{1 - \betay} - \alphay = \overline{\pnd}      %= \pnd_{\,\xi=1}
\end{equation}
The gravitational field follows from the integration of the simplified \REq{gfd.s.nd}
\begin{equation} \label{gfd.s.nd.N=0}
   \pd{}{\gammay}{\etay} + \frac{2}{\etay} \gammay  + 3 \simeq 0
\end{equation}
that can be contracted
\begin{equation}\label{gfd.s.nd.N=0.c}
    \frac{1}{\etay^{2}}\pd{}{}{\etay}\left( \etay^{2} \gammay \right) + 3 \simeq 0
\end{equation}
and easily integrated
\begin{equation}\label{gfd.s.nd.N=0.c.i}
    \etay^{2} \gammay + \etay^{3} \simeq K''
\end{equation}
The peripheral boundary condition [\REq{bc.r=a.nd}] requires the vanishing of the integration constant $K''$ and establishes the linear profile
\begin{equation}\label{gfd.s.nd.N=0.c.i.l}
    \gammay \simeq - \etay
\end{equation}
of the gravitational field.

This analytical solution represents a benchmark to test the numerical algorithms.
For that purpose, similarly to what we did in \ma, we began the series of calculations with \itm{N=10^{-5}} and indeed we found the analytical profiles of \REq{xisv.N=0}, \REq{ndap.N=0} and \REq{gfd.s.nd.N=0.c.i.l} reliably reproduced numerically, as evidenced in \Rfi{N=0-3rd}, \Rfi{N=0-3rp} and \Rfi{N=0-3rgf}, respectively.
\begin{figure}[b!]
  \includegraphics[keepaspectratio=true, trim= 9ex 5ex 1.5ex 6ex , clip , width=\columnwidth]{\figdir/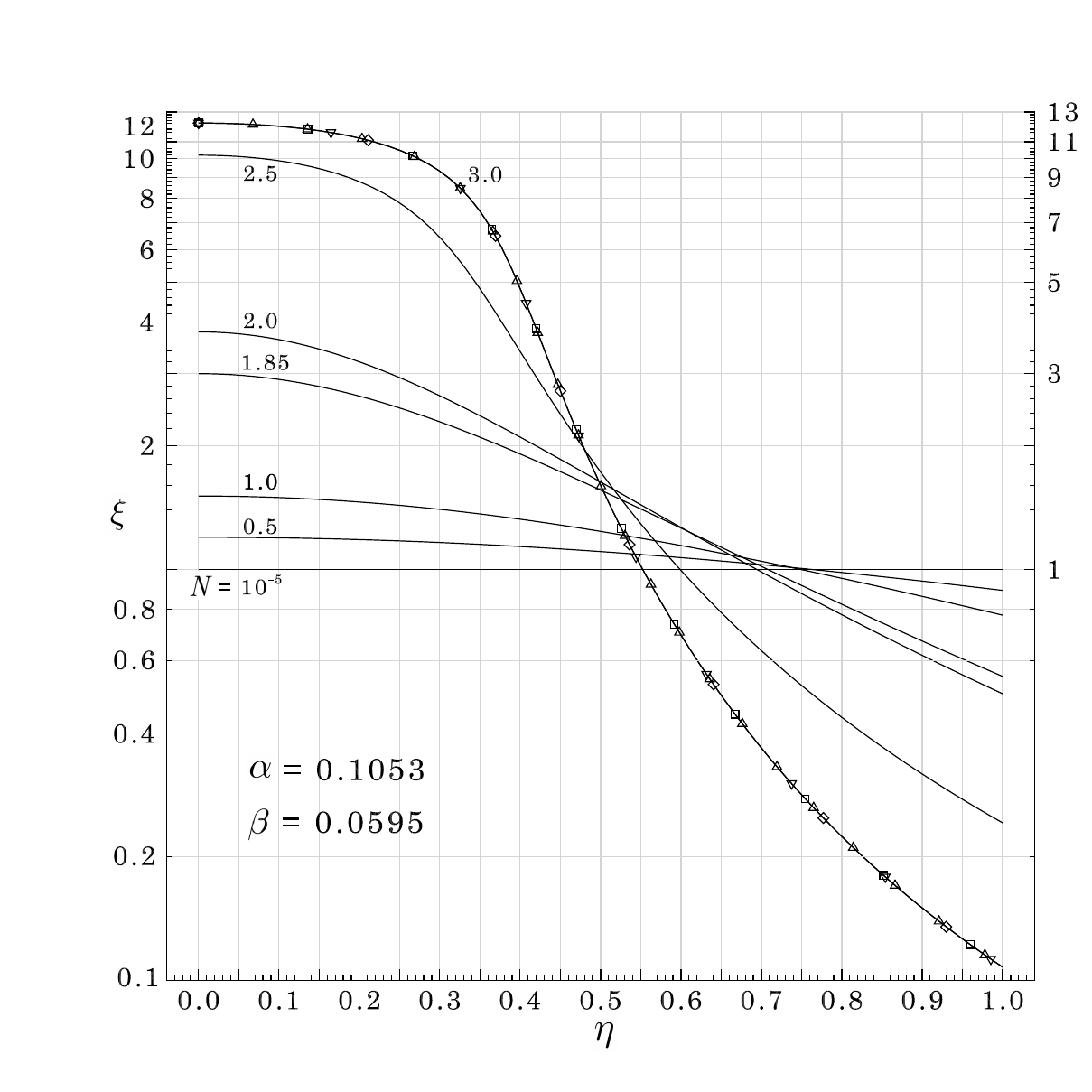} % 9ex 6ex 3ex 6ex
  \caption{Radial profiles of density at selected values of the gravitational number for the vdW model.\hfill\ }\label{N=0-3rd}
\end{figure}
\begin{figure}[t]
  \includegraphics[keepaspectratio=true, trim= 9ex 5ex 1.5ex 6ex , clip , width=\columnwidth]{\figdir/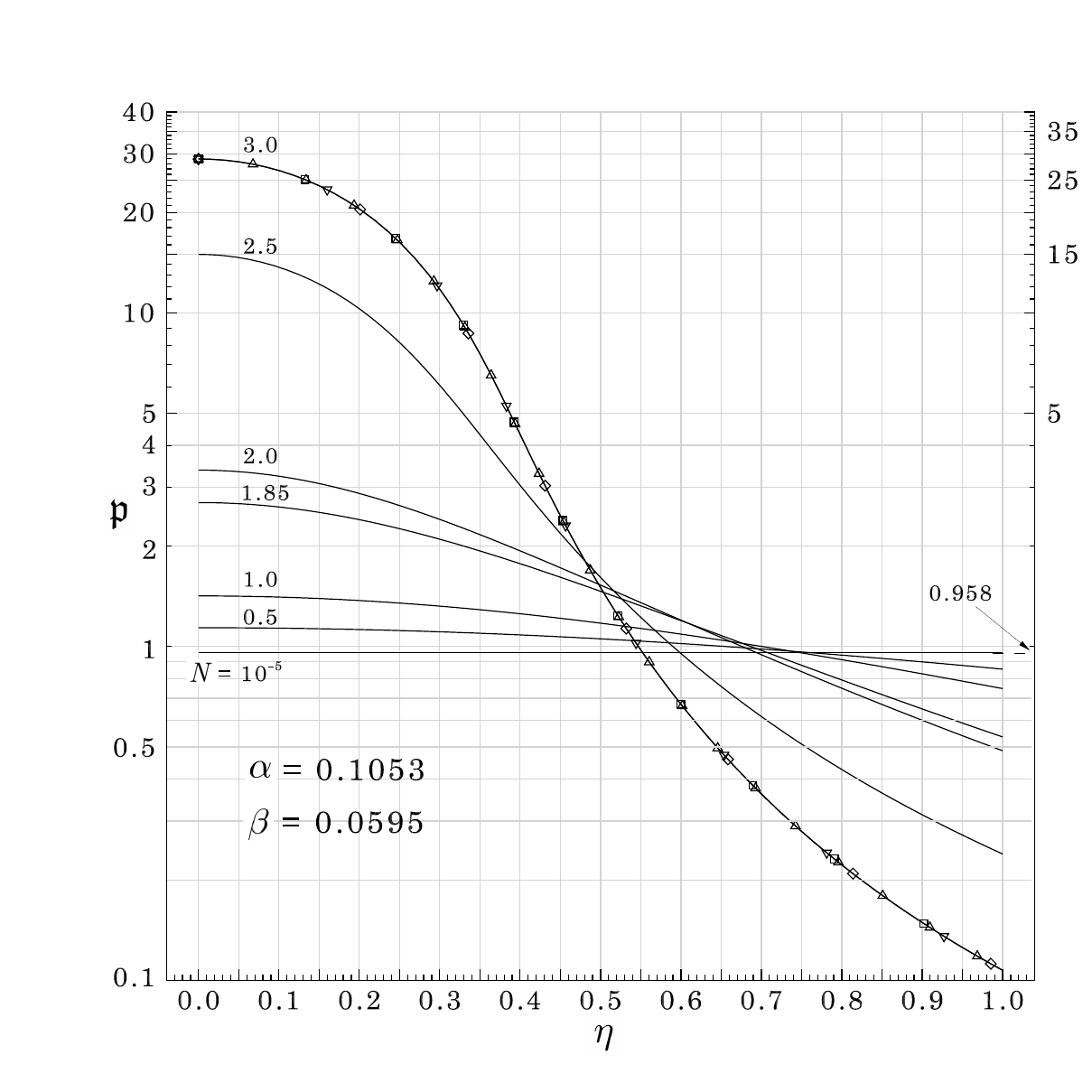}
  \caption{Radial profiles of pressure at selected values of the gravitational number for the vdW model.\hfill\ }\label{N=0-3rp}
\end{figure}
\begin{figure}[b!]
  \includegraphics[keepaspectratio=true, trim= 9ex 5ex 1.5ex 6ex , clip , width=\columnwidth]{\figdir/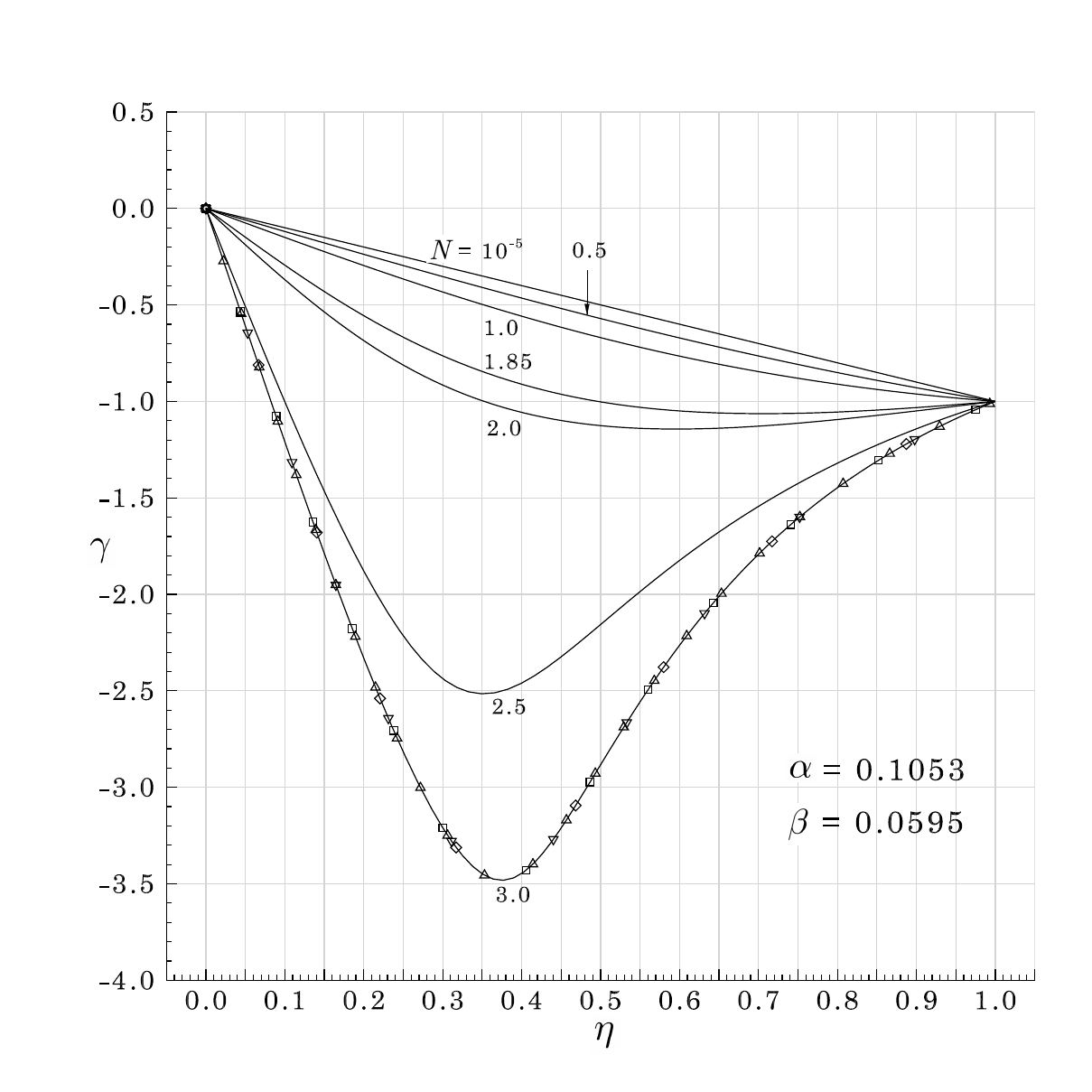}
  \caption{Radial profiles of gravitational field at selected values of the gravitational number for the vdW model.\hfill\ }\label{N=0-3rgf}
\end{figure}

\subsubsection{Finite values of the gravitational number\label{fvgn}}

Following in the footsteps of \ma, we considered next the cases \itm{N=0.5} and \itm{N=1} mainly with the purpose in mind to detect and to put in evidence differences between vdW-model and perfect-gas results without worrying for the existence of multiple solutions, a concern that, we remind the reader, kicks in at \itm{N\simeq1.84} for a perfect gas
[see \Rma{Figs. (5)} and \Rma{(6)}].
%in a zone safely away from the appearance of multiple solutions
We describe only the case \itm{N=0.5}; the case \itm{N=1} is basically similar.
The radial profiles of density, pressure and gravitational field are illustrated in \Rfid{N=0.5rpd}{N=0.5rgf} and echoed in \Rfis{N=0-3rd}{N=0-3rgf}.
The perfect-gas curve in \Rfi{N=0.5rpd} represents both density and pressure because they coincide in their nondimensional form [\Rma{Eqs. (14b)} and \Rma{(47)}]; the density and pressure curves are distinct for the vdW model due to the equation of state [\REq{vdwse.nd.s}].
The density profile acquires steepness and the pressure profile lies lowlier with respect to the perfect gas'.
Thus, the vdW-model sphere is denser in the central region and more rarefied in the peripheral region with an overall less intense pressure distribution, to maintain the fluid- and gravito-static equilibrium, with respect to the perfect-gas sphere.
The gravitational field (\Rfi{N=0.5rgf}) becomes slightly more intense in the case of the vdW model.
These differences grow more prominently with an increase of either the characteristic number $\alphay$ or the gravitational number $N$.
%It is of value to project the profiles of \Rfi{N=0.5rpd}
%In view of forthcoming discussion of the cases with possible presence of phase equilibrium [\Rse{ppe}]
%\begin{figure}[H]
%  \includegraphics[keepaspectratio=true, trim= 9ex 5ex 1.5ex 6ex , clip , width=\columnwidth]{\figdir/N=0-3rp.pdf}
%  \caption{Radial profiles of pressure for selected values of the gravitational number.\hfill\ }\label{N=0-3rp}
%\end{figure}
%\begin{figure}[H]
%  \includegraphics[keepaspectratio=true, trim= 9ex 5ex 1.5ex 6ex , clip , width=\columnwidth]{\figdir/N=0-3rgf.pdf}
%  \caption{Radial profiles of gravitational field for selected values of the gravitational number.\hfill\ }\label{N=0-3rgf}
%\end{figure}

Next, we turned attention to the peculiar existence of multiple solutions in the case of perfect gas.
We considered the cases \itm{N=1.85,\,2,\,2.5}.
With reference to \Rma{Fig. 5}, also reproduced in \Rfi{xi1}, the first value is slightly above the multiple-solution threshold and gives rise to three solutions; the second value corresponds to the center of the spiral characterised by infinite solutions; the third value is slightly below the mysterious upper bound \itm{N_{m}} and produces two solutions.
In the case of the vdW model, the calculations with the five algorithms went smooth and gave systematically unique solutions regardless of the initial-guess conditions assumed to start the numerical machinery.
As illustrated in \Rfis{N=0-3rd}{N=0-3rgf} and in comparison with \Rma{Figs. (2)} and \Rma{(3)} for the cases \itm{N=2,\,2.5}, the density and pressure profiles are steeper and the gravitational-field profiles have deeper minima which imply stronger gravitational attraction consistently with the more pronounced steepness of the density profile.

The smoothness of the calculation at \itm{N=2.5} gave us confidence and determination to attempt leaping over the perfect gas' distrust-inspiring barrier at $N_{m}$.
With $N$ set to 3, the algorithms converged flawlessly \footnote{All algorithms we used in our perfect-gas study \ma\ blew off and failed miserably when we attempted the same leap.} again and we were rewarded with the smooth profiles shown in \Rfis{N=0-3rd}{N=0-3rgf}; for them, we have explicitly superposed the results of all five algorithms to give the reader a visual reassurance that the breaking of the barrier at \itm{N=N_{m}} is a reality consistently confirmed by five differently and independently coded numerical methods.

\begin{figure}[h]
  \includegraphics[keepaspectratio=true, trim = 9ex 5ex 3ex 6ex , clip , width=\columnwidth]{\figdir/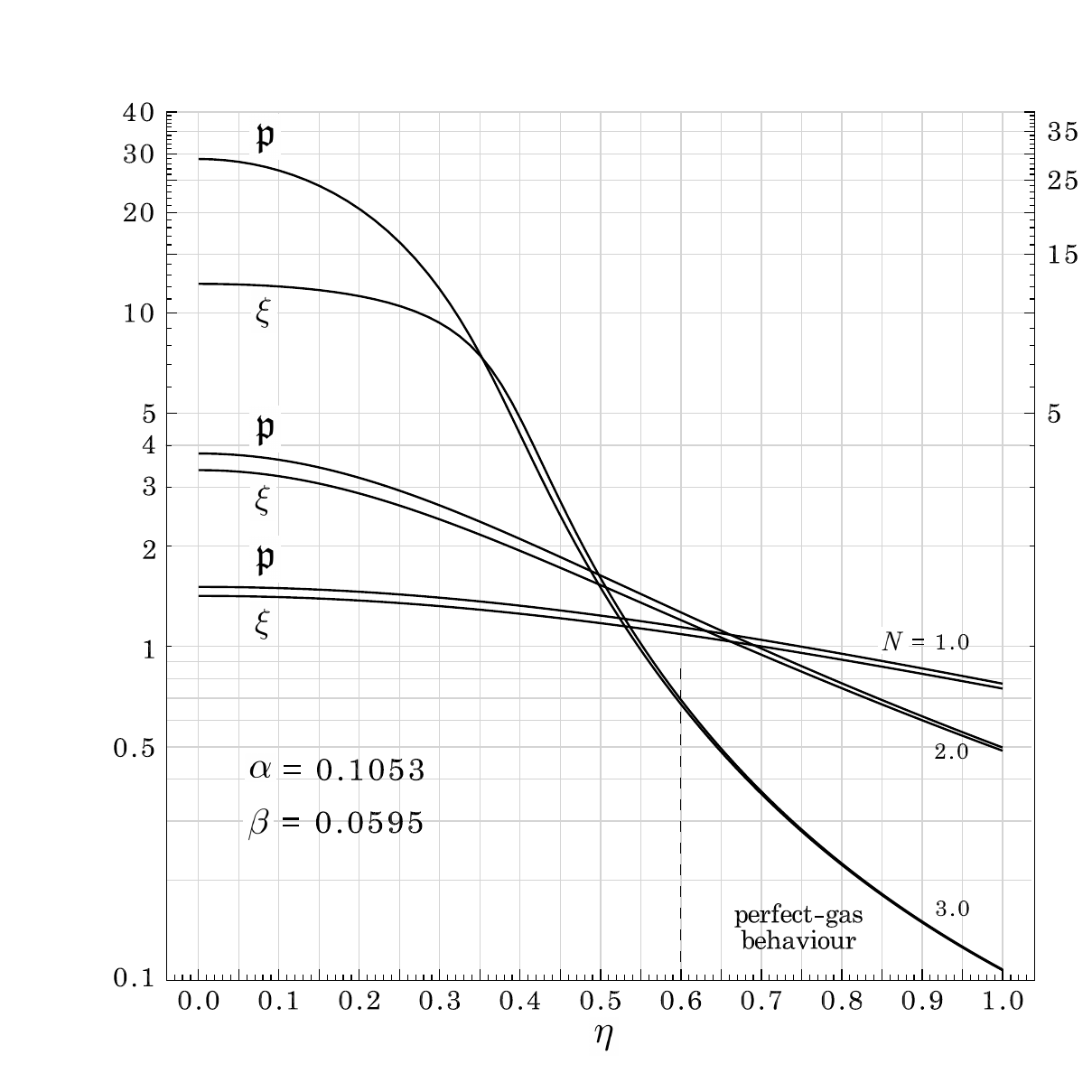}
  \caption{Radial profiles of density and pressure from M$_{2}$ scheme.\lpush}\label{N=3pg-a}
\end{figure}
The superposition of the radial profiles of density and pressure, as shown in \Rfi{N=3pg-a}, gives a visual explanation of the limits restricting the perfect-gas assumption.
We have already touched on this matter when we commented \Rfi{N=0.5rpd}. In general, density and pressure curves are distinct for a given gravitational number; but when the latter increases then the differences tend to vanish in the peripheral zone of the sphere.
Indeed, at \itm{N=3}, the curves are basically coincident, say, for \itm{\etay\ge0.6}; this means that, according to \Rma{Eq. (47)}, perfect-gas behaviour is an accurate approximation in such a layer because the density is sufficiently small so that the terms accounting for molecular attraction \itm{(\alphay\,\xiy^{2})} and size \itm{(\betay\,\xiy)} in the equation of state [\REq{vdwse.nd.s}] can be neglected.
However, density and pressure curves diverge inexorably in the central zone and, by reflection, their divergence implies that the perfect-gas assumption cannot be maintained for all the sphere.

Encouraged by the successful calculation at \itm{N=3}, we repeated the exercise of producing the graphs of peripheral and central densities as functions of the gravitational number as we did in \Rma{Figs. 5} and \Rma{6}; the updated graphs are displayed in \Rfid{xi1}{xi0}.
\begin{figure}[t]   %2ex 5ex 8ex 10ex
  \includegraphics[keepaspectratio=true, trim= 6ex 4ex 3ex 6ex , clip , width=\columnwidth]{\figdir/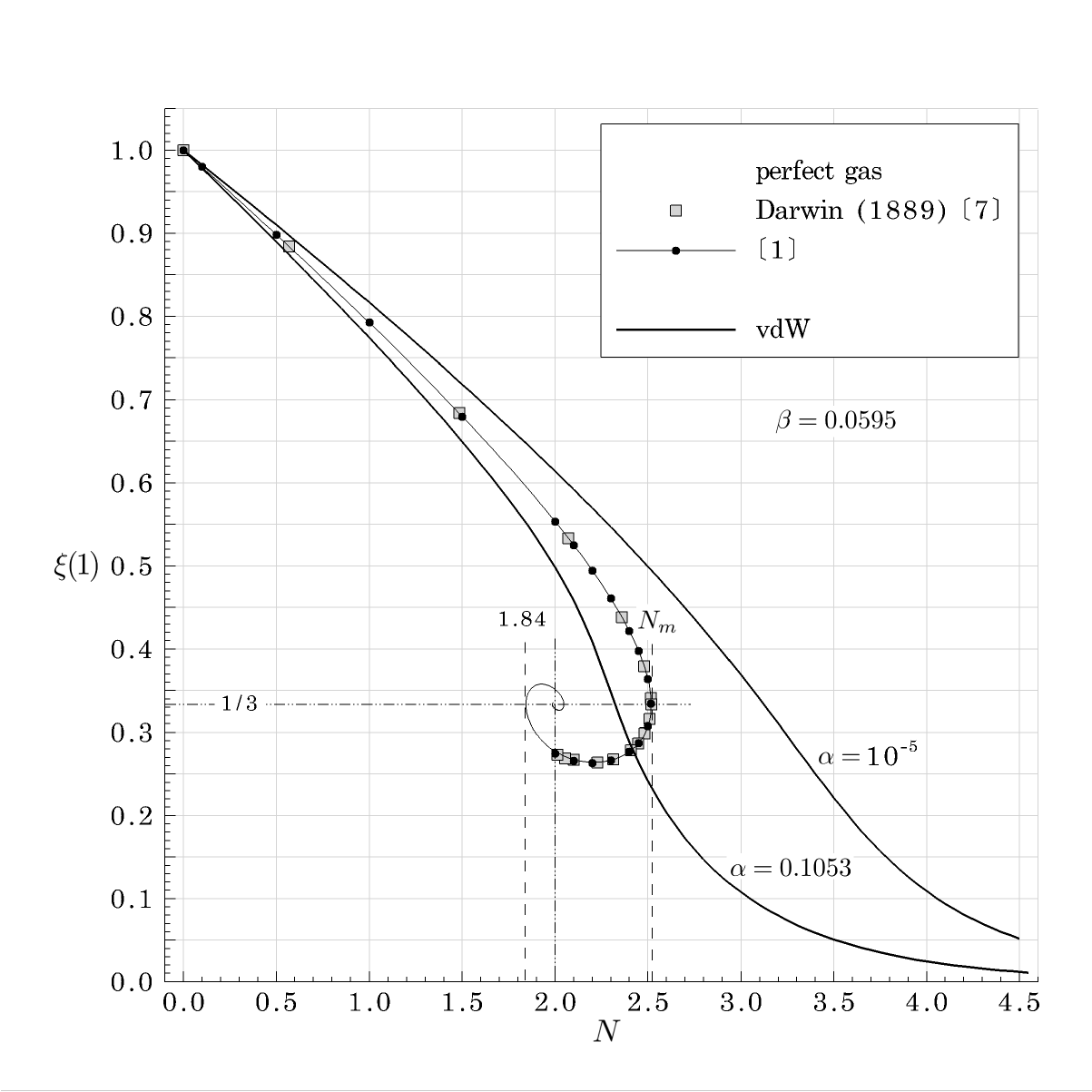}
  \caption{Peripheral density as a function of the gravitational number for perfect-gas and vdW models; updated from \Rma{Fig. 5}.\lpush}\label{xi1}
\end{figure}
\begin{figure}[b!]   %2ex 5ex 8ex 10ex
  \includegraphics[keepaspectratio=true, trim= 6ex 3ex 3ex 6ex , clip , width=\columnwidth]{\figdir/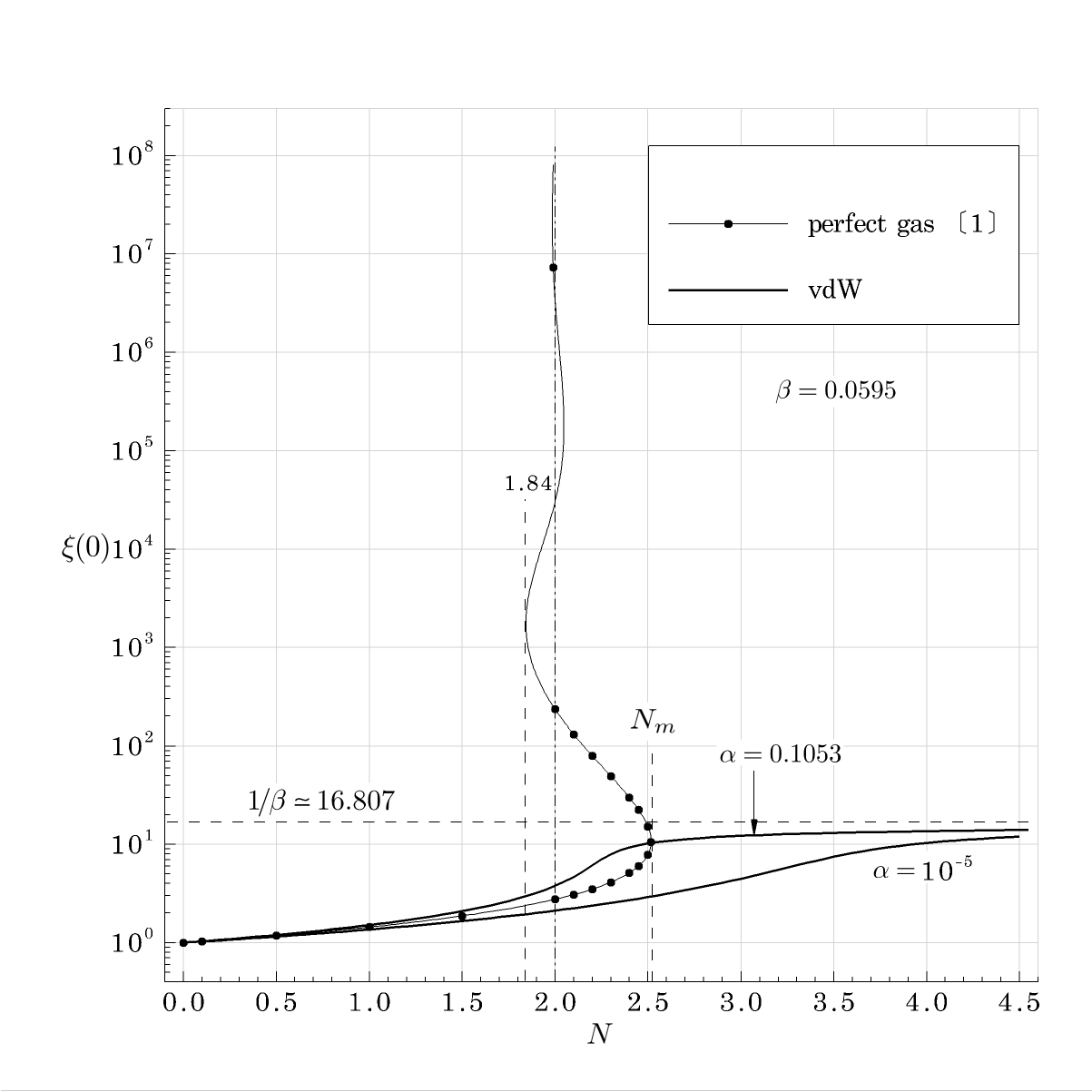}
  \caption{Central density as a function of the gravitational number for perfect-gas and vdW models; updated from \Rma{Fig. 6}.\lpush}\label{xi0}
\end{figure}%
With reference to the vdW-model curves, labelled with \itm{\alphay} values, the peripheral density decreases monotonically and vanishes asymptotically; the central density increases monotonically and approaches asymptotically the density upper bound \itm{1/\betay=16.807} imposed by molecular size [\REq{vdwse.nd.s}].
It is remarkable that the monotonicity is not lost even in the extreme circumstance \itm{\alphay=10^{-5}}, explicit evidence that not only molecular attraction but also finite molecular size plays a role to produce displacement from the perfect-gas curve.
In conclusion, the vdW-model curves offer beyond any doubt a definitely more consistent physical picture that brushes off the underlying spiralling and oscillating behaviours of the perfect-gas curves.
We are aware, of course, that the vdW-model curves in \Rfid{xi1}{xi0} prove only the disappearance of such behaviours for the assumed values of $\alphay$ and $\betay$ but that extrapolating such power to arbitrary values of $\alphay$ and $\betay$ is hazardous; indeed, an investigation regarding existence of absence of values \footnote{Of course, in case of existence, one should still ponder about the physical meaningfulness or realisability of such values.} in correspondence to which spiralling and oscillating behaviours appear also with the vdW model is certainly interesting and attractive but we had to put it in the future-work stack for the time being.

\subsubsection{A brief sidelight on thermodynamics\label{bst}}
Although thermodynamics is outside the scope of the present study, the situation portrayed in \Rfi{xi1} occasions and justifies a short digression.

The peripheral-density spiralling in the case of the perfect gas has far reaching consequences on its thermodynamics, whose details we have thoroughly discussed in \Rma{Sec. 3}.
It affects entropy [\Rma{Eq. (104)}] and energy [\Rma{Eq. (128)}] in a way that gives rise to a zigzag pattern of the fundamental relation (\Rma{Sec. 3.3.3} and \Rma{Fig. 22}) which reveals the famous energy minimum \footnote{There exists also an entropy minimum but, inexplicably, it is hardly mention in the literature. See \Rma{Fig. 22}.} discovered by Antonov \cite{va1985} and interpreted as the brink of a gravothermal catastrophe by Lynden-Bell and Wood \cite{dl1968mnras}.
It generates a gravitational correction to entropy [\Rma{Eqs. (102)} and \Rma{(109a)}] that can introduce thermal instability by flipping the sign of the thermodynamic derivative \itm{\pdst{}{S}{T}{V,m_{g}}} [\Rma{Eq. (110)}] from positive to negative (\Rma{Fig. 16}) and that, in so doing, sparks the interesting, although intriguing, debate regarding the existence of negative specific heats \cite{dl1977mnras,dl1999p,wt1970zfp,ih1978ptp,wt2003prl}.
It is inherited by the isotherms in the pressure-volume plane (\Rma{Fig. 24}) and makes appear zones of mechanical instability by flipping the sign of the thermodynamic derivative \itm{\pdst{}{E}{V}{S,m_{g}}}.
%Well, these preoccupations, i.e. gravothermal catastrophe, thermal instability and negative specific heats, mechanical instability, pushed into the physical picture by the perfect-gas model lose significance  with the vdW-gas model because, with it, the spiralling behaviour disappears.
Well, these preoccupations pushed into the physical picture by the perfect-gas model lose significance  with the vdW model because, with it, the spiralling behaviour disappears.
As \Rfi{xi1} indicates, the vdW peripheral density shows a monotonic trend that vanishes asymptotically and that monotonicity should be expected to propagate also to other thermodynamic properties.
As preliminary evidence supporting such an expectation, we adduce the graph of \Rfi{ient} which shows the gravitational correction to entropy presented in \Rma{Fig. 15} for the perfect-gas model updated with the curve corresponding to the vdW model.
\begin{figure}[t]
  \includegraphics[keepaspectratio=true, trim = 5ex 3ex 3ex 6ex , clip , width=\columnwidth]{\figdir/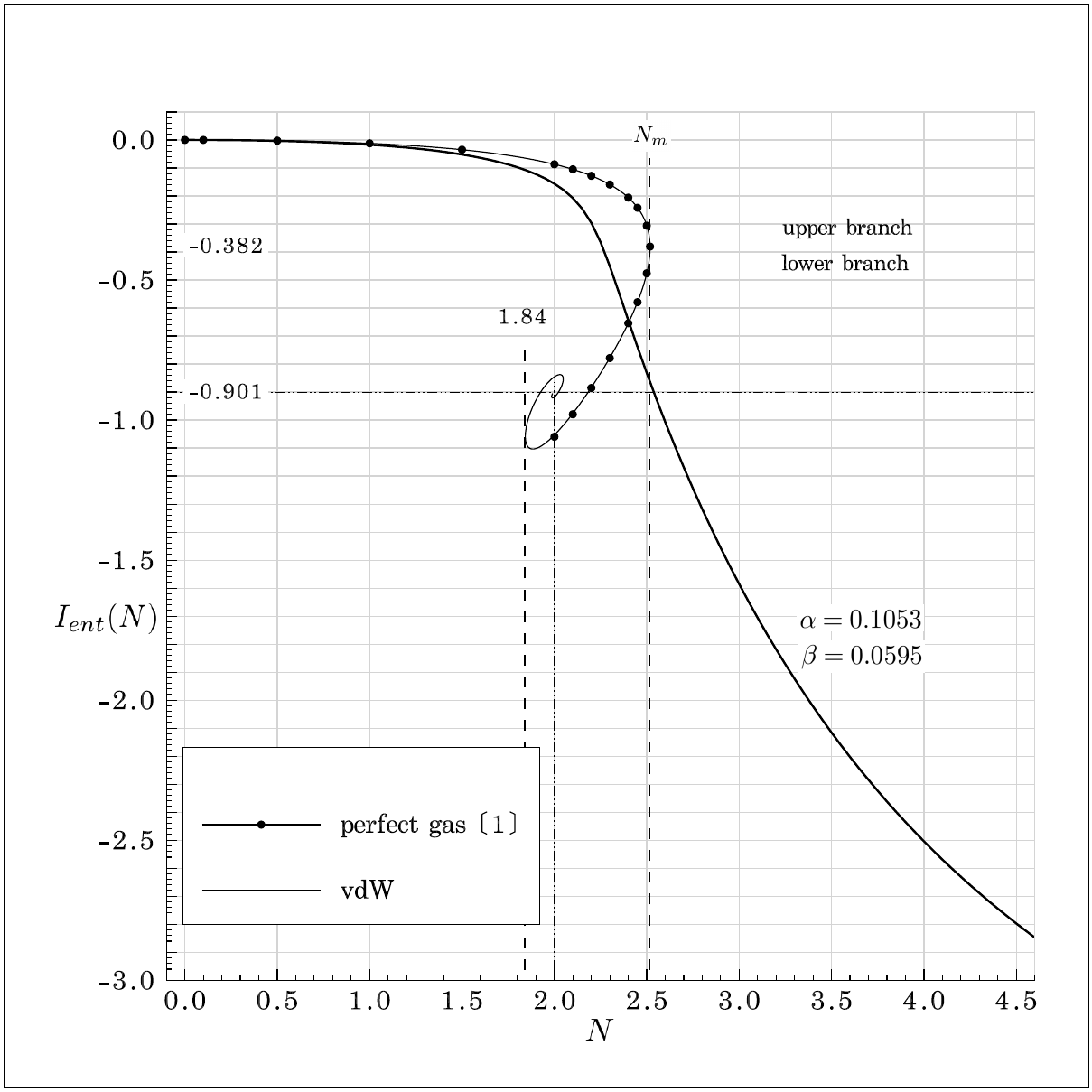}
  \caption{Gravitational correction to entropy as a function of the gravitational number for perfect-gas and vdW models; updated from \Rma{Fig. 15}.\lpush}\label{ient}
\end{figure}
The curves are obtained from \Rma{Eq. (102)} for the perfect gas and from the numerical integration of
\begin{equation}\label{ient.vdw}
   I_{ent} = \int_{0}^{1}\hspace*{-0.5em}3\etay^{2}\xiy\,\ln\frac{1-\betay\xiy}{\xiy\left(1-\betay\right)}\,d\etay
\end{equation}
for the vdW model \footnote{\REqb{ient.vdw} is obtained from \Rma{Eq. (96)} by substituting into it the vdW-model specific entropy  \[ s_{g} = K + \frac{3}{2}R\,\ln T +R\,\ln\left(v-B\right)\] rather than the perfect-gas specific entropy given in \Rma{Eq. (98)}. It reduces to \Rma{Eq. (102)} if \itm{\,\protect\betay\rightarrow0}.}.
The unconditional negativeness of the vdW-model curve's slope secures the positiveness of the thermodynamic derivative \itm{\pdst{}{S}{T}{V,m_{g}}} and safeguards the fluid-static equilibrium against thermal instabilities; there are no negative specific heats with the vdW model.

On the basis of this short digression, we believe that thermodynamics with self-gravitation deserves a refreshing as well as restorative revision starting from a bedrock that outdoes the idea of the perfect-gas model.

\subsubsection{Alternative attempts beyond perfect gas\label{aabpg}}
As anticipated in the end of \Rse{intro}, we have familiarised ourselves with the papers of Aronson and Hansen \cite{ea1972aj} and Stahl et al. \cite{bs1995pss} in search of possible similarities with our findings, particularly in connection with the conclusion drawn at the end of the \Rse{fvgn} regarding uniqueness of solutions, and of opportunities to validate even more the numerical performance of our algorithms.%\footnote{The more the better.}
\begin{table}[h]
  \caption{Molecular parameters selected by Aronson and Hansen \cite{ea1972aj}; notation and units as in original paper.\lpush\label{ah1}} \vspace*{-2ex}
  \centering
  \includegraphics[keepaspectratio=true, trim= 0ex 0ex 0ex 0ex , clip , width=.60\columnwidth]{\figdir/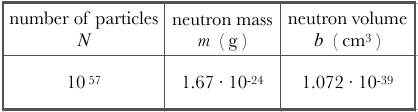}
\end{table}

Aronson and Hansen opted for the vdW equation of state [\REq{vdwse.nd.sv}] deprived of the molecular-attraction term \itm{\alphay/\zetay^{2}}; therefore, their approach is embedded in ours and, we thought, their results should be easily retrievable by setting systematically \itm{\alphay=0} in our schemes.
With a view to neutron stars, those authors considered a gas composed by neutrons and carried out numerical calculations with the physical parameters specified at page 148 of their paper, just below Eq.~(15), that we have collected in \Rtad{ah1}{ah2} for the readers' convenience.
The gravitational number $N$ [\REq{gn}] and the characteristic number $\betay$ [\REq{cn.b}] follow straightforwardly from the parameters selected by Aronson and Hansen via the conversion formulae
\begin{table}[t]%
  \caption{On grey background: container sizes and gas-sphere temperatures selected by Aronson and Hansen \cite{ea1972aj}; notation and units as in original paper. On white background: average density $\protect\brhoy$, characteristic number $\protect\betay$ and gravitational number $N$ derived [\REqq{cn.ah}] from Aronson and Hansen's parameters; central densities obtained from calculations based on our M$_{2}$ scheme. Dimensional densities are given in CGS units for consistency with original paper. We did not calculate the 30 km case. \lpush\label{ah2}}
\begin{minipage}{\columnwidth}%
  \setlength{\footnotesep}{2ex}
  \renewcommand{\footnoterule}{\hspace*{-0\columnwidth}\vspace*{-0.ex}\noindent\rule{0.1\columnwidth}{0.2pt}\vspace*{0cm}}
  %\centering
  \begin{tabular}{c} {\color{white}temperature\footnote{\scriptsize Aronson and Hansen comply with the statistical-thermodynamics standard notation \itm{\betayf=1/kT}; we use here the right-hand side of that definition to avoid conflict and ambiguity with our characteristic number $\betayf$ in the third column from left.}} \end{tabular} \\[-1.75\baselineskip]   %-1.75
  \includegraphics[keepaspectratio=true, trim= 0ex 0ex 0ex 0ex , clip , width=\columnwidth]{\figdir/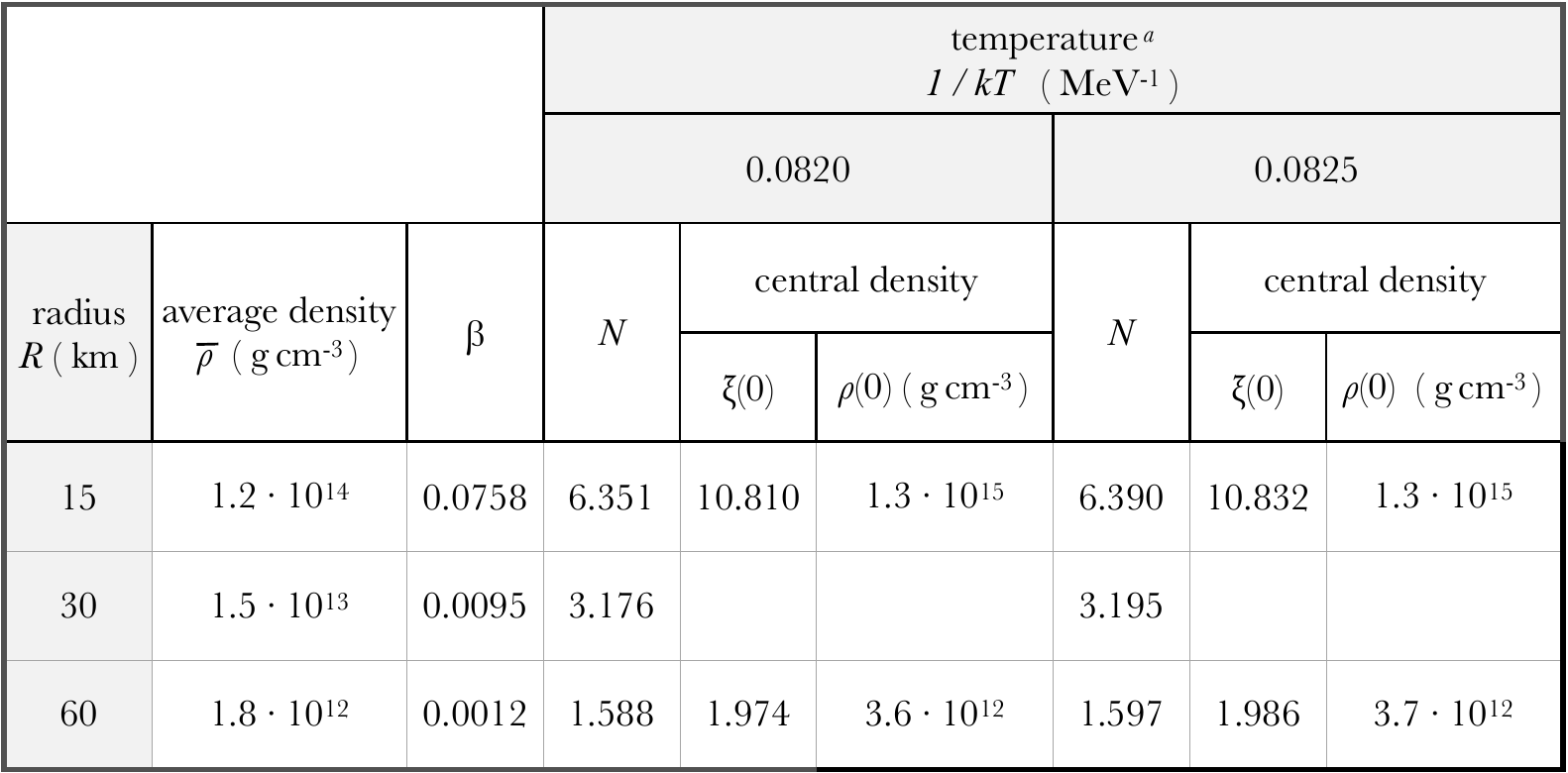} \\[-2.5\baselineskip]
\end{minipage}%
\end{table}%
\begin{subequations}\label{cn.ah}
\begin{align}
  N      = \frac{G m_{f}}{aR\oT} \; & \rightarrow \; \left( G N m^{2} \frac{1}{R} \frac{1}{kT} \right)_{\mbox{\scriptsize \cite{ea1972aj}}}  \label{cn.ah.gn} \\
  \betay =  B \brhoy             \; & \rightarrow \; \left( \frac{N b}{\frac{4}{3} \piy R^{3}}  \right)_{\mbox{\scriptsize \cite{ea1972aj}}} \label{cn.ah.b}
\end{align}
\end{subequations}%
whose rightmost-hand sides are expressed with their notation, obviously.
With the characteristic numbers in hand, we settled to replicate the results shown in Fig.~1 of \cite{ea1972aj}, reproduced here in \Rfi{ah.rhoonrhoc}; the shallow curve is reconstructed satisfactorily well but we remained a bit puzzled by the differences we found for the steep curve.
\begin{figure}[t]
  \includegraphics[keepaspectratio=true, trim = 5ex 3ex 3ex 6ex , clip , width=.97\columnwidth]{\figdir/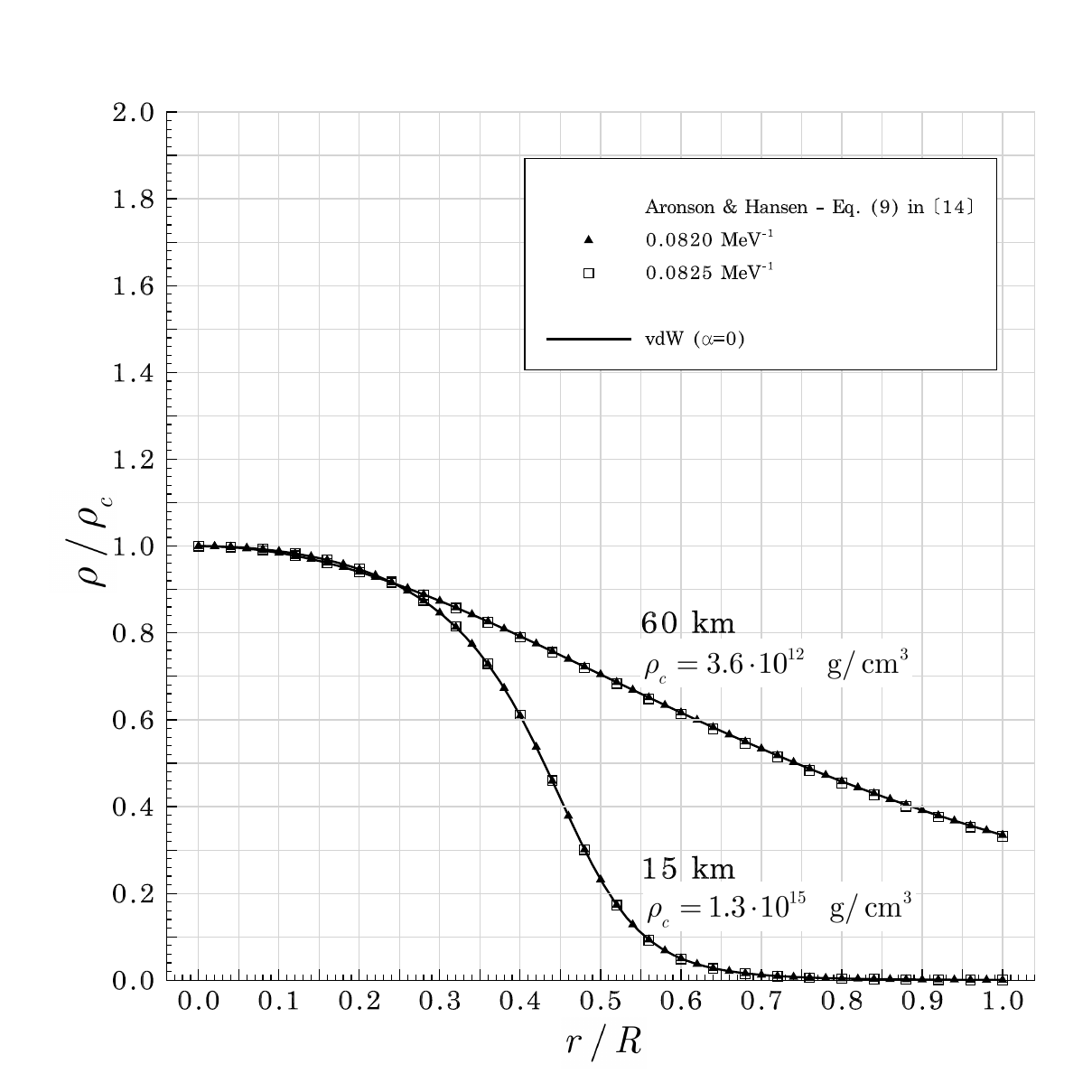}
  \caption{Reproduction of Fig.~1 of \cite{ea1972aj}. Aronson and Hansen's results have been obtained by modifying our M$_{2}$ scheme with the implementation of Eq.~(9) of \cite{ea1972aj}. We conform to the notation of the original paper.\lpush}\label{ah.rhoonrhoc}
\end{figure}%
In the original text as well as in the caption of their Fig.~1, Aronson and Hansen explicitly declared that both curves correspond to the 60 km case and left implied to the reader that the huge difference between them should be imputed to the tiny difference between the two values of \itm{1/kT}.
Our results lead to different conclusions.
For the 60 km case, the calculations with \itm{1/kT=0.0820} MeV$^{-1}$ and with \itm{1/kT=0.0825} MeV$^{-1}$ returned only slight differences and basically the same central density; this is to be expected because the corresponding values of the gravitational number \itm{N=1.588} and \itm{N=1.597} (see \Rta{ah2}) differ slightly and the corresponding characteristic number $\betay=0.0012$ is the same for both cases.
%We did find the same curve and same central density for the 60 km case with \itm{1/kT=0.0820} MeV$^{-1}$ but the calculation with \itm{1/kT=0.0825} MeV$^{-1}$ returned only slight differences from the previous curve and basically the same central density; this is to be expected because the corresponding values (1.588 and 1.597; see \Rta{ah2}) of the gravitational number differ slightly and the characteristic number $\betay$ remains the same (0.0012).
We did find the central density that labels the steep curve in their Fig.~1 but for the 15 km case; yet, our curve is not as steep as the former and, again, only slight differences exist between the results with \itm{0.0820} MeV$^{-1}$ and with \itm{0.0825} MeV$^{-1}$, respectively.
Intrigued by such a situation and motivated by desire to find out the reason for divergence, we revised formulae in both camps.
As a matter of fact, we detected the absence of a ``+1'' between the square brackets in the expression of the local chemical potential in Eq.~(9) of \cite{ea1972aj} but its absence turns out to be irrelevant because any constant would be brushed aside by the radial derivatives of Eq.~12 of \cite{ea1972aj}.
Apart that, we did not find any inconsistency.
On the other hand, Padmanabhan \cite{tp1990pr} referenced Aronson and Hansen's work and dealt with the same problem in his \mbox{Sec.~4.5}; indeed, he showed in his Fig.~4.9 the same steep curve of Aronson and Hansen's Fig.~1 but made reference to a rather irretrievable internal report for the details of the calculations.
Thus, we were not able to find out the explanation of the divergence between Aronson and Hansen's and our steep curve, unfortunately.

With the purpose in mind of better characterising the formation of condensed bodies, Stahl et al. \cite{bs1995pss} chose the equation of state developed by Carnahan and Starling \cite{nc1969tjocp} for \textit{non-attracting} rigid spheres, that we have rephrased to the equivalent, but more convenient for us, nondimensional form
\begin{subequations}\label{ca-st}
\begin{equation}\label{ca-st.se}
%  p = \frac{1}{v - B} - \frac{A}{v^{2}} = \frac{\rhoy RT}{1 - B\rhoy} - A\rhoy^{2}
  \pnd = \frac{\thetay}{\zetay - \betay}\cdot\Phi(y)
\end{equation}
with \footnote{Stahl et al. use the symbol $\protect\etayf$ which would conflict with our nondimensional radial coordinate [\REq{ndv}]; for this reason we reverted to the original symbol $y$ used by Carnahan and Starling.}
\begin{equation}\label{ca-st.psi}
  \Phi(y) = \frac{1+y+y^{2}-y^{3}}{\left(1-y\right)^{2}} = 1 - y + \frac{2y}{1-y} + \frac{2y}{\left(1-y\right)^{2}}
\end{equation}
and
\begin{equation}\label{ca-st.y}
  y = \frac{\betay}{\zetay} \le y^{\ast}\simeq 0.49
\end{equation}
\end{subequations}
The limit of applicability indicated in \REq{ca-st.y} was explicitly declared and explained by Stahl et al.;
we conform to it and refer the reader to their paper for the details.
The reduced chemical potential corresponding to the equation of state formulated in \REq{ca-st.se} can be easily derived via the standard methods of thermodynamics; it reads
\begin{subequations}\label{rcp.cast}
\begin{equation}\label{rcp.cast.nd}
   \phi(\thetay=1,\zetay) = - \ln ( \zetay - \betay ) + \frac{\zetay}{\zetay - \betay}\cdot\Phi(y) + \phiy(y)
\end{equation}
with
\begin{equation}\label{rcp.cast.phi}
  \phiy(y) = \ln\left( 1 - y \right) + \frac{2}{1-y} + \frac{1}{\left(1-y\right)^{2}} - 3
\end{equation}
\end{subequations}
\begin{table}[h]%
  \caption{Physical parameters selected by Stahl et al. \cite{bs1995pss}; notation and units as in original paper. According to their indication, the H-atom volume was calculated as \itm{v=4\protect\piy\,\bohrr^{3}/3}.\lpush\label{stahl1}} \vspace*{-2ex}
  \centering
  \includegraphics[keepaspectratio=true, trim= 0ex 0ex 0ex 0ex , clip , width=.75\columnwidth]{\figdir/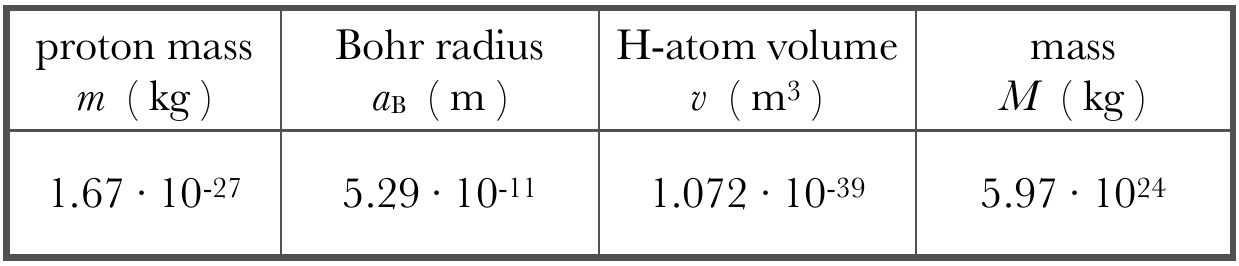}
\end{table}%
These authors considered a gas composed by hydrogen atoms.
The physical parameters chosen for their calculations are indicated at page 275 of their paper, in the middle of the left column, and are collected here in \Rta{stahl1}.
The gravitational number $N$ [\REq{gn}] and the characteristic number $\betay$ [\REq{cn.b}] coincide with the dimensionless parameters $A, B$ defined in Eqs.~(15)~and~(16) of \cite{bs1995pss}
\begin{subequations}\label{cn.stahl}
\begin{align}
  N      = \frac{G m_{f}}{aR\oT} \; & \rightarrow \; \left( \frac{G M m}{ R \kb T } = A \right)_{\mbox{\scriptsize \cite{bs1995pss}}} \label{cn.stahl.gn} \\
  \betay =  B \brhoy             \; & \rightarrow \; \left( \frac{M}{m}{\frac{v}{4\piy R^{3}/3} = B }  \right)_{\mbox{\scriptsize \cite{bs1995pss}}} \label{cn.stahl.b}
\end{align}
\end{subequations}
Our attention was attracted particularly by Sec.~5 of \cite{bs1995pss} dedicated to systems with prescribed temperature because of its connection with our study.
In Fig.~9 of that section, the authors show the impact of their equation of state on the central density as a function of the gravitational number with \itm{\betay=10^{-7}}; the curve follows faithfully the perfect-gas curve of \Rfi{xi0} up to \itm{\xiy(0)\simeq 10^{6}} but there the finite molecular size kicks in and dramatically introduces multiple solutions even below the perfect-gas threshold \itm{N=1.84}.
According to the authors, the new threshold is pushed down to \itm{N\simeq 0.15}, a rather small value of the gravitational number.
Seen from the perspective of our motivations explicitly stated in the beginning of \Rse{intro}, the proliferation of multiple solutions represents a discouraging setback.
Nevertheless, we took advantage of the radial profile of density presented in Fig.~11 of \cite{bs1995pss} as ``quasi-uniform state'', reproduced in \Rfi{bs.rhoonarho}, as further test of algorithm validation.
%\begin{figure}[t]
%  \includegraphics[keepaspectratio=true, trim = 0ex 3ex 3ex 6ex , clip , width=\columnwidth]{\figdir/bs.rhoonarho.pdf}
%  \caption{Partial reproduction of Fig.~11 of \cite{bs1995pss}. Stahl et alii's profile, labelled ``quasi-uniform state'' in the original caption, has been obtained by modifying our M$_{2}$ scheme with the implementation of \REqd{rcp.cast.nd}{rcp.cast.phi}. The original diagram is in logarithmic scale to give visual relevance to the ``collapsed state'' so that the ``quasi-uniform state'' appears as a flat line. We have used a linear scale to put in evidence the small differences belonging to the ``quasi-uniform state'' between sphere centre and sphere wall; we did not attempt to reproduce the ``collapsed state''. We conform to the notation of the original paper.\lpush}\label{bs.rhoonarho}
%\end{figure}

\begin{figure}[h]%
  \includegraphics[keepaspectratio=true, trim = 0ex 3ex 3ex 6ex , clip , width=\columnwidth]{\figdir/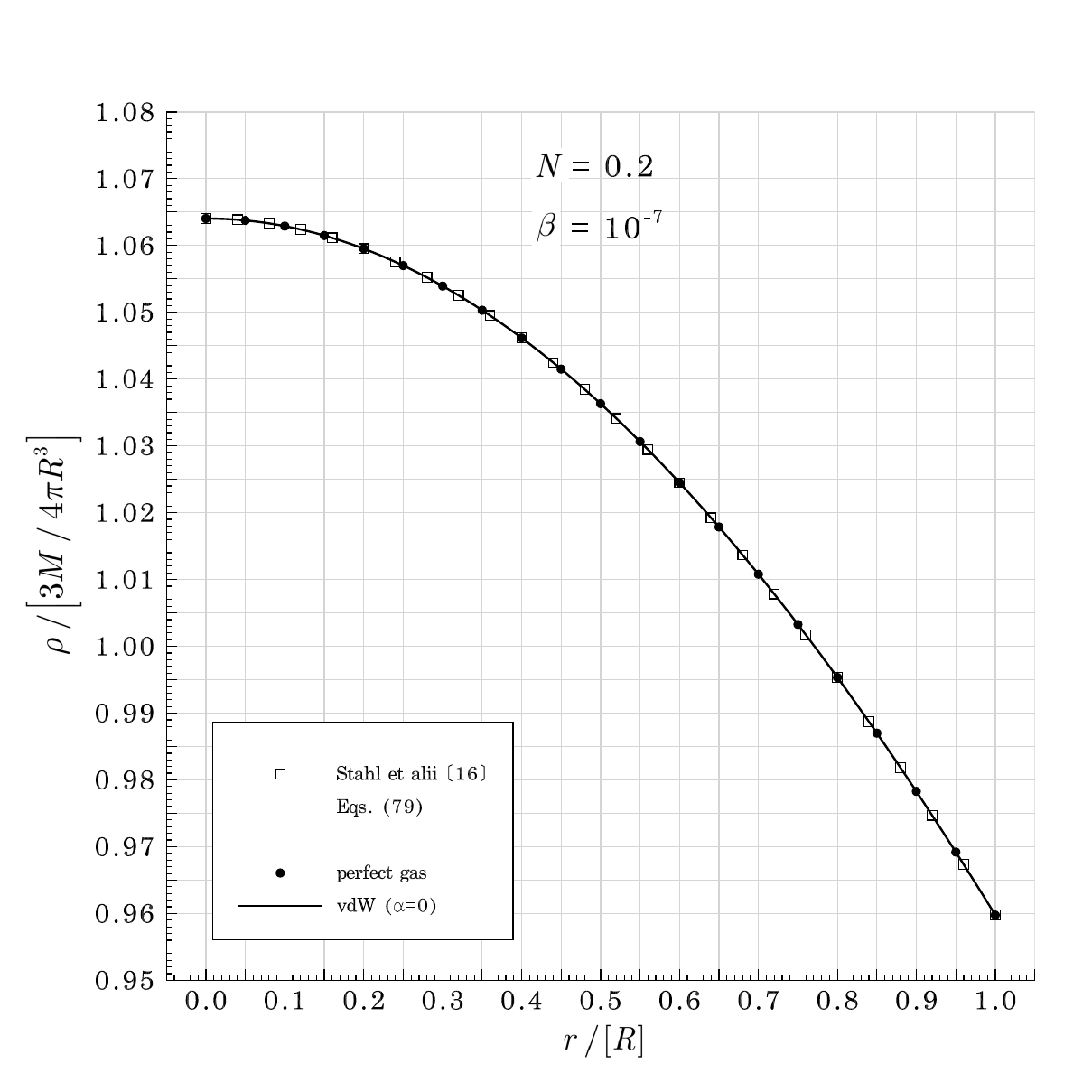}
  \caption{Partial reproduction of Fig.~11 of \cite{bs1995pss}. Stahl et al.'s profile, labelled ``quasi-uniform state'' in the original caption, has been obtained by implementing \REqq{rcp.cast} into our M$_{2}$ scheme. The original diagram is in logarithmic scale to give visual relevance to the ``collapsed state'' so that the ``quasi-uniform state'' appears as a flat line; we have used a linear scale to put in evidence the small differences between sphere centre and sphere wall for the ``quasi-uniform state''. We did not attempt to reproduce the ``collapsed state''. We conform to the axes' labels of the original paper.\lpush}\label{bs.rhoonarho}
\end{figure}%
\begin{figure}[h]
  \includegraphics[keepaspectratio=true, trim= 9ex 6ex 3ex 6ex , clip , width=\columnwidth]{\figdir/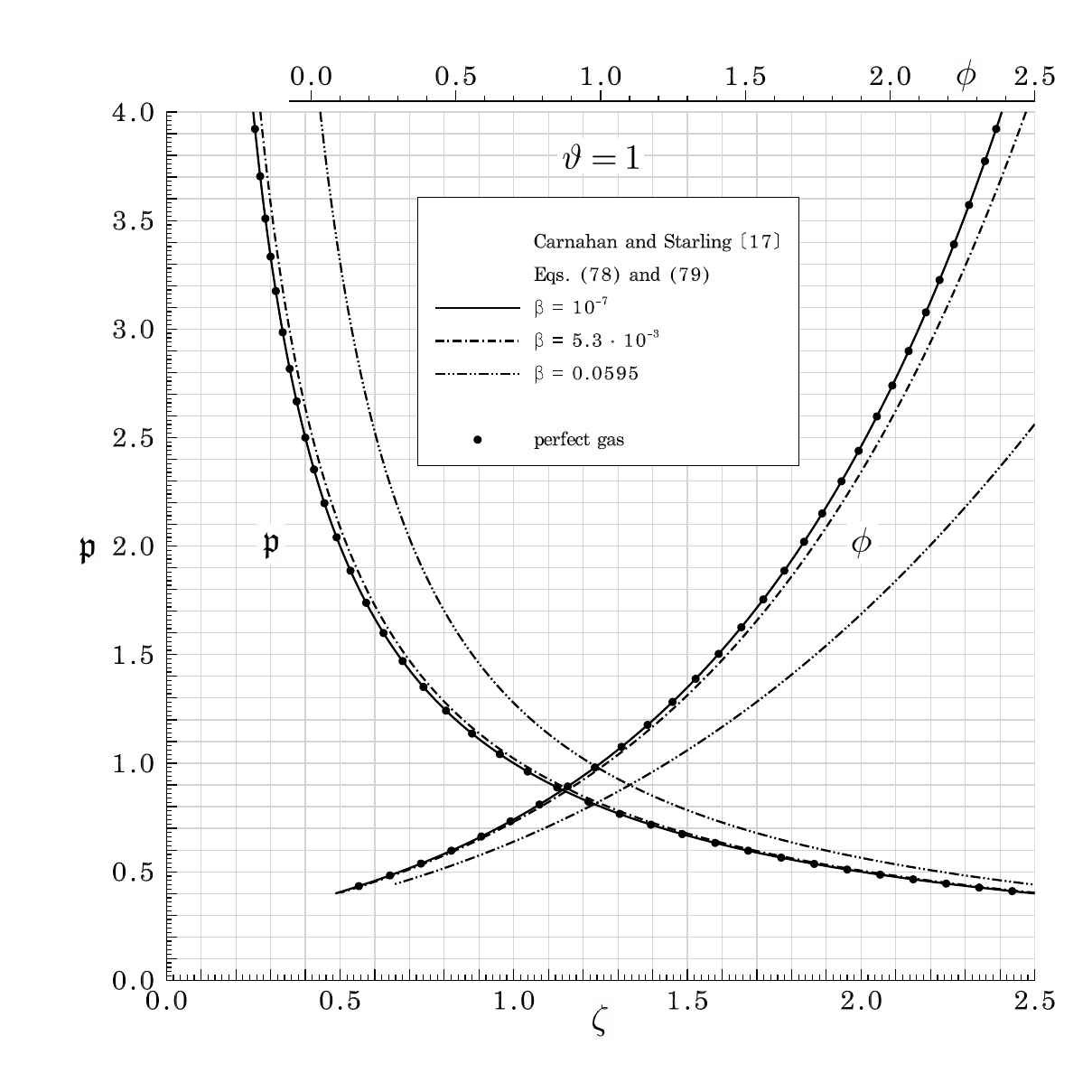}  % 9ex 6ex 3ex 6ex
  \caption{Graphical visualisation of the inability of the thermodynamic model based on the Carnahan and Starling's equation of state \cite{nc1969tjocp} to predict phase equilibrium.\hfill\ }\label{cast-npe}
\end{figure}
To conclude this section, we wish to remark that neither the vdW model deprived of the molecular-attraction term selected by Aronson and Hansen nor the Carnahan and Starling model selected by Stahl et al. possess the built-in capability to account for phase equilibrium.
The deficiency is explicitly evident for the former model because the top condition in \REq{pes} prevails if \itm{\alphay=0}.
Its evidence for the latter model would require the prerequisite study of the solutions of the thermodynamic-equilibrium equations [\REqq{c.tpmu}] but we followed the graphical shortcut of drawing a thermodynamic diagram, shown in \Rfi{cast-npe}, similar to the one in \Rfi{phaseeq} but using the Carnahan and Starling's equation of state [\REqq{ca-st}] and the associated reduced chemical potential [\REqq{rcp.cast}]
in correspondence to two values of $\betay$ considered by Stahl et al. and the value [\REq{cn.b.he}] we selected for our study.
The graph delineates the absence of minima and maxima in the pressure's isotherms as well as of any closed loop in the reduced chemical potential's isotherms and confirms unambiguously the remark we made in the beginning of this paragraph.
\begin{figure}[h]
  \includegraphics[keepaspectratio=true, trim= 9ex 6ex 3ex 6ex , clip , width=\columnwidth]{\figdir/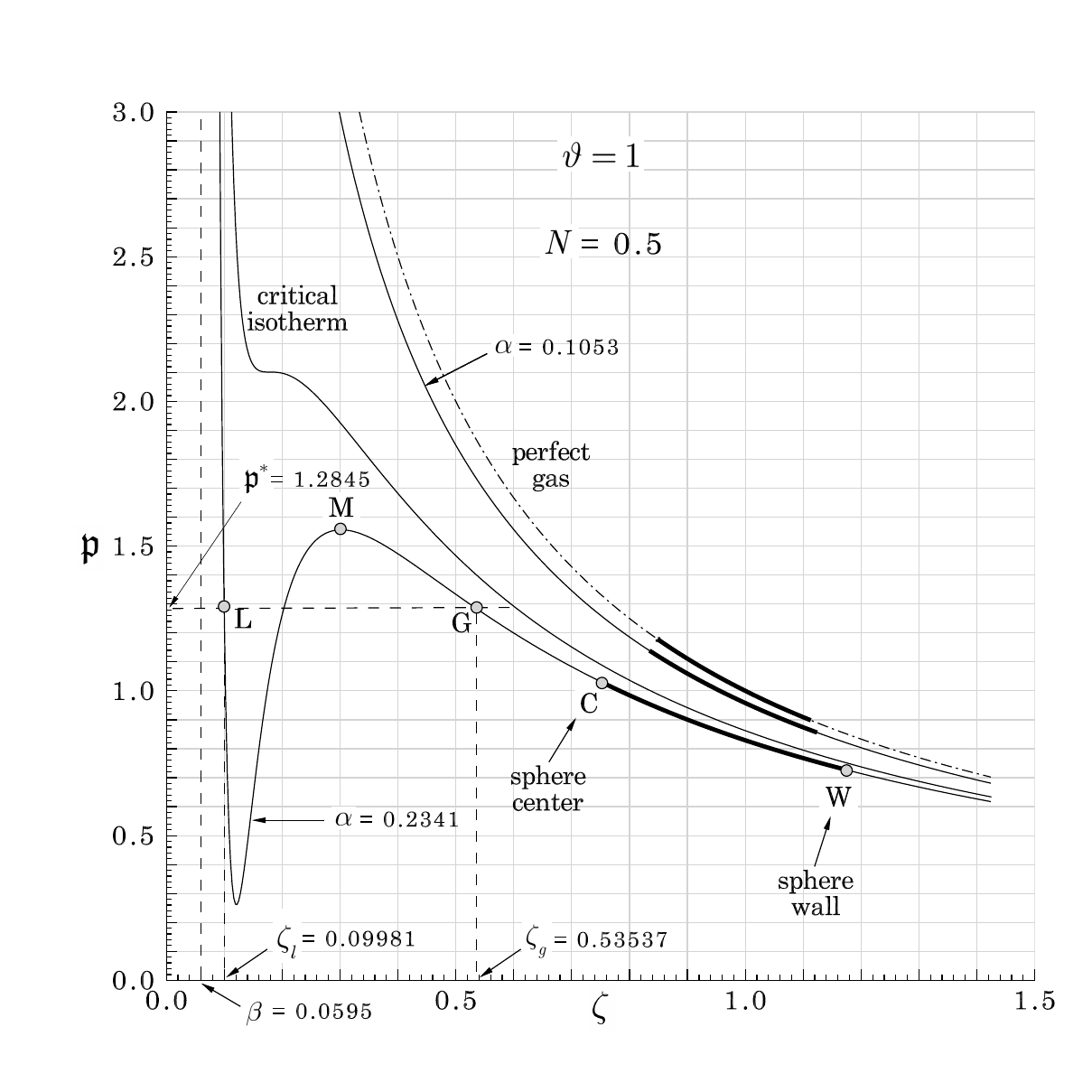}
  \caption{The radial profiles of \Rfi{N=0.5rpd} overlaid (bold segments) on the isotherms of \Rfi{vdwisot}.\hfill\ }\label{N=0.5isoT}
\end{figure}
\begin{figure}[h]%
  \includegraphics[keepaspectratio=true, trim= 9ex 6ex 3ex 6ex , clip , width=\columnwidth]{\figdir/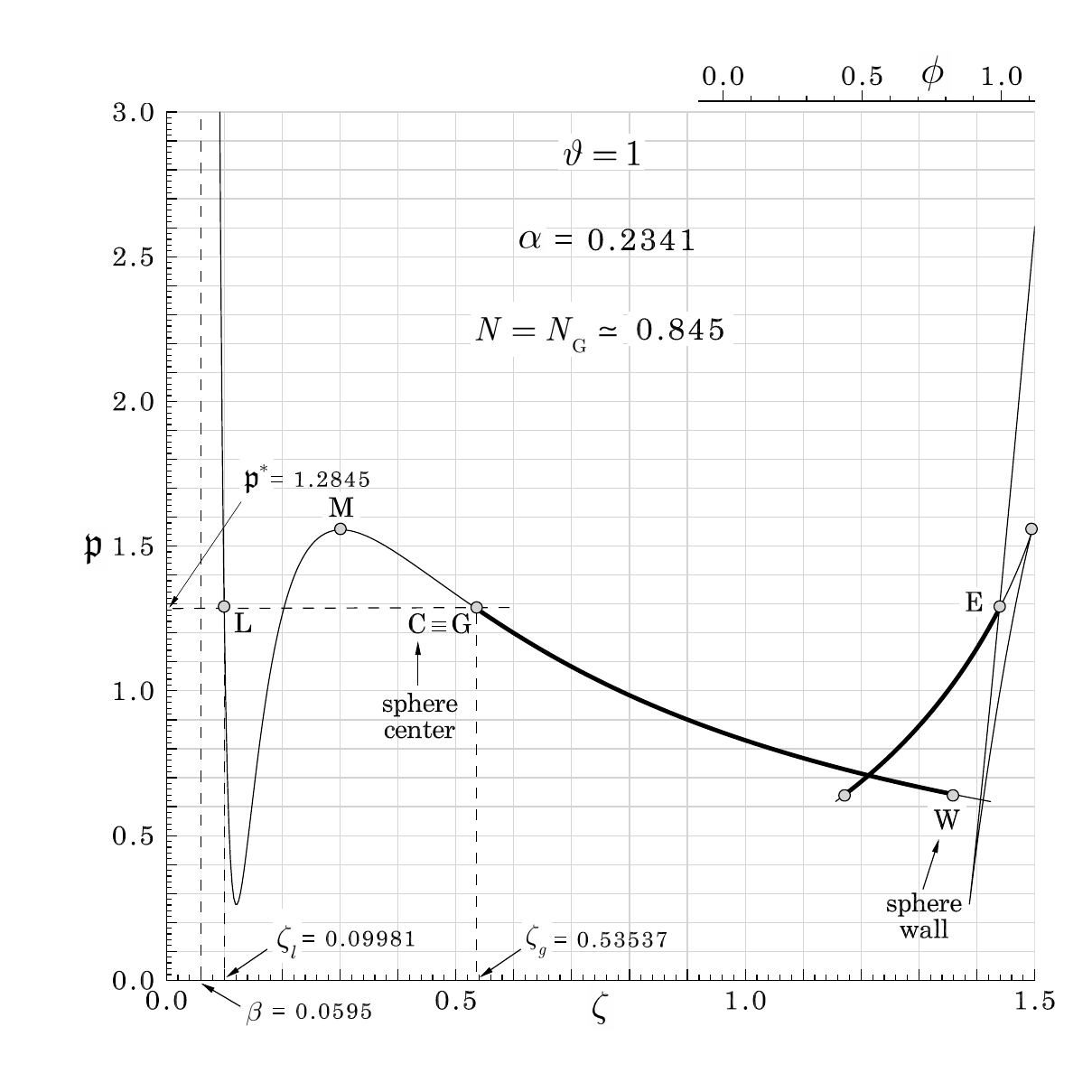}
  \caption{Incipient formation of either gas' metastable states or liquid-gas phase equilibrium at \itm{N=\Ni}.\hfill\ }\label{N=0.845isoT}
\end{figure}%

\subsection{Fluid-sphere temperature below critical temperature\label{ppe}}

\subsubsection{General considerations}
A fluid-sphere temperature below the critical temperature [\REq{pes} bottom] presupposes the possible existence of a phase equilibrium.
A convenient introduction to the study of this circumstance begins with the overlay of the radial profiles of \Rfi{N=0.5rpd} onto the isotherms of \Rfi{vdwisot}, an operation that generates the graph shown in \Rfi{N=0.5isoT}.
%With reference to the isotherm below the critical one, if the gravitational number attains a moderate value, such as \itm{N=0.5} in \Rfi{N=0.5isoT}, then the profile lies completely at the right of G.
With reference to the isotherm below the critical one, the projection of the profiles represented by the bold segment lies completely at the right of G for the moderate value \itm{N=0.5} of the gravitational number.
An increase of $N$ results in a shift of W to the right and, more importantly, a shift of C to the left towards G; there must, therefore, exist a value $\Ni$, for the specified couple $\alphay$ and $\betay$, in correspondence to which C comes to coincide with G.
This condition, illustrated in \Rfi{N=0.845isoT}, represents the incipient formation of either gas' metastable states or a liquid-gas phase equilibrium.
The value $\Ni$ constitutes an additional unknown and its calculation requires a slight modification of the nondimensional computational schemes described at the end of \Rse{ndf}.
In the M$_{2}$ scheme, for example, \REq{bc.rcp.r=a.nd} loses its role of boundary condition and turns into an equation for the new unknown
\begin{equation}\label{eq.rcp.r=a.nd}
     {\left. \pd{}{\phi}{\etay} \right|}_{\etays=1} + \Ni  =  0
\end{equation}
but the knowledge of the thermodynamic state (\Rta{phaseeqv}) at the sphere centre restores the lost boundary condition in terms of either specific volume's
\begin{equation}\label{bc.rcp.r=a.nd.zeta}
   \zetay(0) = \zetay_{g}
\end{equation}
or reduced chemical potential's
\begin{equation}\label{bc.rcp.r=a.nd.rcp}
   \phi(0) = \phi^{\ast}
\end{equation}
prescription.
Apart these slight adaptations, the numerical algorithms do not require any drastic change.
We obtained \itm{\Ni\simeq0.845} for our choice of \itm{\alphay=0.2341} and \itm{\betay=0.0595}.
%For our choice of \itm{\alphay=0.2341} we found \itm{\Ni\simeq0.845}.
Obviously, a further increase of the gravitational number above $\Ni$ leads to the presence of either of the two mentioned physical situations.

\begin{figure}[h]
  \includegraphics[keepaspectratio=true, trim= 9ex 6ex 3ex 6ex , clip , width=\columnwidth]{\figdir/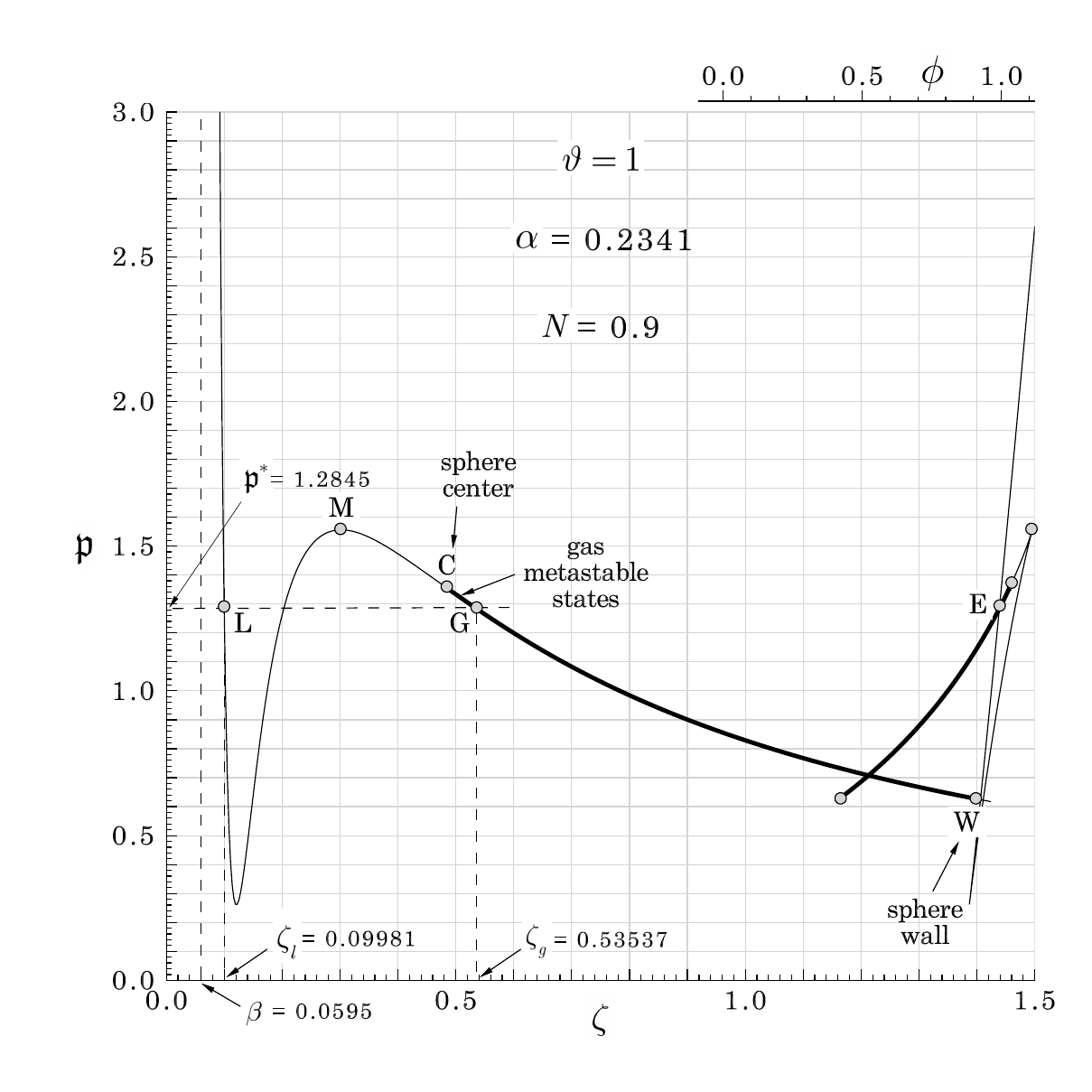}
  \caption{Formation of a central core of metastable gas at \itm{N=0.9}.\lpush}\label{N=0.9mssisoT}
\end{figure}
The former situation occurs if C shifts along the metastable branch of the isotherm towards M; then a central core of metastable gas settles in, as shown in \Rfi{N=0.9mssisoT} for \itm{N=0.9}.%
\begin{figure}[h]
  \includegraphics[keepaspectratio=true, trim= 9ex 6ex 3ex 6ex , clip , width=\columnwidth]{\figdir/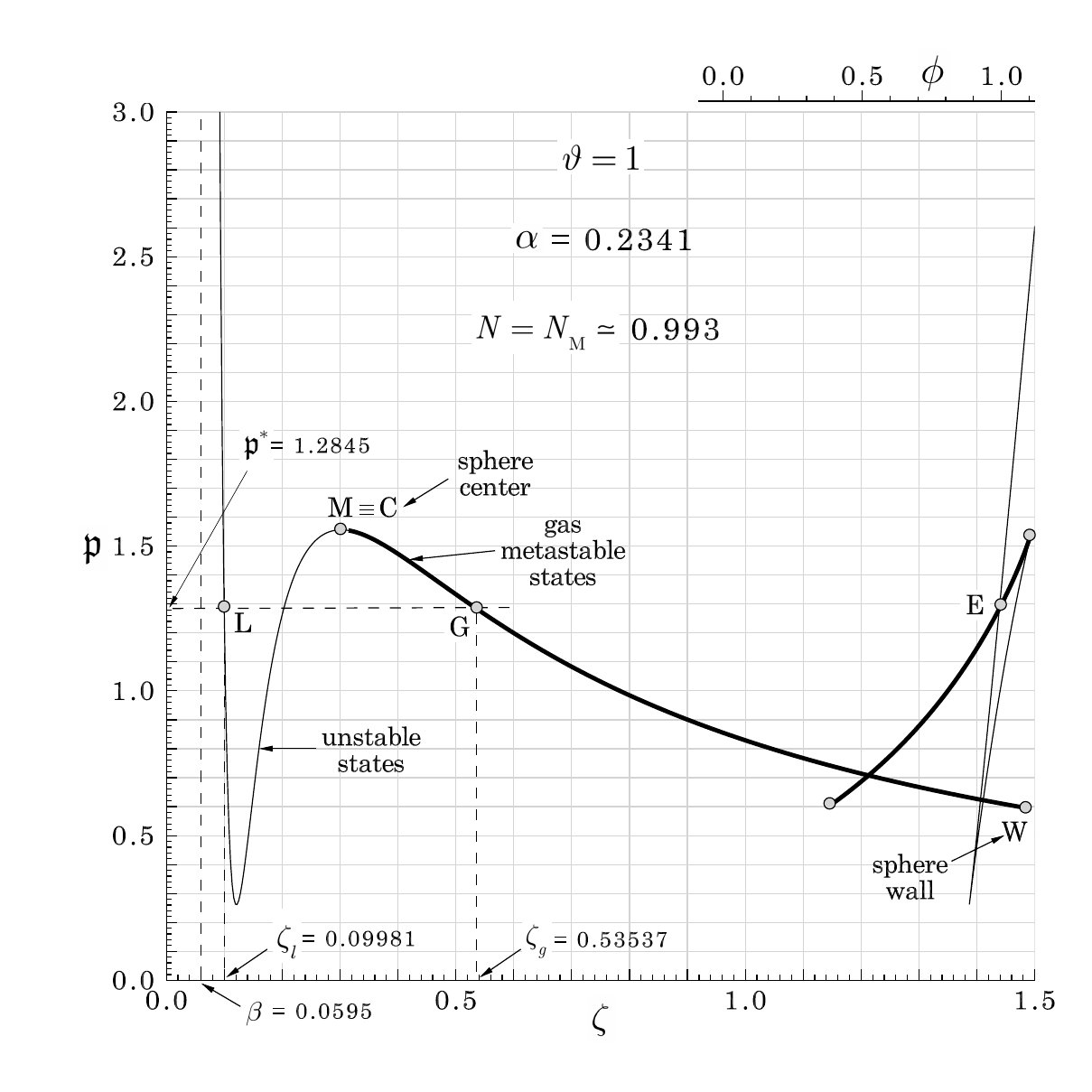}
  \caption{The greatest central core of metastable gas at \itm{N=\NM}.\lpush}\label{N=0.993mssisoT}
\end{figure}%
\begin{figure}[b!]
  \includegraphics[keepaspectratio=true, trim= 9ex 5ex 3ex 6ex , clip , width=\columnwidth]{\figdir/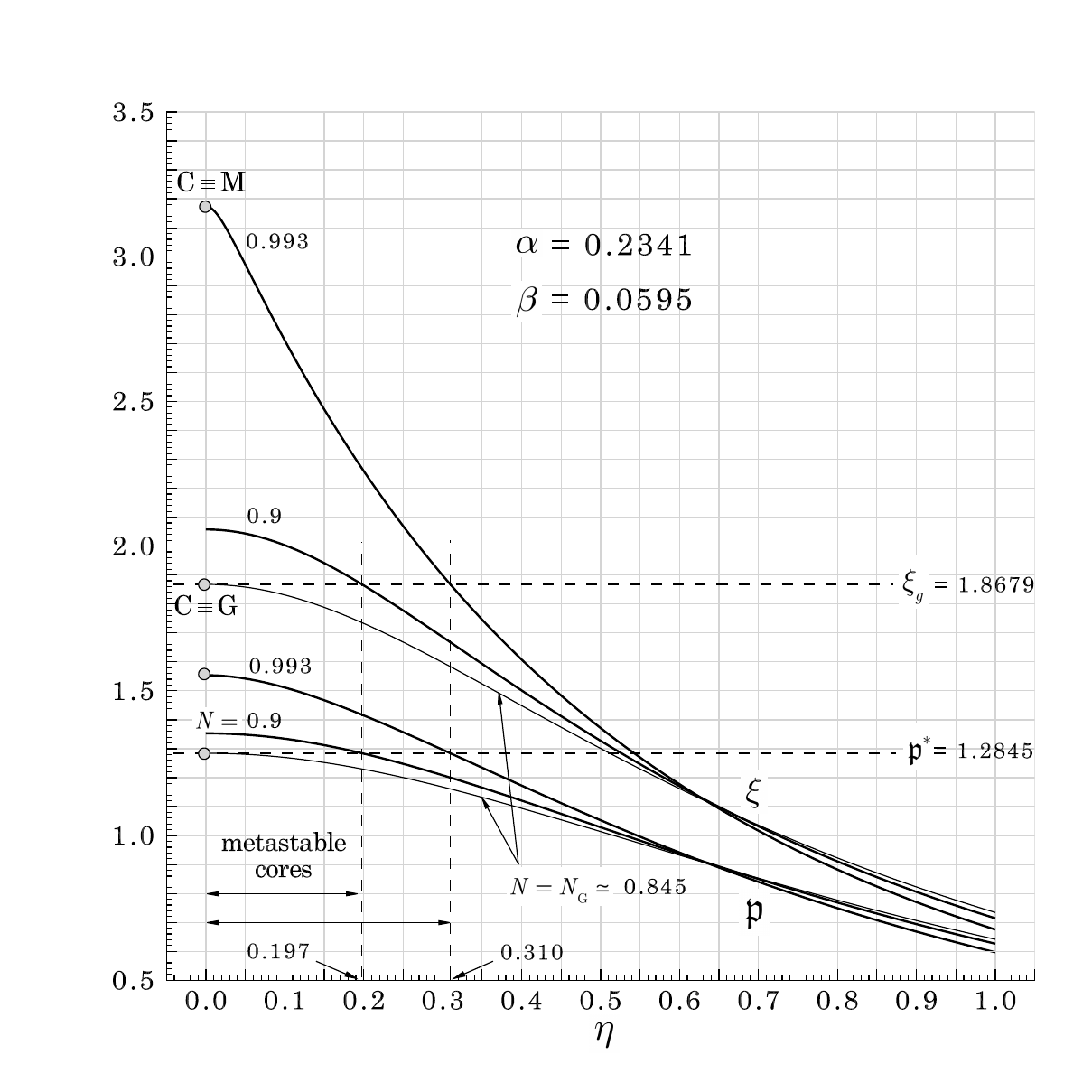}
  \caption{Radial profiles of density and pressure with presence of gas' metastable states for selected values of the gravitational number. The profiles corresponding to \itm{N=\Ni} (thinner lines) are included as reference baseline.\hfill\ }\label{rpd-mss}
\end{figure}
As clearly portrayed in \Rfi{N=0.993mssisoT}, this situation can persist until C reaches the last metastable state M, an unpassable physical limit because the unstable states are irrealisable.
The superposition C$\equiv$M occurs at another noticeable value of the gravitational number which we calculated to be \itm{\NM\simeq0.993} for \itm{\alphay=0.2341} and \itm{\betay=0.0595}.
The radial profiles of density and pressure with the presence of gas' metastable states corresponding to \Rfid{N=0.9mssisoT}{N=0.993mssisoT} are shown in \Rfi{rpd-mss}.
%The profiles corresponding to \itm{N=\Ni\simeq0.845} (thinner lines) are included for reference as starting baseline.
As indicated therein, the radial extent of the metastable cores is identified by the intersections of the density and pressure profiles with the horizontal lines \itm{\xiy=\xiy_{g}} and \itm{\pnd=\pnd^{\ast}}, respectively.
With \itm{\alphay=0.2341} and \itm{\betay=0.0595}, the greatest core at \itm{N=\NM\simeq0.993} extends to about 31\% of the sphere radius.

The phase-equilibrium situation takes place if C jumps from G to L and then shifts along the liquid branch of the isotherm %establishing a central core of liquid surrounded by stable gas,
as outlined thermodynamically in \Rfi{N=0.9peisoT}, again for \itm{N=0.9}, which portrays%
\begin{figure}[h]
  \includegraphics[keepaspectratio=true, trim= 9ex 6ex 3ex 6ex , clip , width=\columnwidth]{\figdir/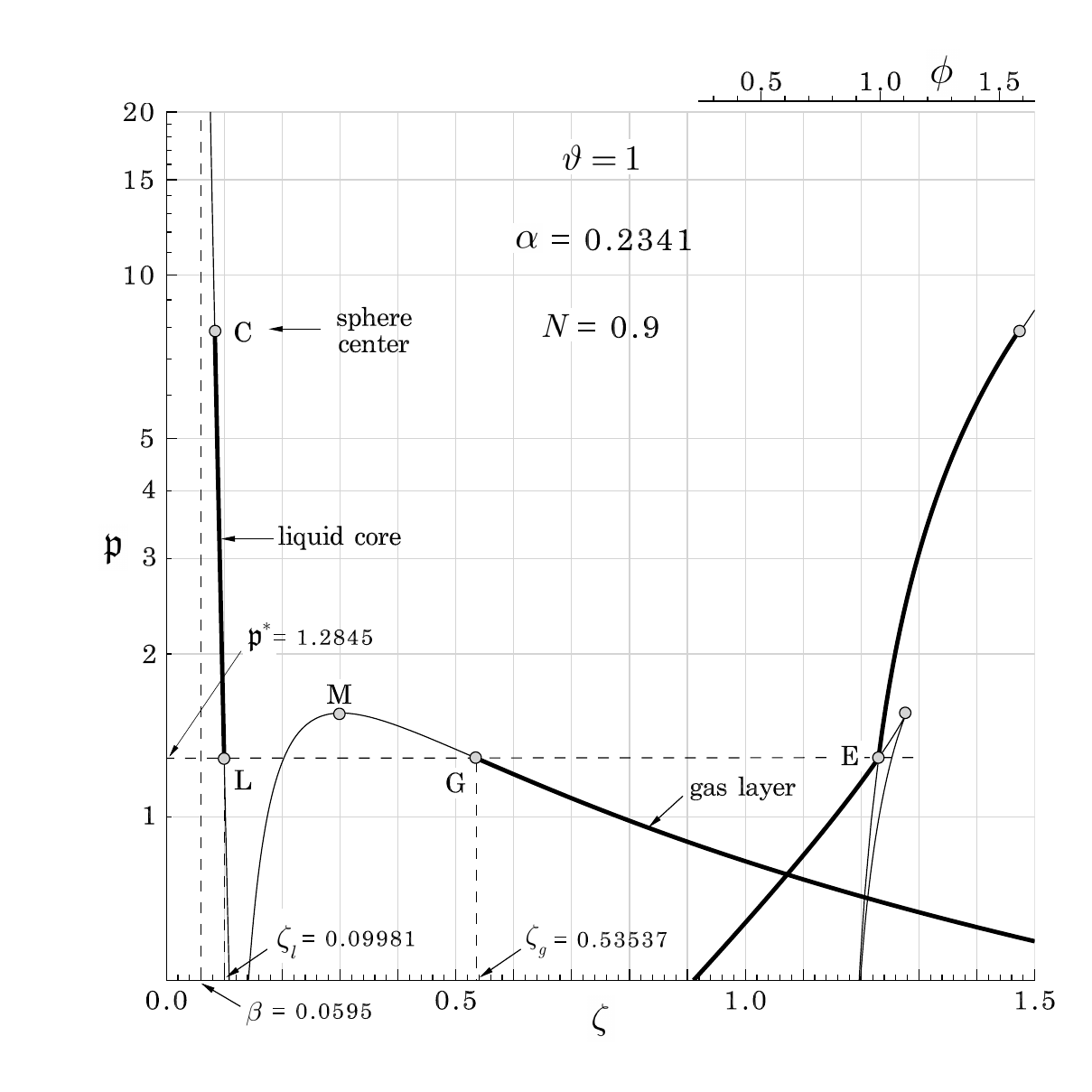}
%  \caption{Formation of a central liquid core in equilibrium with a peripheral gas layer at \itm{N=0.9}.\hfill\ }\label{N=0.9peisoT}
  \caption{Formation of a central liquid core in equilibrium with a peripheral gas layer at \itm{N=0.9}.\lpush}\label{N=0.9peisoT}
\end{figure}
the coexistence of a liquid core surrounded by a peripheral gas layer; they are separated by a spherical interface whose radial location $\etai$ constitutes an additional unknown.
The thermodynamic state on both liquid and gas sides of the interface is known (\Rta{phaseeqv}); specific volume and density are discontinuous across the interface but temperature, pressure and reduced chemical potential are continuous [\REqq{c.tpmu.vdw}].
The gravitational field is also continuous and the proof of its continuity is rather straightforward.
In order to distinguish variables of the liquid zone from those of the gas zone, we mark them with the accents $\,\check{}\,$ and $\,\hat{}\,$, respectively.
A simple proof is based on Gauss' theorem.
By taking advantage of the spherical symmetry, it suffices to apply the theorem to two spheres of radii \itm{\etai\pm\epsilony} to obtain the gravitational field on the liquid side \itm{\check{\gammay}(\etai-\epsilony)} and on the gas side \itm{\hat{\gammay}(\etai+\epsilony)} in terms of the masses contained in the spheres; then the passage to the limit for vanishing $\epsilony$ must give
\begin{equation}\label{gfc}
%   \lim_{\epsilony \rightarrow 0} \check{\gammay}(\etai-\epsilony) = \lim_{\epsilony \rightarrow 0} \hat{\gammay}(\etai+\epsilony) \rightarrow \gammay(\etai) = \gammay^{\ast}
   \check{\gammay}(\etai) = \lim_{\epsilony \rightarrow 0} \check{\gammay}(\etai-\epsilony) = \lim_{\epsilony \rightarrow 0} \hat{\gammay}(\etai+\epsilony) = \hat{\gammay}(\etai)
\end{equation}
because the mass contained in a sphere of radius $\etay$ is continuous for $\etay$ ranging in the interval \itm{[0,1]} even in the presence of the interface.
A more formal proof starts from \REq{sode.ss.vdw.nd}.
First, we need to introduce the density \itm{\xiy=1/\zetay} and to multiply by \itm{\etay^{2}}; then we can proceed to integrate the adapted equation on the interval [0,1] by taking into account the presence of the interface at \itm{\etay=\etai} and the gravitational boundary conditions given by \REqq{bc.rcp.nd}.
In this way, we obtain the gravitational-field continuity
%\begin{equation}\label{gfcp}
%   \frac{{\etai}^{2}}{N} \left[ \pds{}{\check{\phi}}{\etay}{\ssub{0.75}{\etay=\etai}} - \pds{}{\hat{\phi}}{\etay}{\ssub{0.75}{\etay=\etai}} \right] = 1 - 3\int_{0}^{1}\etay^{2}\xiy d\etay = 0
%\end{equation}
\begin{equation}\label{gfcp}
    \pds{}{\check{\phi}}{\etay}{\!\!\etays=\etays^{\ast}} - \pds{}{\hat{\phi}}{\etay}{\!\!\etays=\etays^{\ast}} = \frac{N}{{\etai}^{2}} \left[1 - 3\int_{0}^{1}\etay^{2}\xiy d\etay\right] = 0
\end{equation}
as a consequence of the normalisation condition given by \Reqma{43}.
%\textbf{more formal proof?}

\subsubsection{Adaptation of the governing equations in the presence of a phase-equilibrium interface\label{agepei}}
The phase-equilibrium situation is beyond the reach of the computational schemes P, M$_{1}$ and M$_{2}$ as described in \Rse{fsgsf} because there is no physical information in them from which the phase equilibrium pops up automatically and self-consistently.
In other words, the schemes apply to continuous phases and, as a matter of fact, they produce the radial profiles with metastable states of \Rfi{rpd-mss} if put at work for \itm{N>\Ni}; so, the schemes need to be told about this new circumstance.
Now, the specific volume's and density's discontinuities across the interface perturb the differential equations in all computational schemes [\REqs{meq.s.nd}{nmeq.s.nd}, \REq{sode.ss.vdw.nd}] and oblige to apply them in the liquid/gas zones \textit{separately}.
Thus, the boundary conditions at the sphere centre [\REq{bc.r=0.nd}, \REq{bc.rcp.r=0.nd}] become attached to the liquid zone and those at the sphere wall [\REq{bc.r=a.nd}, \REq{bc.rcp.r=a.nd}] are consigned to the gas zone.
Zone-matching boundary conditions must be imposed at the interface and, obviously, descend from the knowledge of the thermodynamic state (\Rta{phaseeqv}).
%in order to meet the need to distinguish variables of the liquid zone from those of the gas zone, we mark them with the accents $\,\check{}\,$ and $\,\hat{}\,$, respectively.
The P scheme requires the imposition of the density's discontinuity
\begin{subequations}\label{bc.i.P}
  \begin{align}
     \check{\xiy}(\etai) & = \xiy_{l}    \label{bc.i.P.liq}  \\[.5\baselineskip]
     \hat{\xiy}(\etai)   & = \xiy_{g}    \label{bc.i.P.gas}
  \end{align}
\end{subequations}
The M$_{1}$ scheme calls for the imposition of the pressure's continuity
\begin{subequations}\label{bc.i.M1}
  \begin{align}
     \check{\pnd}(\etai) & = \pnd^{\ast}  \label{bc.i.M1.liq}  \\[.5\baselineskip]
     \hat{\pnd}(\etai)   & = \pnd^{\ast}  \label{bc.i.M1.gas}
  \end{align}
\end{subequations}
The M$_{2}$ scheme demands the imposition of the reduced chemical potential's continuity
\begin{subequations}\label{bc.i.M2}
  \begin{align}
     \check{\phi}(\etai) & = \phi^{\ast}  \label{bc.i.M2.liq}  \\[.5\baselineskip]
     \hat{\phi}(\etai)   & = \phi^{\ast}  \label{bc.i.M2.gas}
  \end{align}
\end{subequations}
The unknown position $\etai$ of the interface is fixed by the gravitational field's continuity that applies in the form of \REq{gfc} in the P, M$_{1}$ schemes \footnote{\color{black} We made an interesting numerical experiment in the ``bvp'' variant of the P scheme by introducing a fictitious \textit{differential} equation for $\,\protect\etayf^{\ast}$ that, after a suitable change of coordinate, we have solved together with the others on the interval $[0,1]$ with the boundary conditions defined by \REqq{bc.nd}, \REq{gfc}, \REqq{bc.i.P}.} but that translates to the more explicit form
\begin{equation}\label{bc.i.gfc.M2}
  \pds{}{\check{\phi}}{\etay}{\!\!\etays=\etays^{\ast}} - \pds{}{\hat{\phi}}{\etay}{\!\!\etays=\etays^{\ast}} = 0
\end{equation}
in the M$_{2}$ scheme, according to \REq{gfcp}.
\begin{figure}[t]
  \includegraphics[keepaspectratio=true, trim= 9ex 5ex 3ex 6ex , clip , width=\columnwidth]{\figdir/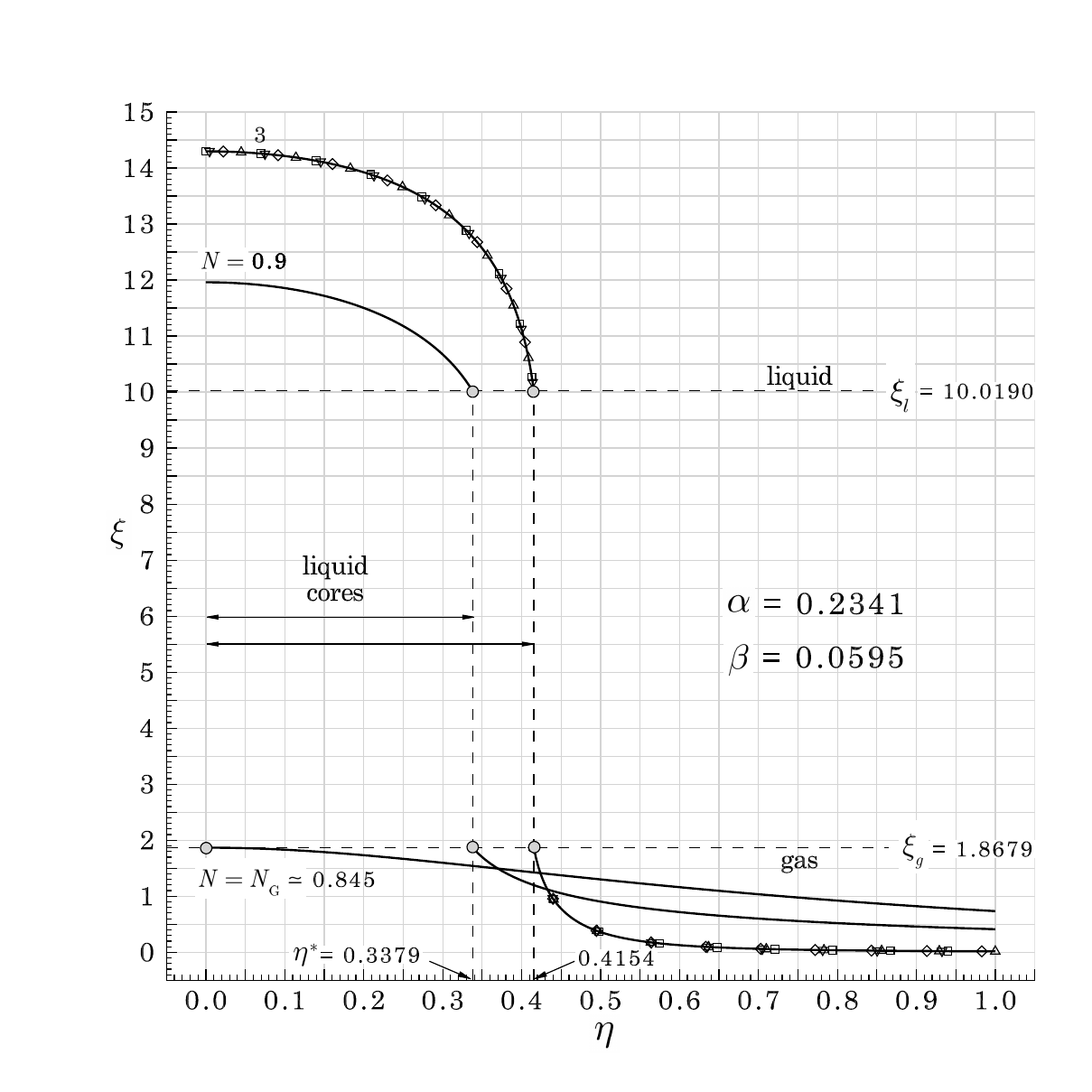}
  \caption{Radial profiles of density with presence of phase equilibrium at selected values of the gravitational number. The radial profile at \itm{N=\Ni} is included for convenience with a view to remarks forthcoming in \Rse{ysim}.\hfill\ }\label{rd-pe}
\end{figure}
The boundary-condition complexity introduced into the computational schemes by the presence of the interface is absorbed without difficulties by the robustness of the numerical algorithms.
Typical radial profiles of density, pressure, reduced chemical potential and gravitational field are shown in \Rfis{rd-pe}{rgf-pe}.
The profiles of reduced chemical potential (\Rfi{rrcp-pe}) are definitely aesthetically elegant due to the function's and its first derivative's continuity; the disturbance produced by the second derivative's discontinuity
\begin{equation}\label{sdd}
   \frac{1}{3N} \left[ \pds{2}{\hat{\phi}}{\etay}{\!\!\etays=\etays^{\ast}} - \pds{2}{\check{\phi}}{\etay}{\!\!\etays=\etays^{\ast}} \right] = \xiy_{l} - \xiy_{g} > 0
\end{equation}
at the interface is virtually invisible in \Rfi{rrcp-pe} but appears fiercely in \Rfi{rgf-pe}.
Visual comparison between \Rfi{rpd-mss} and \Rfid{rd-pe}{rp-pe} for \itm{N=0.9} indicates that liquid cores are substantially more extended than metastable cores and that the profiles are much steeper in the presence of phase equilibrium than of gas' metastable states.
%\textbf{striking features ...}
\begin{figure}[t]%
  \includegraphics[keepaspectratio=true, trim= 9ex 5ex 3ex 6ex , clip , width=\columnwidth]{\figdir/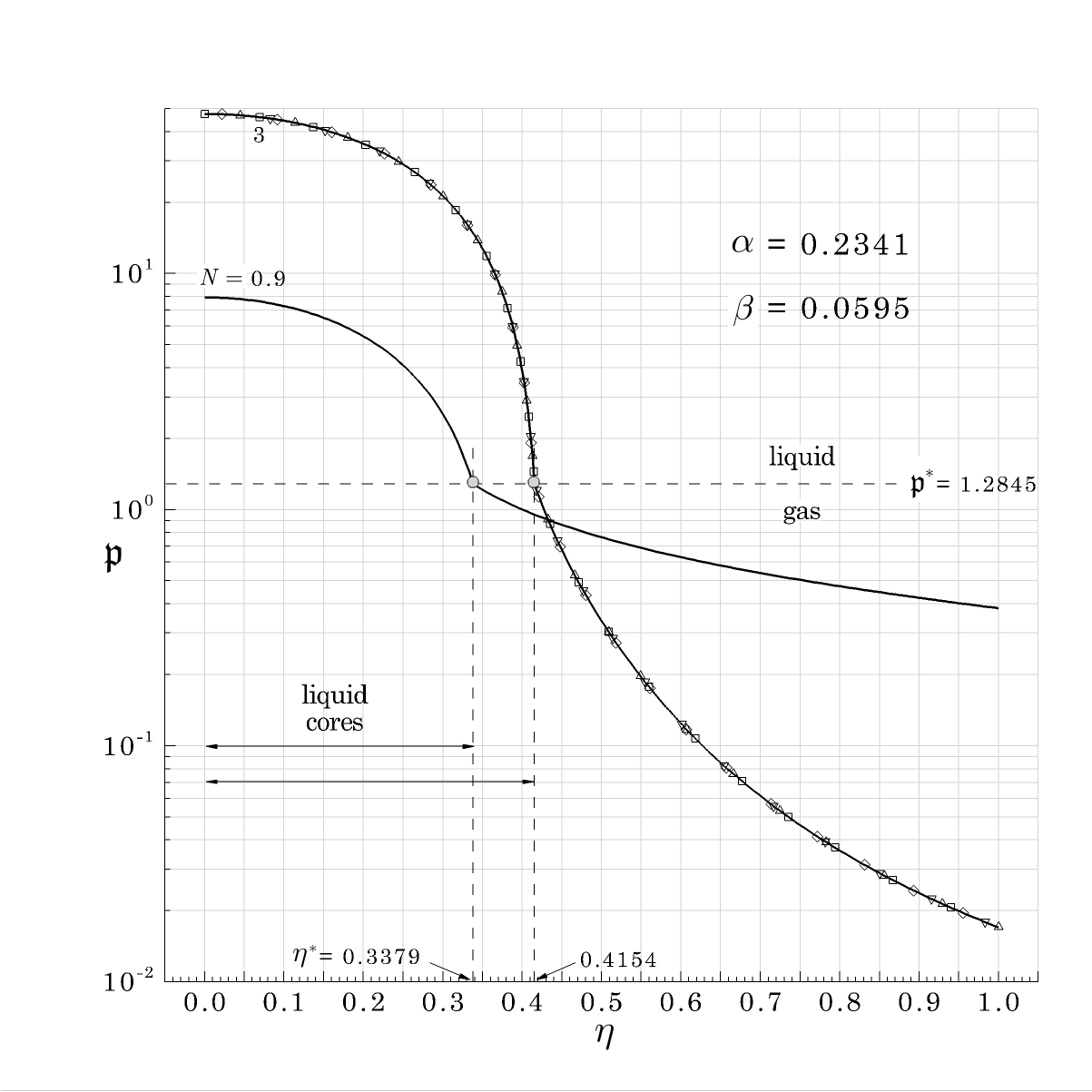}
  \caption{Radial profiles of pressure with presence of phase equilibrium at selected values of the gravitational number.\hfill\ }\label{rp-pe}
\end{figure}%
\begin{figure}[b!]%
  \includegraphics[keepaspectratio=true, trim= 9ex 5ex 3ex 6ex , clip , width=\columnwidth]{\figdir/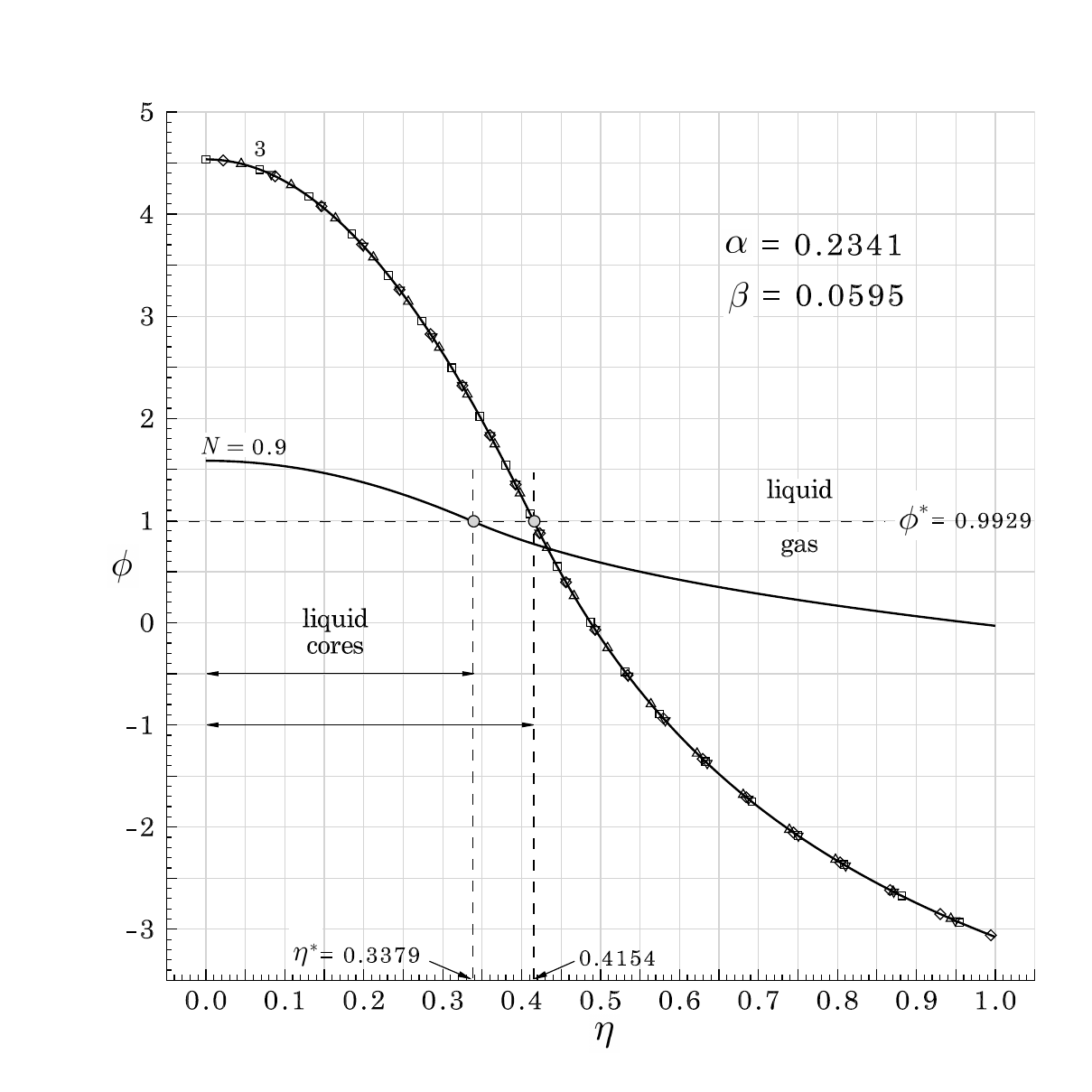}
  \caption{Radial profiles of reduced chemical potential with presence of phase equilibrium at selected values of the gravitational number.\hfill\ }\label{rrcp-pe}
\end{figure}%
\begin{figure}[h]%
  \includegraphics[keepaspectratio=true, trim= 9ex 5ex 3ex 6ex , clip , width=\columnwidth]{\figdir/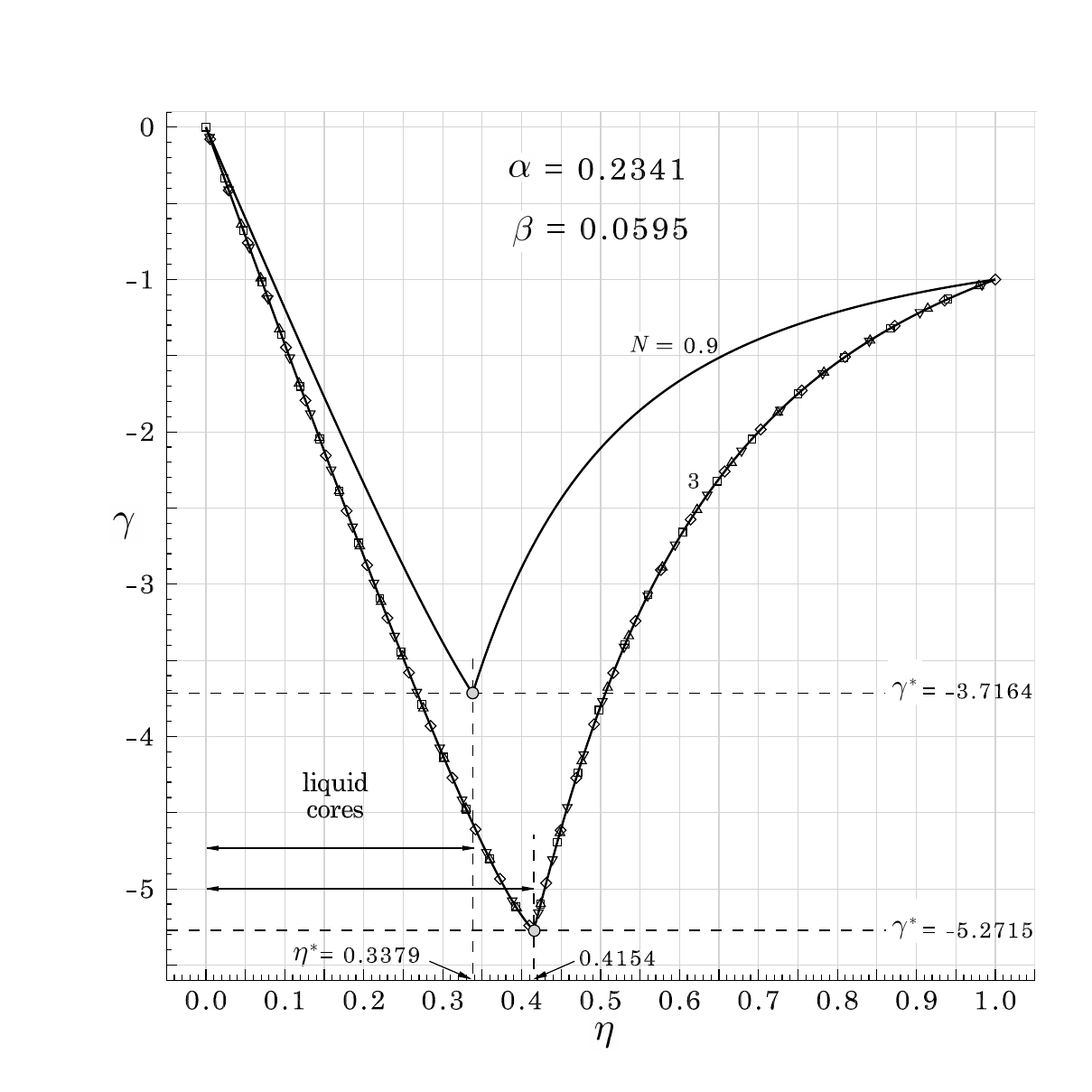}
  \caption{Radial profiles of gravitational field with presence of phase equilibrium at selected values of the gravitational number.\hfill\ }\label{rgf-pe}
\end{figure}%
Similarly to what we experienced with \itm{\alphay=0.1053}, also the calculations with \itm{\alphay=0.2341} did not reveal any unexpected surprise: they consistently attested uniqueness of solutions for both \itm{N\le\Ni}, via the algorithms without knowledge of phase-equilibrium, and \itm{N>\Ni}, via the algorithms adapted to the presence of phase equilibrium.
The perfect gas' barrier at $N_{m}$ was crossed without any numerical flinch, as testified by the smooth radial profiles at \itm{N=3} in \Rfis{rd-pe}{rgf-pe}; here again the results of the five independent algorithms are superposed for the same reason we mentioned in the case \itm{\alphay=0.1053}.

\subsubsection{Yet something is missing\label{ysim}}
The description of the circumstances with phase equilibrium given in \Rse{agepei} sounds physically convincing but leaves implied that phase equilibria reside above the threshold set by \itm{\Ni}.
% corresponding to the specified couple \itm{\alphay, \betay}.
Yet, a more attentive look at \Rfi{rd-pe} reveals a detail suggesting that perhaps that description may be incomplete and something may be missing in it.
The incipient phase-equilibrium formation at \itm{\Ni\simeq0.845} we have encountered in \Rfi{N=0.845isoT}, and whose radial profile of density is also shown for convenience in \Rfi{rd-pe}, presupposes \itm{\etai=0}; on the other hand, the phase equilibrium at \itm{N=0.9}, decidedly a value of the gravitational number not far from $\Ni$, presents the interface at \itm{\etai=0.3379} which happens to be a rather substantial distance from the sphere centre.
On the basis of the description in \Rse{agepei}, it seems intuitively reasonable to expect that there must exist a slight increase of the gravitational number above $\Ni\simeq0.845$, say \itm{N=0.85} for example, which should give rise to a phase equilibrium with interface located rather near the sphere centre, say \itm{\etai<0.1} for example.
Well, all our efforts to extract solutions corresponding to such a situation turned out to be in vain.

\subsubsection{Phase equilibrium with vanishingly small gravitational number\label{as.N=0.pe}}
Our first move in the pursuit of an explanation for the mysterious absence of solutions was inspired by the idea investigated in \Rse{as.N=0}.
We ask: are there solutions \textit{with phase equilibrium} for vanishingly small gravitational number?
We go back to \REq{sode.ss.vdw.nd.N=0} to obtain \textit{separately} the approximated analytical solutions in the liquid zone \itm{[0,\etai]}
\begin{subequations}\label{pe.N=0.l}
  \begin{align}
    \check{\phi}(\etay) & \simeq \phii  \label{pe.N=0.l.rcp} \\[.5\baselineskip]
    \check{\xiy}(\etay) & \simeq \xil   \label{pe.N=0.l.xi}
  \end{align}
\end{subequations}
and in the gas zone \itm{[\etai,1]}
\begin{subequations}\label{pe.N=0.g}
  \begin{align}
    \hat{\phi}(\etay) & \simeq \phii  \label{pe.N=0.g.rcp} \\[.5\baselineskip]
    \hat{\xiy}(\etay) & \simeq \xig   \label{pe.N=0.g.xi}
  \end{align}
\end{subequations}
For the determination of $\etai$, we should then go back to \REq{gfd.s.nd}, solve it again separately in the liquid and gas zones with due account of the corresponding densities' uniformity [\REqd{pe.N=0.l.xi}{pe.N=0.g.xi}], and impose the gravitational-field continuity [\REq{gfc}].
The integral of \REq{gfd.s.nd} with uniform density $\xiy$ is similar to \REq{gfd.s.nd.N=0.c.i} and reads
\begin{equation}\label{gf.nd.N=0.ud}
   \etay^{2} \gammay + \etay^{3} \xiy \simeq K''
\end{equation}
In the liquid zone, the evaluation of \REq{gf.nd.N=0.ud} at \itm{\etay=0} yields a vanishing constant $K''$ so that the equation reduces to a linear dependence
\begin{equation}\label{gf.nd.N=0.ud.l}
   \check{\gammay}(\etay)  \simeq - \xil\,\etay
\end{equation}
In the gas zone, the imposition of the peripheral boundary condition [\REq{bc.r=a.nd}] fixes the constant to the value
\begin{equation}\label{Kgas}
  K'' = - \left( 1 - \xig \right)
\end{equation}
and \REq{gf.nd.N=0.ud} becomes
\begin{equation}\label{gf.nd.N=0.ud.g}
   \hat{\gammay}(\etay)  \simeq - \frac{1 - \xig}{\etay^{2}} - \xig\,\etay
\end{equation}
The requirement of gravitational-field continuity at $\etai$
\begin{equation}\label{gf.nd.N=0.ud.gfc}
  - \xil\,\etai = - \frac{1 - \xig}{{\etai}^{2}} - \xig\,\etai
\end{equation}
yields the interface's location
\begin{equation}\label{peil}
   \etai = \left(\frac{1 - \xig}{\xil - \xig}\right)^{1/3}
\end{equation}
We could also have taken the shortcut offered by the normalisation condition indicated in \Reqma{43} because, given the uniformity of the densities [\REqd{pe.N=0.l.xi}{pe.N=0.g.xi}], its integrals are readily obtained
\begin{equation}\label{nc.pe}
%   3\int_{0}^{1}\etay^{2}\xiy d\etay \simeq 3\int_{0}^{\etai}\etay^{2}\xil d\etay + 3\int_{\etai}^{1}\etay^{2}\xig d\etay = 1
   \begin{split}
     1 = 3\int_{0}^{1}\etay^{2}\xiy d\etay & \simeq 3\int_{0}^{\etays^{\ast}}\etay^{2}\xil d\etay + 3\int_{\etays^{\ast}}^{1}\etay^{2}\xig d\etay \\[.35\baselineskip]
                                       & =      {\etai}^{3} \xil + \left( 1 - {\etai}^{3} \right) \xig
   \end{split}
\end{equation}
and the extraction of $\etai$ from \REq{nc.pe} returns more swiftly the same expression [\REq{peil}] obtained from the gravitational-field continuity.
Incidentally, the imprint of Gauss' theorem is evident in the analytical solution we have just obtained for the gravitational field.
It transpires from the linearity through the uniform-density liquid core expressed by \REq{gf.nd.N=0.ud.l}.
In the gas zone, it surfaces limpidly after the slight rearrangement of \REq{gf.nd.N=0.ud.g} to the form
\begin{equation}\label{gf.nd.N=0.ud.g.gf}
   \hat{\gammay}(\etay) \simeq - \frac{{\etai}^{3}}{\etay^{2}}\xil  - \frac{\etay^{3} - {\etai}^{3}}{\etay^{2}}\xig
\end{equation}
whose right-hand side reveals the central-field contributions of the liquid core followed by that of the gas contained in the volume limited by the external sphere of radius $\etay$ and the internal sphere of radius $\etai$; \REq{gf.nd.N=0.ud.g.gf} is obtained by extracting \itm{(1-\xig)} from \REq{peil} and by substituting the extracted expression in the numerator of the first term on the right-hand side of \REq{gf.nd.N=0.ud.g}.

\begin{figure}[h]
  \includegraphics[keepaspectratio=true, trim= 6ex 3ex 3ex 6ex , clip , width=\columnwidth]{\figdir/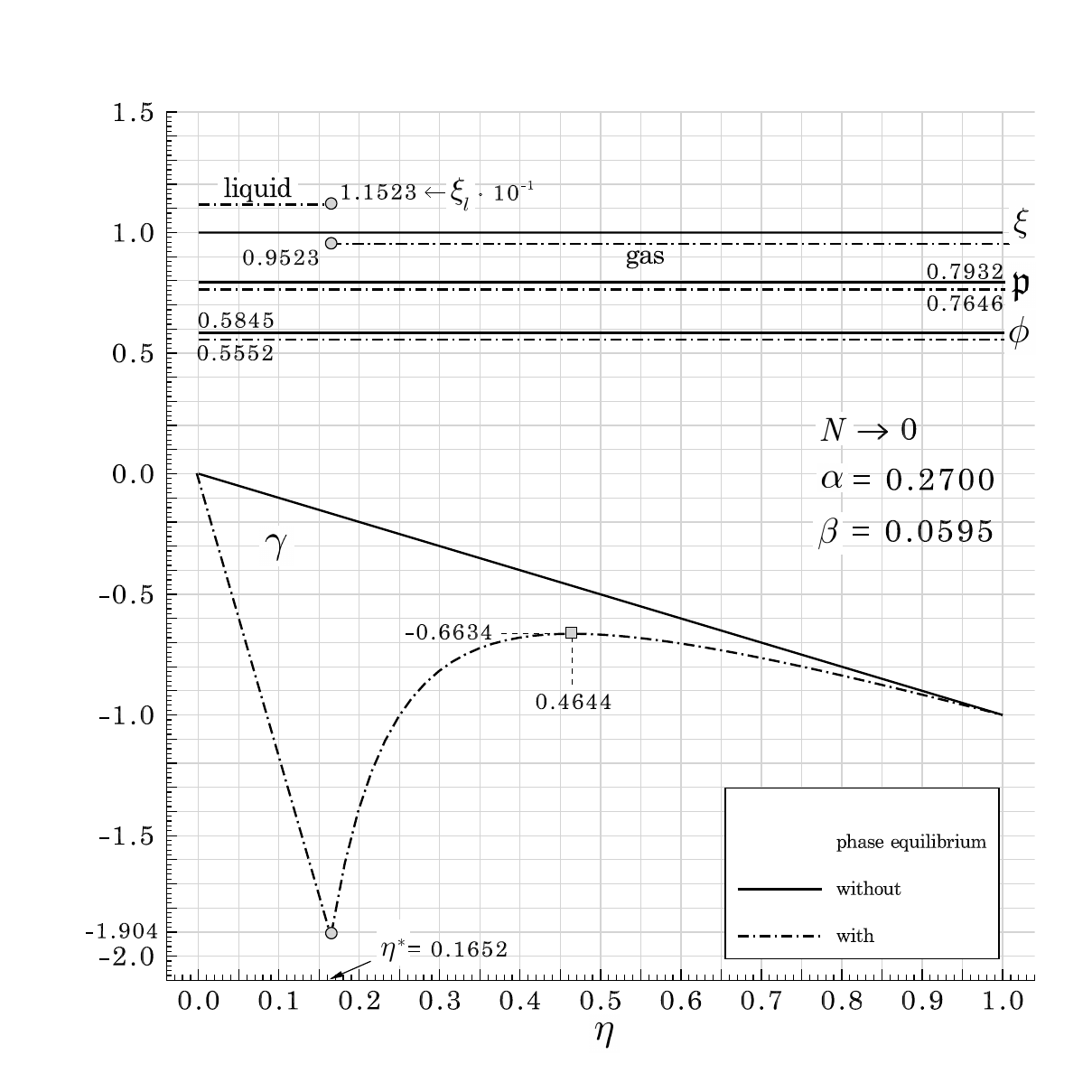}
  \caption{The two analytical solutions, without and with phase equilibrium, for vanishing gravitational number and with \itm{\protect\alphay=0.2700} and $\protect\betay=0.0595$. The saturated liquid density is scaled by a factor $10^{-1}$ to keep in good view the gravitational-field radial profiles.\hfill\ }\label{asN=0alpha=0.27}
\end{figure}
\REqb{peil} confirms that phase equilibria can indeed exist for vanishingly small gravitational number but their existence
%\begin{equation}\label{pee}
%   0 \le \etai \le 1
%\end{equation}
is subjected to the fulfilment of the condition
\begin{equation}\label{pe.N=0.cond}
   \xig \le 1 \le \xil
\end{equation}
because only then $\etai$ happens to belong to [0,1].
This existence condition is not fulfilled with \itm{\alphay=0.2341} and \itm{\betay=0.0595} because \itm{\xig=1.868>1} (\Rta{phaseeqv}) but it is easy to find couples of values for which the existence condition turns out to be verified.
For example, with \itm{\alphay=0.2700} and \itm{\betay=0.0595} the saturated gas and liquid densities become \itm{\xig\simeq0.955<1} and \itm{\xil\simeq11.514>1} and the interface is positioned at \itm{\etai\simeq0.1652} \footnote{By the way, in the circumstance of vanishing gravitational number, the configuration consisting of a liquid core surrounded by a gas layer is not the only possible one; the other configuration, gas core in \itm{[0,\protect\etayf^{\ast}]} surrounded by liquid layer in \itm{[\protect\etayf^{\ast},1]}, is also possible. The corresponding formulae are obtained from \REqq{pe.N=0.l}, \REqq{pe.N=0.g}, \REqs{gf.nd.N=0.ud.l}{gf.nd.N=0.ud.g} and \REq{nc.pe} by swapping the subscripts $l,g$ and the accents $\,\check{}\, , \hat{}\,$ wherever they occur. \REqb{peil} becomes
\[\protect\etayf^{\ast} = \left(\frac{\protect\xiyf_{l} - 1}{\protect\xiyf_{l} - \protect\xiyf_{g}}\right)^{1/3}\]
and the existence condition [\REq{pe.N=0.cond}] remains unaltered.
In the case of the numerical example of the text, the \textit{gas-liquid} interface's location would be \itm{\protect\etayf^{\ast} \simeq 0.9999}; thus, there would be an extremely thin liquid layer separating the gas core from the container.}.
The radial profiles of the solutions without and with phase equilibrium are shown in \Rfi{asN=0alpha=0.27}.
%\begin{figure}[t]
%  \includegraphics[keepaspectratio=true, trim= 6ex 3ex 3ex 6ex , clip , width=\columnwidth]{\figdir/asN=0alpha=0.27.pdf}
%  \caption{The two analytical solutions, without and with phase equilibrium, for vanishing gravitational number and with \itm{\alphayc=0.27} and $\betayc=0.0595$. The saturated liquid density is scaled by a factor $10^{-1}$ to keep in good view the gravitational-field radial profiles.\hfill\ }\label{asN=0alpha=0.27}
%\end{figure}

\subsubsection{A better understanding of phase-equilibrium occurrence\label{abu}}

The exercise of \Rse{as.N=0.pe} delivers an important message: there can definitely exist phase equilibria for vanishingly small values of the gravitational number.
And that occurrence clearly also suggests that maybe the existence of phase equilibria below the threshold set by $\Ni$ should not intuitively be ruled out; therefore, we looked for and found them by applying the phase-equilibrium adapted algorithms with \itm{N\le\Ni}.
The interesting results we obtained are well rendered by the interdependence between interface's location and gravitational number illustrated in \Rfi{nvetas}.%
\begin{figure}[h]%
  \includegraphics[keepaspectratio=true, trim= 6ex 3ex 3ex 6ex , clip , width=\columnwidth]{\figdir/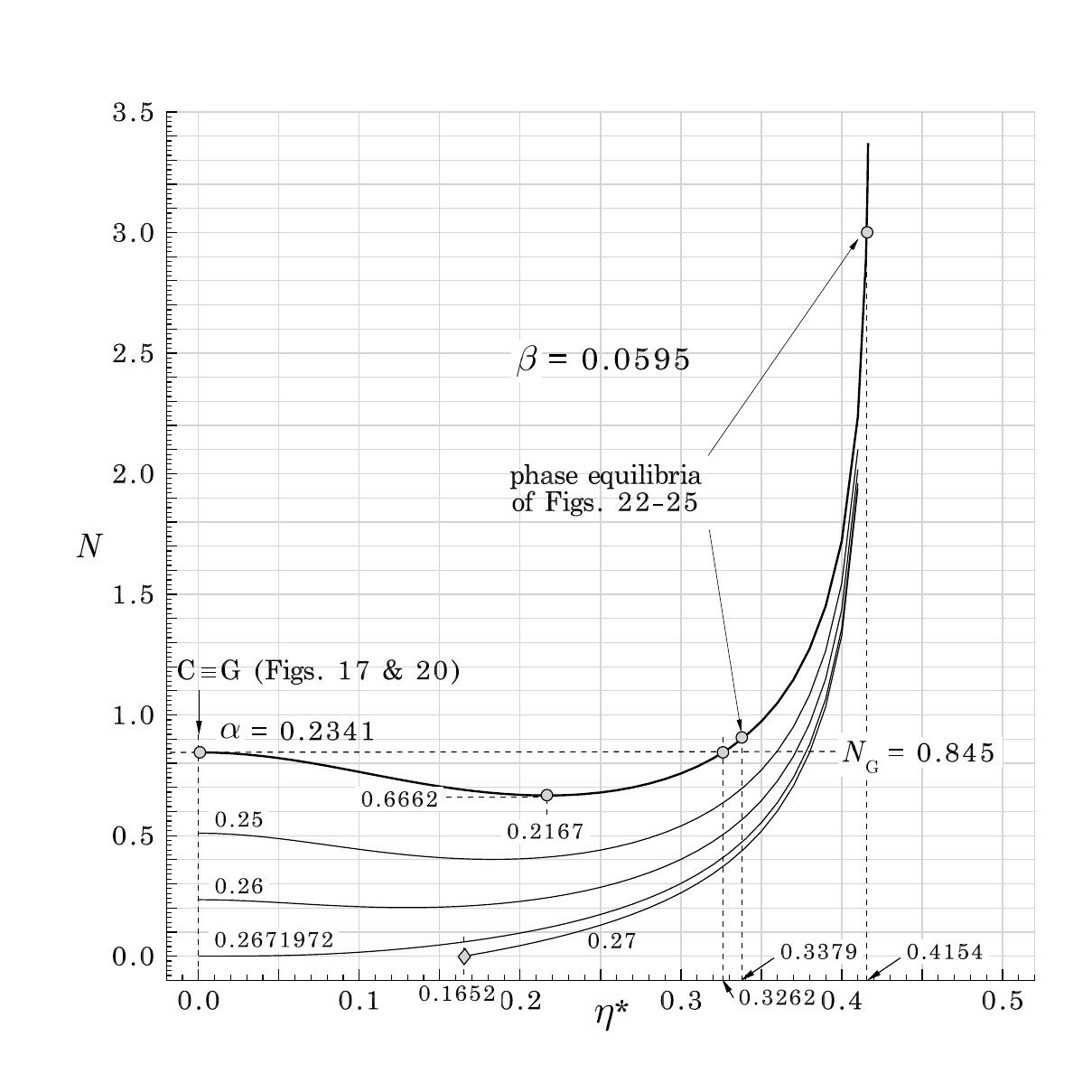}
  \caption{The interdependence between the phase-equilibrium interface's position $\protect\etay^{\ast}$ and the gravitational number $N$ for several values of $\protect\alphay$ with \itm{\protect\betay=0.0595}.\lpush}\label{nvetas}
\end{figure}
The lowest curve corresponds to the numerical example considered at the end of \Rse{as.N=0.pe} with \itm{\alphay=0.2700} and \itm{\betay=0.0595}; the existence condition [\REq{pe.N=0.cond}] is fulfilled and there exists a phase equilibrium at \itm{N\simeq0} whose interface's position is located at \itm{\etai\simeq0.1652} (lozenge symbol).
The dependence of $\etai$ on $N$ is monotonic on this curve so that there is one solution with phase equilibrium that accompanies the solution without, obviously not visible in this graph.
%are two solutions for a specified gravitational number: one without and one with phase equilibrium.
If $\alphay$ decreases then the curve and its intersection with the horizontal axis shift to the left and the interface's position moves towards the sphere centre until a limit value is reached, for example \itm{\alphay\simeq0.2671972} for \itm{\betay=0.0595}, in correspondence to which the curve originates from \itm{\etai=0}; this is the last possibility for the existence of phase equilibrium for vanishingly small values of the gravitational number.
With reference to this limiting curve, we call the reader's attention to an important detail that reinforces the suggestion regarding the existence of phase equilibria below $\Ni$: the slope \itm{\tdst{}{N}{{\etai}}{\etays^{\ast}=0}} vanishes.
This is the hallmark of an incoming minimum; and indeed, if $\alphay$ lowers below the limit value, i.e. \itm{\alphay<0.2671972}, then a minimum appears in the curves.
The existence condition [\REq{pe.N=0.cond}] is now infringed and there are no more phase equilibria for vanishingly small gravitational number.
The shape of the curves proves beyond any doubt the existence of phase equilibria below $\Ni$ for a given $\alphay$ and resolves the mystery of why we were not able to detect solutions slightly above $\Ni$ with the interface's position in the neighbourhood of the sphere centre.

%\begin{figure}%[h]
%  \includegraphics[keepaspectratio=true, trim= 6ex 3ex 3ex 6ex , clip , width=\columnwidth]{\figdir/rd-pe-NG.pdf}
%%  \caption{The interdependence between the phase-equilibrium interface's position $\etayc^{\ast}$ on the gravitational number $N$ for several values of \alphayc.\hfill\ }\label{rd-pe-NG}
%  \caption{The radial profiles of density without and with presence of phase equilibrium at \itm{N=\Ni\simeq0.845}. No phase equilibrium is possible in between.\hfill\ }\label{rd-pe-NG}
%\end{figure}
\begin{figure}[h]%
  \includegraphics[keepaspectratio=true, trim= 6ex 3ex 3ex 6ex , clip , width=\columnwidth]{\figdir/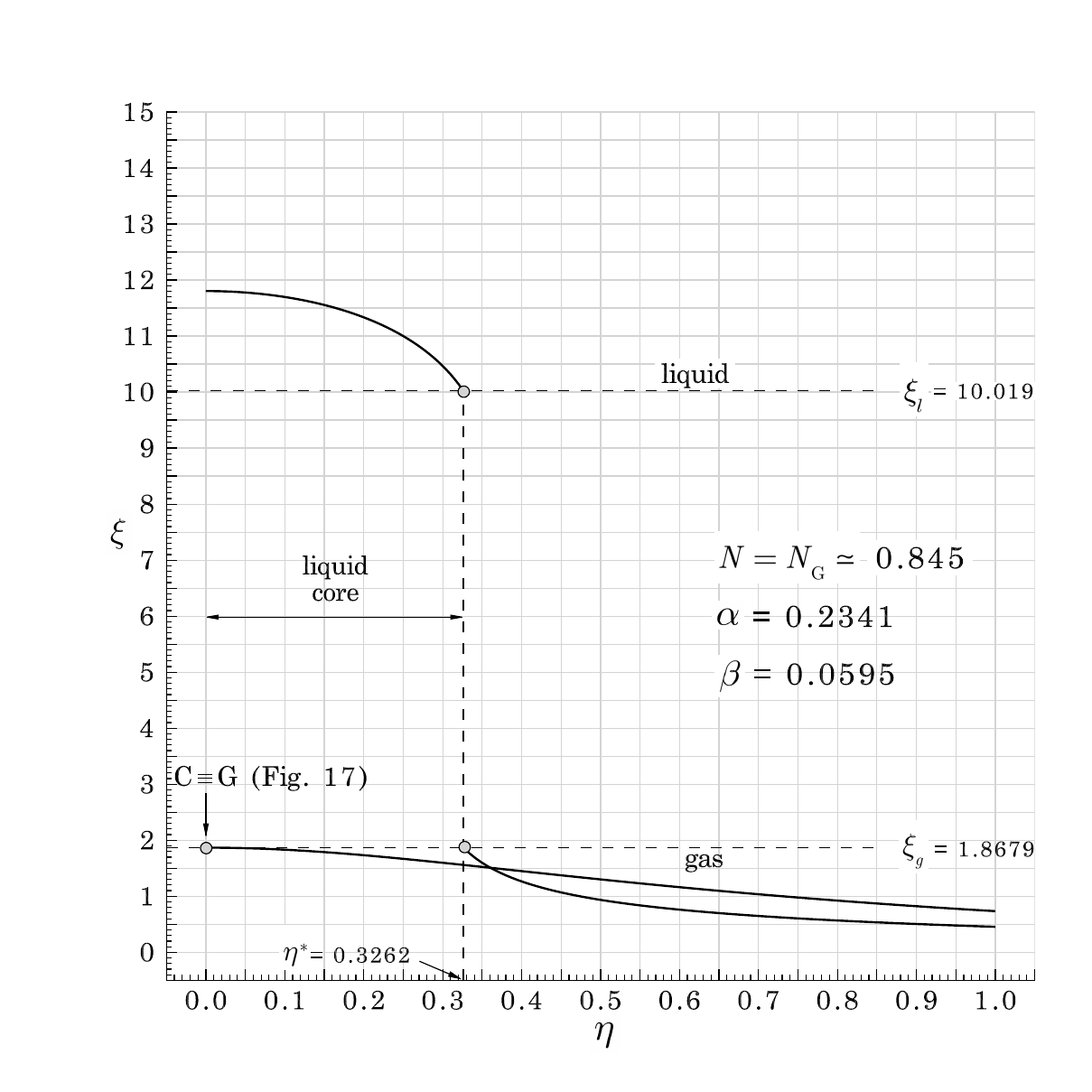}
%  \caption{The interdependence between the phase-equilibrium interface's position $\etayc^{\ast}$ on the gravitational number $N$ for several values of \alphayc.\hfill\ }\label{rd-pe-NG}
  \caption{The radial profiles of density without and with presence of phase equilibrium at \itm{N=\Ni\simeq0.845}. No phase equilibrium is possible in between.\hfill\ }\label{rd-pe-NG}
\end{figure}
\Rfib{nvetas} certainly delivers a better understanding of phase-equilibrium occurrence.
Let us concentrate on the curve with \itm{\alphay=0.2341}.
If the gravitational number is below the minimum, \itm{N<0.6662}, then there is only the solution without phase equilibrium, such as the one shown in \Rfid{N=0.5rpd}{N=0.5rgf} with \itm{N=0.5}.
If the gravitational number coincides with the minimum, \itm{N=0.6662}, then a solution with phase equilibrium whose interface is positioned at \itm{\etaim=0.2167} becomes possible and accompanies the solution without phase equilibrium.
So, phase equilibrium does not appear at the sphere centre but inside the sphere at a location $\etaim$ determined by $\alphay$ and $\betay$.
Further increase of the gravitational number introduces two possible solutions with phase equilibrium, one to the left and one to the right of the interface's location $\etaim$ corresponding to the minimum;
%The left solution's interface moves towards the sphere centre and reaches it when \itm{N=\Ni}; the right solution's interface moves towards the sphere wall and attains a specific position, \itm{\etai=0.3262}, when \itm{N=\Ni}.
the left solution's interface moves towards the sphere centre while the right solution's interface moves towards the sphere wall.
When \itm{N=\Ni\simeq0.845}, the former interface reaches the sphere centre and the latter attains a specific position, \itm{\etai=0.3262}; the radial profiles of density corresponding to this situation are shown in \Rfi{rd-pe-NG}.
Hence, if the gravitational number ranges between the minimum and $\Ni$ then there are three solutions: one without and two with phase equilibrium.
Above $\Ni$, the metastable-state solutions settle in up to \itm{\NM\simeq0.993} (\Rfi{rpd-mss}) and accompany the single solution with phase equilibrium.
Above \itm{\NM} there is only one solution with phase equilibrium.
%\begin{figure}[t]%
%  \includegraphics[keepaspectratio=true, trim= 6ex 3ex 3ex 6ex , clip , width=\columnwidth]{\figdir/rd-pe-NG.pdf}
%%  \caption{The interdependence between the phase-equilibrium interface's position $\etayc^{\ast}$ on the gravitational number $N$ for several values of \alphayc.\hfill\ }\label{rd-pe-NG}
%  \caption{The radial profiles of density without and with presence of phase equilibrium at \itm{N=\Ni\simeq0.845}. No phase equilibrium is possible in between.\hfill\ }\label{rd-pe-NG}
%\end{figure}

The asymptotic behaviour of the curves in \Rfi{nvetas} is easily explainable with the help of the normalisation condition given by \Rma{Eq. (43)} adapted to the circumstance of phase equilibrium
\begin{equation}\label{nc.pe.a}
     1 = 3\int_{0}^{1}\etay^{2}\xiy d\etay =  3\int_{0}^{\etays^{\ast}}\etay^{2}\xiy d\etay + 3\int_{\etays^{\ast}}^{1}\etay^{2}\xiy d\etay
\end{equation}
The visual inspection of the density profile corresponding to \itm{N=3} in \Rfi{rd-pe} indicates that if \itm{N \rightarrow \infty}  then the profile seemingly tends towards a step function with the interface positioned at a rightmost value $\etay^{\ast}_{\ssub{0.5}{\mbox{a}}}$, about 0.42 in \Rfi{nvetas}; we could (mathematically) imagine that the liquid portion of the profile tends to flatten at the density value \itm{\xiy=1/\betay} while the gas portion of the profile tends to flatten at the density value \itm{\xiy=0}.
Well, almost but not exactly because at \itm{\etay=\etay^{\ast}_{\ssub{0.5}{\mbox{a}}}} the liquid density is anchored at the value $\xil$ and the gas density is anchored at the value $\xig$;
so, there will always exist an however small left neighbourhood of $\etay^{\ast}_{\ssub{0.5}{\mbox{a}}}$ in which the liquid density plunges from the top of the step down to $\xil$ and an however small right neighbourhood of $\etay^{\ast}_{\ssub{0.5}{\mbox{a}}}$ in which the gas density surges from the bottom of the step up to $\xig$.
Therefore the top of the step must settle at an extremely low amount below \itm{1/\betay} and the bottom of the step must settle at an extremely low amount above 0.
With this picture in mind, we can now process the integrals in \REq{nc.pe.a}.
First, we change the integration variable
\begin{subequations}\label{nc.pe.a.e}
  \begin{equation}\label{nc.pe.a.e.iv}
       1 =  \int_{0}^{\etays^{\ast}}\xiy\,d(\etay^{3}) + \int_{\etays^{\ast}}^{1}\xiy\,d(\etay^{3})
  \end{equation}
  and then apply the mean-value theorem
  \begin{equation}\label{nc.pe.a.e.mvt}
       1 = \check{\xiy}_{\ssub{0.5}{\mbox{mean}}}\cdot(\etay^{\ast})^{3} + \hat{\xiy}_{\ssub{0.5}{\mbox{mean}}}\cdot\left[ 1 - (\etay^{\ast})^{3} \right]
  \end{equation}
  Now, if \itm{N\rightarrow\infty} then
  \begin{equation}\label{nc.pe.a.e.eta}
     \etay^{\ast}\rightarrow\etay^{\ast}_{\ssub{0.5}{\mbox{a}}}
  \end{equation}
  and, in accordance with the picture described a few lines ago, we are permitted to set
  \begin{align}
     \check{\xiy}_{\ssub{0.5}{\mbox{mean}}} & \rightarrow \frac{1}{\betay} \left( 1 - \check{\epsilony} \right) \label{nc.pe.a.e.mt}\\
       \hat{\xiy}_{\ssub{0.5}{\mbox{mean}}} & \rightarrow 0 + \hat{\epsilony} \label{nc.pe.a.e.mb}
  \end{align}
  with both the offsets \itm{\check{\epsilony},\hat{\epsilony}\ll 1}.
We can then easily substitute \REqs{nc.pe.a.e.eta}{nc.pe.a.e.mb} into \REq{nc.pe.a.e.mvt}, linearise to first order the resulting equation thanks to the smallness of the offsets, and solve for the asymptotic position of the interface
\begin{equation}\label{nc.pe.a.e.etaa}
   \etay^{\ast}_{\ssub{0.5}{\mbox{a}}} \simeq \sqrt[3]{\betay}\cdot \left[ 1 + \frac{\check{\epsilony}- \left(1-\betay\right)\hat{\epsilony}}{3}   \right]
\end{equation}
\end{subequations}
Thus, \itm{\sqrt[3]{\betay}} is the zeroth-order estimate; it equals 0.39 in the case of our \itm{\betay=0.0595} which is acceptably near the value 0.42 indicated in \Rfi{nvetas}.

This is, clearly, the situation in which the mass in the gas zone is negligible and the fluid is just a liquid core of radius $\etay^{\ast}_{\ssub{0.5}{\mbox{a}}}$; the presence of the container becomes irrelevant.

\section{Conclusions\label{concl}}

The assumption that self-gravitating gas spheres behave according to the perfect-gas thermodynamic model traces back to the first investigations \cite{hl1870ajs,eb1880inc,ar1882adp,wt1887pma,gh1888aom,gd1889ptrs,re1907} on the subject matter and was/is taken by glossing over a substantial inconsistency: that a thermodynamic model built on the fundamental idea of non-interacting particles, on one side, be capable to describe situations in which gravitational attraction among particles acquires the primary role as driving mechanism to determine fluid-static spatial distributions of thermodynamic properties and gravitational field, on the other side.
There is, of course, no blame in taking that assumption, for obvious reasons.
The perfect-gas model is characterised by mathematical simplicity, particularly when injected into fluid dynamics or, better, statics in our case, that favours easiness of calculations.
As a matter of fact, there exist other physical situations in which the perfect-gas model is put to work accompanied by the same inconsistency but the latter does not surface enough to invalidate results.
For example, chemically reacting flows are modelled as mixtures of perfect gases in many applications even if, strictly speaking, the absence of interactions among particles inhibits the occurrence of chemical processes.
%chemical processes in a perfect gas are impeded by the absence of interactions among particles.
%the perfect gas cannot react chemically due to the absence of interactions among particles.
Nevertheless, the assumption fits the circumstances in which the fraction of molecules that do indeed react, and in so doing do not behave as perfect-gas molecules, is negligible with respect to the total number of molecules so that thermodynamic properties are still predicted within an acceptable approximation \footnote{Chemical reaction rates, though, are affected by that tiny fraction of molecules and their determination requires more sophisticated approaches.}.
Unfortunately in the case of self-gravitating gas spheres, the mentioned inconsistency does come to surface and the hallmarks denouncing it have been in front of our eyes since long time: spiralling behaviour of the peripheral density (\Rfi{xi1}), oscillations of the central density (\Rfi{xi0}), existence of multiple solutions for the same gravitational number (\Rma{Figs. 7} and \Rma{8}), and existence of the gravitational number's upper bound \itm{N_{m}\simeq2.5175}.

Motivated by dissatisfaction and skepticism about those hallmarks and armed with scientific curiosity, we went a step further by selecting the vdW model and accepted, in so doing, the compromise of more complex mathematics in exchange of physical consistency, although admittedly more qualitative rather than quantitative, brought in by the terms accounting for molecular attraction \itm{A\,\rhoy^{2}} and molecules' finite size \itm{B\,\rhoy} [\REq{vdwse}].
We have formulated the resulting governing equations according to three different schemes of differential equations, two of first order (P, M$_{1}$) and one of second order (M$_{2}$), and applied five different and independently coded numerical algorithms to solve them.
The algorithms have returned systematically concurring results, which we take to be a solid proof of confidence. %, that we have described and commented in the previous sections.
We consider most significative among them those portrayed in the graphs of peripheral and central densities (\Rfid{xi1}{xi0}) because they indicate unequivocally the more physics-compliant behaviour of the vdW-model curves.
They are monotonic with respect to the gravitational number which is free to increase indefinitely; then, asymptotically, the peripheral density vanishes so that the presence of the container becomes practically irrelevant while the central density approaches the geometrical limit \itm{1/\betay} imposed by the finiteness of the molecules' size which conveys the picture of a core of packed molecules rather than the perfect gas' central point of infinite density.
The hallmarks listed at the end of the previous paragraph that have boggled scientific minds for almost 150 years and generated by the misuse of the perfect-gas model when put at work within a self-gravitating environment disappear unambiguously as a consequence of the physical consistency built in the vdW state equation with respect to molecular attraction.
%We are aware, of course, that the vdW-model curves of \Rfid{xi1}{xi0} do not qualify as solid proof ruling out unconditionally the existence of the mentioned unphysical results for arbitrary values of $\alphay$ and $\betay$; and, indeed, the study regarding existence or absence of values in correspondence to which the perfect-gas unphysical results appear also for the vdW-model is a rather interesting one but we have put it in the future-work stack for the time being.

%Although thermodynamics is not in the central focus of this study, we consider worth to express a few considerations.
%\Rma{Eq. (104)} \Rma{Eqs. (109a)}\Rma{\,and (110)} \Rma{Eq. (113)-gas} \Rma{Eq. (117c)}\Rma{\,or Eq. (122)}

We have described briefly in \Rse{bst} how the sanitising action of the vdW model also extends to perfect-gas based thermodynamics with self-gravitation and invalidates concepts belonging to it such as prophesied gravothermal catastrophes, intriguing negative specific heats, and thermal/mechanical instabilities. %analyses \cite{dg2019ejmb} [and references cited therein after \Rma{Eq. (112)} at page 76 and after \Rma{Eq. (146)} at page 82].
%In this regard, we believe that thermodynamics with self-gravitation certainly requires a refreshing, restorative revision.

The built-in capability of the vdW model to account for liquid-gas phase equilibria turns out to be an added bonus that permits to study the circumstances under which those equilibria are gravitationally induced (\Rse{ppe}).
Central cores of either metastable gas states or stable liquid states surrounded by stable gas states are predicted flawlessly by the perfectly matched governing equations of thermodynamics (\Rse{lgpe}) and of gravito-fluid statics formulated in the computational schemes described at the end of \Rse{ndf}.
We cannot refrain but wonder about what opinion Eddington and Jeans would have had regarding these results within the perspective of their famous strong debate \cite{jj1926mnras,jj1927.03mnras,jj1927.07mnras,jj1928.02n,ae1928.02n,ae1928.03mnras,jj1928.03mnras,ae1928.03n,jj1928.03n,jj1929} and also that of other scientists who touched on this matter, whom we became aware of by reading the interesting survey of Robitaille \cite{pmr2011.Ipip,pmr2011.IIpip}.

Finally, we wish to close our concluding remarks with a thought- as well as future-work provoking reflection.
The important message that springs from the results we have presented in the preceding sections is loud and clear: perfect gas and self-gravitation are incompatible concepts within the Newtonian scheme of gravity.
But a perfect gas can certainly exist in curved spacetime where metric tensor and fluid static configurations must be determined within the general-relativity scheme of gravity \cite{ks1916skpaw,rt1939pr,ro1939pr,cm1970,sw1972,rw1984,bs2009,sc2019}; then, may the curved spacetime become flat?
If yes, to which simplified form does the general-relativity static solution with a perfect gas tend?

%\section*{Acknowledgments}

%\raggedbottom
\bibliographystyle{apsrev4-1}

\bibliography{/Users/dg/Library/texmf/bibtex/bib/mybibreflibrary.bib,../1/ffp.bib,./fm.bib}

\begin{thebibliography}{78}%
\makeatletter
\providecommand \@ifxundefined [1]{%
 \@ifx{#1\undefined}
}%
\providecommand \@ifnum [1]{%
 \ifnum #1\expandafter \@firstoftwo
 \else \expandafter \@secondoftwo
 \fi
}%
\providecommand \@ifx [1]{%
 \ifx #1\expandafter \@firstoftwo
 \else \expandafter \@secondoftwo
 \fi
}%
\providecommand \natexlab [1]{#1}%
\providecommand \enquote  [1]{``#1''}%
\providecommand \bibnamefont  [1]{#1}%
\providecommand \bibfnamefont [1]{#1}%
\providecommand \citenamefont [1]{#1}%
\providecommand \href@noop [0]{\@secondoftwo}%
\providecommand \href [0]{\begingroup \@sanitize@url \@href}%
\providecommand \@href[1]{\@@startlink{#1}\@@href}%
\providecommand \@@href[1]{\endgroup#1\@@endlink}%
\providecommand \@sanitize@url [0]{\catcode `\\12\catcode `\$12\catcode
  `\&12\catcode `\#12\catcode `\^12\catcode `\_12\catcode `\%12\relax}%
\providecommand \@@startlink[1]{}%
\providecommand \@@endlink[0]{}%
\providecommand \url  [0]{\begingroup\@sanitize@url \@url }%
\providecommand \@url [1]{\endgroup\@href {#1}{\urlprefix }}%
\providecommand \urlprefix  [0]{URL }%
\providecommand \Eprint [0]{\href }%
\providecommand \doibase [0]{http://dx.doi.org/}%
\providecommand \selectlanguage [0]{\@gobble}%
\providecommand \bibinfo  [0]{\@secondoftwo}%
\providecommand \bibfield  [0]{\@secondoftwo}%
\providecommand \translation [1]{[#1]}%
\providecommand \BibitemOpen [0]{}%
\providecommand \bibitemStop [0]{}%
\providecommand \bibitemNoStop [0]{.\EOS\space}%
\providecommand \EOS [0]{\spacefactor3000\relax}%
\providecommand \BibitemShut  [1]{\csname bibitem#1\endcsname}%
\let\auto@bib@innerbib\@empty
%</preamble>
\bibitem [{\citenamefont {Giordano}\ \emph {et~al.}(2019)\citenamefont
  {Giordano}, \citenamefont {Amodio}, \citenamefont {Iavernaro}, \citenamefont
  {Labianca}, \citenamefont {Lazzo}, \citenamefont {Mazzia},\ and\
  \citenamefont {Pisani}}]{dg2019ejmb}%
  \BibitemOpen
  \bibfield  {author} {\bibinfo {author} {\bibfnamefont {D.}~\bibnamefont
  {Giordano}}, \bibinfo {author} {\bibfnamefont {P.}~\bibnamefont {Amodio}},
  \bibinfo {author} {\bibfnamefont {F.}~\bibnamefont {Iavernaro}}, \bibinfo
  {author} {\bibfnamefont {A.}~\bibnamefont {Labianca}}, \bibinfo {author}
  {\bibfnamefont {M.}~\bibnamefont {Lazzo}}, \bibinfo {author} {\bibfnamefont
  {F.}~\bibnamefont {Mazzia}}, \ and\ \bibinfo {author} {\bibfnamefont
  {L.}~\bibnamefont {Pisani}},\ }\href@noop {} {\bibfield  {journal} {\bibinfo
  {journal} {European Journal of Mechanics / B Fluids}\ }\textbf {\bibinfo
  {volume} {78}},\ \bibinfo {pages} {66} (\bibinfo {year} {2019})}\BibitemShut
  {NoStop}%
\bibitem [{\citenamefont {Lane}(1870)}]{hl1870ajs}%
  \BibitemOpen
  \bibfield  {author} {\bibinfo {author} {\bibfnamefont {H.~J.}\ \bibnamefont
  {Lane}},\ }\href@noop {} {\bibfield  {journal} {\bibinfo  {journal} {American
  Journal of Science}\ }\textbf {\bibinfo {volume} {50}},\ \bibinfo {pages}
  {57} (\bibinfo {year} {1870})}\BibitemShut {NoStop}%
\bibitem [{\citenamefont {Betti}(1880)}]{eb1880inc}%
  \BibitemOpen
  \bibfield  {author} {\bibinfo {author} {\bibfnamefont {E.}~\bibnamefont
  {Betti}},\ }\href@noop {} {\bibfield  {journal} {\bibinfo  {journal} {Il
  Nuovo Cimento}\ }\textbf {\bibinfo {volume} {7}},\ \bibinfo {pages} {26}
  (\bibinfo {year} {1880})}\BibitemShut {NoStop}%
\bibitem [{\citenamefont {Ritter}(1882)}]{ar1882adp}%
  \BibitemOpen
  \bibfield  {author} {\bibinfo {author} {\bibfnamefont {A.}~\bibnamefont
  {Ritter}},\ }\href@noop {} {\bibfield  {journal} {\bibinfo  {journal}
  {Annalen der Physik und Chemie}\ }\textbf {\bibinfo {volume} {252}},\
  \bibinfo {pages} {166} (\bibinfo {year} {1882})}\BibitemShut {NoStop}%
\bibitem [{\citenamefont {Thomson}(1887)}]{wt1887pma}%
  \BibitemOpen
  \bibfield  {author} {\bibinfo {author} {\bibfnamefont {W.}~\bibnamefont
  {Thomson}},\ }\href@noop {} {\bibfield  {journal} {\bibinfo  {journal}
  {Philosophical Magazine Series 5}\ }\textbf {\bibinfo {volume} {23}},\
  \bibinfo {pages} {287} (\bibinfo {year} {1887})}\BibitemShut {NoStop}%
\bibitem [{\citenamefont {Hill}(1888)}]{gh1888aom}%
  \BibitemOpen
  \bibfield  {author} {\bibinfo {author} {\bibfnamefont {G.~W.}\ \bibnamefont
  {Hill}},\ }\href@noop {} {\bibfield  {journal} {\bibinfo  {journal} {Annals
  of Mathematics}\ }\textbf {\bibinfo {volume} {4}},\ \bibinfo {pages} {19}
  (\bibinfo {year} {1888})}\BibitemShut {NoStop}%
\bibitem [{\citenamefont {Darwin}(1889)}]{gd1889ptrs}%
  \BibitemOpen
  \bibfield  {author} {\bibinfo {author} {\bibfnamefont {G.~H.}\ \bibnamefont
  {Darwin}},\ }\href@noop {} {\bibfield  {journal} {\bibinfo  {journal}
  {Philosophical Transactions of the Royal Society of London A}\ }\textbf
  {\bibinfo {volume} {180}},\ \bibinfo {pages} {1} (\bibinfo {year}
  {1889})}\BibitemShut {NoStop}%
\bibitem [{\citenamefont {Emden}(1907)}]{re1907}%
  \BibitemOpen
  \bibfield  {author} {\bibinfo {author} {\bibfnamefont {R.}~\bibnamefont
  {Emden}},\ }\href@noop {} {\emph {\bibinfo {title} {Gaskugeln}}}\ (\bibinfo
  {publisher} {Teubner},\ \bibinfo {address} {Leipzig, Germany},\ \bibinfo
  {year} {1907})\BibitemShut {NoStop}%
\bibitem [{Note1()}]{Note1}%
  \BibitemOpen
  \bibinfo {note} {The credit for the idea goes to {M. Sz\"{u}cs}.}\BibitemShut
  {Stop}%
\bibitem [{\citenamefont {{van der Waals}}(1873)}]{jvdw1873phd}%
  \BibitemOpen
  \bibfield  {author} {\bibinfo {author} {\bibfnamefont {J.~D.}\ \bibnamefont
  {{van der Waals}}},\ }\emph {\bibinfo {title} {Over de continuiteit van den
  gas- en vloeistoftoestand}},\ \href@noop {} {\bibinfo {type} {Academisch
  proefschrift}},\ \bibinfo  {school} {Hoogeschool te Leiden} (\bibinfo {year}
  {1873})\BibitemShut {NoStop}%
\bibitem [{\citenamefont {Saslaw}(1987)}]{ws1987}%
  \BibitemOpen
  \bibfield  {author} {\bibinfo {author} {\bibfnamefont {W.~C.}\ \bibnamefont
  {Saslaw}},\ }\href@noop {} {\emph {\bibinfo {title} {Gravitational physics of
  stellar and galactic systems}}},\ Cambridge monographs on mathematical
  physics\ (\bibinfo  {publisher} {Cambridge University Press},\ \bibinfo
  {address} {Cambridge, UK},\ \bibinfo {year} {1987})\BibitemShut {NoStop}%
\bibitem [{\citenamefont {{van Kampen}}(1964)}]{nvk1964pr}%
  \BibitemOpen
  \bibfield  {author} {\bibinfo {author} {\bibfnamefont {N.~G.}\ \bibnamefont
  {{van Kampen}}},\ }\href@noop {} {\bibfield  {journal} {\bibinfo  {journal}
  {Physical Review}\ }\textbf {\bibinfo {volume} {135}},\ \bibinfo {pages}
  {A362} (\bibinfo {year} {1964})}\BibitemShut {NoStop}%
\bibitem [{\citenamefont {Ornstein}(1908)}]{lso1908phd}%
  \BibitemOpen
  \bibfield  {author} {\bibinfo {author} {\bibfnamefont {L.~S.}\ \bibnamefont
  {Ornstein}},\ }\emph {\bibinfo {title} {Toepassing der statistische mechanica
  van {G}ibbs op molekulair-theoretische vraagstukken}},\ \href@noop {}
  {\bibinfo {type} {Academisch proefschrift}},\ \bibinfo  {school}
  {Rijks-Universiteit te Leiden} (\bibinfo {year} {1908})\BibitemShut {NoStop}%
\bibitem [{Note2()}]{Note2}%
  \BibitemOpen
  \bibinfo {note} {Ornstein's thesis, produced under the supervision of
  Lorentz, is really a valuable piece of scientific work. In §23 of chapter
  III, Ornstein re-derived vdW equation of state as first-order solution [Eq.
  (58)] of his method based, according to the thesis title, on the application
  of Gibbs' statistical mechanics to molecular-theoretical problems. In §24,
  he achieved an improvement by obtaining the second-order solution [Eq. (59)],
  with due acknowledgment to the same formula presented by Boltzmann, together
  with alternative equations of state, in his famous lectures on gas theory
  \cite [Eq. (156) and §54]{lb1986}. In §25, another interesting
  generalisation [Eq. (69)] of vdW equation of state, that boils down to the
  introduction of a temperature dependence of its coefficients, was obtained by
  Ornstein with the help of microcanonical-ensemble theory. Regrettably, the
  thesis was never published; fortunately, it is publicly available online at
  {\relax \protect \fontsize {8}{9.5pt}\protect \selectfont \abovedisplayskip
  6\p@ plus2\p@ minus4\p@ \belowdisplayskip \abovedisplayskip
  \abovedisplayshortskip \z@ plus\p@ \belowdisplayshortskip 3\p@ plus\p@
  minus2\p@ \def \leftmargin \leftmargini \parsep 4\p@ plus2\p@ minus\p@
  \topsep 8\p@ plus2\p@ minus4\p@ \itemsep 4\p@ plus2\p@ minus\p@ {\leftmargin
  \leftmargini \topsep 3\p@ plus\p@ minus\p@ \parsep 2\p@ plus\p@ minus\p@
  \itemsep \parsep
  }https://web.archive.org/web/20090729212141/http://igitur-archive.library.uu.nl/phys/2006-0117-200056/UUindex.html}
  but, obviously, the contents are accessible only to Dutch-speaking
  readers.}\BibitemShut {Stop}%
\bibitem [{\citenamefont {Aronson}\ and\ \citenamefont
  {Hansen}(1972)}]{ea1972aj}%
  \BibitemOpen
  \bibfield  {author} {\bibinfo {author} {\bibfnamefont {E.}~\bibnamefont
  {Aronson}}\ and\ \bibinfo {author} {\bibfnamefont {C.}~\bibnamefont
  {Hansen}},\ }\href@noop {} {\bibfield  {journal} {\bibinfo  {journal} {The
  Astrophysical Journal}\ }\textbf {\bibinfo {volume} {177}},\ \bibinfo {pages}
  {145} (\bibinfo {year} {1972})}\BibitemShut {NoStop}%
\bibitem [{Note3()}]{Note3}%
  \BibitemOpen
  \bibinfo {note} {We thank our colleague R. Trasarti-Battistoni for bringing
  to our attention Aronson and Hansen's paper.}\BibitemShut {Stop}%
\bibitem [{\citenamefont {Padmanabhan}(1990)}]{tp1990pr}%
  \BibitemOpen
  \bibfield  {author} {\bibinfo {author} {\bibfnamefont {T.}~\bibnamefont
  {Padmanabhan}},\ }\href@noop {} {\bibfield  {journal} {\bibinfo  {journal}
  {Physics Reports}\ }\textbf {\bibinfo {volume} {188}},\ \bibinfo {pages}
  {285} (\bibinfo {year} {1990})}\BibitemShut {NoStop}%
\bibitem [{\citenamefont {Stahl}\ \emph {et~al.}(1995)\citenamefont {Stahl},
  \citenamefont {Kiessling},\ and\ \citenamefont {Schindler}}]{bs1995pss}%
  \BibitemOpen
  \bibfield  {author} {\bibinfo {author} {\bibfnamefont {B.}~\bibnamefont
  {Stahl}}, \bibinfo {author} {\bibfnamefont {M.}~\bibnamefont {Kiessling}}, \
  and\ \bibinfo {author} {\bibfnamefont {K.}~\bibnamefont {Schindler}},\
  }\href@noop {} {\bibfield  {journal} {\bibinfo  {journal} {Planetary and
  Space Science}\ }\textbf {\bibinfo {volume} {43}},\ \bibinfo {pages} {271}
  (\bibinfo {year} {1995})}\BibitemShut {NoStop}%
\bibitem [{\citenamefont {Carnahan}\ and\ \citenamefont
  {Starling}(1969)}]{nc1969tjocp}%
  \BibitemOpen
  \bibfield  {author} {\bibinfo {author} {\bibfnamefont {N.}~\bibnamefont
  {Carnahan}}\ and\ \bibinfo {author} {\bibfnamefont {K.}~\bibnamefont
  {Starling}},\ }\href@noop {} {\bibfield  {journal} {\bibinfo  {journal} {The
  Journal of Chemical Physics}\ }\textbf {\bibinfo {volume} {51}},\ \bibinfo
  {pages} {635} (\bibinfo {year} {1969})}\BibitemShut {NoStop}%
\bibitem [{Note4()}]{Note4}%
  \BibitemOpen
  \bibinfo {note} {\label {gibbs}The sum \protect \mbox {$f+pv$} is also the
  Legendre transform of the Helmholtz potential with respect to the specific
  volume and defines the Gibbs potential \protect \mbox {$g=g(T,p)$} whose
  natural state parameters are temperature and pressure; in our case, the
  chemical-potential interpretation is preferable because we find more
  convenient to work with temperature and specific volume. \protect \mbox
  {Equation (\ref {cp})} does not strike the eye of well-read fluid dynamicists
  as a surprising novelty because convenient rearrangements of the term
  \protect \mbox {$v\protect \mbox {\relax \protect \mathweight {bold}\protect
  \bfseries $\nabla $}p$} in the momentum balance equation occur from time to
  time in the literature. To mention a few examples, it appeared in Fridman's
  (or Friedmann's) doctoral thesis, defended in 1922, to reformulate the
  right-hand side of the momentum equation for an inviscid ideal gas in terms
  of gradients of temperature and entropy; see Eq.~(94) in the Russian
  publication \cite {af1934} appeared in 1934 and available online at
  http://books.e-heritage.ru/Book/10087382. It played a role in Crocco's
  derivation \cite {lc1937zamm} of his famous theorem in the case of
  homenthalpic flow of ideal gas as well as in Vaszonyi's generalisation \cite
  {av1945qam} of Crocco's theorem to the form, taught nowadays, in terms of
  gradients of enthalpy and entropy applicable to any fluid with two intensive
  thermodynamic degrees of freedom}\BibitemShut {NoStop}%
\bibitem [{Note5()}]{Note5}%
  \BibitemOpen
  \bibinfo {note} {It appears worth to remark that, according to \protect \mbox
  {Eq.~(\ref {gf.g})}, the chemical potential acts as a mechanical potential
  and that can be considered a particular example of the classical-holography
  property of ideal fluids. A thorough elaboration regarding holographic fluids
  has been recently published by V\'{a}n \cite {pv2023pof}.}\BibitemShut
  {Stop}%
\bibitem [{Note6()}]{Note6}%
  \BibitemOpen
  \bibinfo {note} {Indeed, once we independently understood how to obtain
  \protect \mbox {Eqs.~(\ref {gf.g})} and (\ref {gp.g}), the way to break
  through the apparent barrier we saw in \protect \mbox {Eqs.~(\ref {gf})} and
  (\ref {gp}) became immediately clear and we realised that we could not see it
  because we were somehow biased by the mathematical complexity of \protect
  \mbox {Eq.~(\ref {dpdr})}. We leave the mathematical passages as exercise for
  the interested reader.}\BibitemShut {Stop}%
\bibitem [{\citenamefont {Weast}(1978)}]{rw1978}%
  \BibitemOpen
  \bibinfo {editor} {\bibfnamefont {R.~C.}\ \bibnamefont {Weast}},\ ed.,\
  \href@noop {} {\emph {\bibinfo {title} {CRC Handbook of Chemistry and
  Physics}}},\ \bibinfo {edition} {57th}\ ed.\ (\bibinfo  {publisher} {CRC
  Press},\ \bibinfo {address} {Cleveland, OH},\ \bibinfo {year}
  {1978})\BibitemShut {NoStop}%
\bibitem [{\citenamefont {Haynes}\ \emph {et~al.}(2016)\citenamefont {Haynes},
  \citenamefont {Lide},\ and\ \citenamefont {Bruno}}]{wh2016}%
  \BibitemOpen
  \bibinfo {editor} {\bibfnamefont {W.~M.}\ \bibnamefont {Haynes}}, \bibinfo
  {editor} {\bibfnamefont {D.~R.}\ \bibnamefont {Lide}}, \ and\ \bibinfo
  {editor} {\bibfnamefont {T.~J.}\ \bibnamefont {Bruno}},\ eds.,\ \href@noop {}
  {\emph {\bibinfo {title} {CRC Handbook of Chemistry and Physics}}},\ \bibinfo
  {edition} {97th}\ ed.\ (\bibinfo  {publisher} {CRC Press},\ \bibinfo
  {address} {Boca Raton, FL},\ \bibinfo {year} {2016})\BibitemShut {NoStop}%
\bibitem [{Note7()}]{Note7}%
  \BibitemOpen
  \bibinfo {note} {The presence of the thermodynamic derivative $\left
  ({\partial ^{}\protect \mathfrak {p}}/{\partial \protect \xiyf ^{}}\right
  )_{\protect \thetaysf =1}$ in \protect \mbox {Eq.~(\ref {meq.s.nd})} slightly
  compromises this share in the P scheme.}\BibitemShut {Stop}%
\bibitem [{\citenamefont {Giordano}\ and\ \citenamefont {{D}e
  Serio}(2001)}]{dg2001aiaa}%
  \BibitemOpen
  \bibfield  {author} {\bibinfo {author} {\bibfnamefont {D.}~\bibnamefont
  {Giordano}}\ and\ \bibinfo {author} {\bibfnamefont {M.}~\bibnamefont {{D}e
  Serio}},\ }in\ \href@noop {} {\emph {\bibinfo {booktitle} {35th AIAA
  Thermophysics Conference, 11--14 June, Anaheim CA}}},\ \bibinfo {series and
  number} {AIAA 2001-3016}\ (\bibinfo {organization} {American Institute of
  Aeronautics and Astronautics},\ \bibinfo {year} {2001})\BibitemShut {NoStop}%
\bibitem [{\citenamefont {Giordano}\ and\ \citenamefont {{D}e
  Serio}(2002)}]{dg2002jtht}%
  \BibitemOpen
  \bibfield  {author} {\bibinfo {author} {\bibfnamefont {D.}~\bibnamefont
  {Giordano}}\ and\ \bibinfo {author} {\bibfnamefont {M.}~\bibnamefont {{D}e
  Serio}},\ }\href@noop {} {\bibfield  {journal} {\bibinfo  {journal} {Journal
  of Thermophysics and Heat Transfer}\ }\textbf {\bibinfo {volume} {16}},\
  \bibinfo {pages} {261} (\bibinfo {year} {2002})}\BibitemShut {NoStop}%
\bibitem [{\citenamefont {Planck}(1903)}]{mp1903}%
  \BibitemOpen
  \bibfield  {author} {\bibinfo {author} {\bibfnamefont {M.}~\bibnamefont
  {Planck}},\ }\href@noop {} {\emph {\bibinfo {title} {Treatise on
  thermodynamics}}}\ (\bibinfo  {publisher} {Longmans, Green and Co.},\
  \bibinfo {address} {London, UK},\ \bibinfo {year} {1903})\BibitemShut
  {NoStop}%
\bibitem [{\citenamefont {Callen}(1963)}]{hc1963}%
  \BibitemOpen
  \bibfield  {author} {\bibinfo {author} {\bibfnamefont {H.}~\bibnamefont
  {Callen}},\ }\href@noop {} {\emph {\bibinfo {title} {Thermodynamics}}}\
  (\bibinfo  {publisher} {John Wiley \& Sons},\ \bibinfo {address} {New York
  NY},\ \bibinfo {year} {1963})\ \bibinfo {note} {first publication in
  1960}\BibitemShut {NoStop}%
\bibitem [{\citenamefont {Tisza}(1966)}]{lt1966}%
  \BibitemOpen
  \bibfield  {author} {\bibinfo {author} {\bibfnamefont {L.}~\bibnamefont
  {Tisza}},\ }\href@noop {} {\emph {\bibinfo {title} {Generalized
  thermodynamics}}}\ (\bibinfo  {publisher} {The M.I.T. Press},\ \bibinfo
  {address} {Cambridge, MA},\ \bibinfo {year} {1966})\BibitemShut {NoStop}%
\bibitem [{\citenamefont {Callen}(1985)}]{hc1985}%
  \BibitemOpen
  \bibfield  {author} {\bibinfo {author} {\bibfnamefont {H.}~\bibnamefont
  {Callen}},\ }\href@noop {} {\emph {\bibinfo {title} {Thermodynamics and an
  introduction to thermostatistics}}},\ \bibinfo {edition} {2nd}\ ed.\
  (\bibinfo  {publisher} {John Wiley \& Sons},\ \bibinfo {address} {New York
  NY},\ \bibinfo {year} {1985})\BibitemShut {NoStop}%
\bibitem [{Note8()}]{Note8}%
  \BibitemOpen
  \bibinfo {note} {A convenient conceptual switch to the interpretation of the
  reduced chemical potential as Gibbs potential; see \cite
  {Note4}.}\BibitemShut {Stop}%
\bibitem [{\citenamefont {Shampine}\ and\ \citenamefont
  {Kierzenka}(2008)}]{bvp5c}%
  \BibitemOpen
  \bibfield  {author} {\bibinfo {author} {\bibfnamefont {L.}~\bibnamefont
  {Shampine}}\ and\ \bibinfo {author} {\bibfnamefont {J.}~\bibnamefont
  {Kierzenka}},\ }\href@noop {} {\bibfield  {journal} {\bibinfo  {journal}
  {Journal of Numerical Analysis, Industrial and Applied Mathematics}\ }\textbf
  {\bibinfo {volume} {3}},\ \bibinfo {pages} {27} (\bibinfo {year}
  {2008})}\BibitemShut {NoStop}%
\bibitem [{\citenamefont {Cash}\ \emph {et~al.}(2013)\citenamefont {Cash},
  \citenamefont {Hollevoet}, \citenamefont {Mazzia},\ and\ \citenamefont
  {Nagy}}]{Cash2013}%
  \BibitemOpen
  \bibfield  {author} {\bibinfo {author} {\bibfnamefont {J.}~\bibnamefont
  {Cash}}, \bibinfo {author} {\bibfnamefont {D.}~\bibnamefont {Hollevoet}},
  \bibinfo {author} {\bibfnamefont {F.}~\bibnamefont {Mazzia}}, \ and\ \bibinfo
  {author} {\bibfnamefont {A.}~\bibnamefont {Nagy}},\ }\href@noop {} {\bibfield
   {journal} {\bibinfo  {journal} {ACM Transactions on Mathematical Software}\
  }\textbf {\bibinfo {volume} {39}} (\bibinfo {year} {2013})}\BibitemShut
  {NoStop}%
\bibitem [{\citenamefont {Mazzia}\ \emph {et~al.}(2006)\citenamefont {Mazzia},
  \citenamefont {Sestini},\ and\ \citenamefont {Trigiante}}]{Mazzia20061954}%
  \BibitemOpen
  \bibfield  {author} {\bibinfo {author} {\bibfnamefont {F.}~\bibnamefont
  {Mazzia}}, \bibinfo {author} {\bibfnamefont {A.}~\bibnamefont {Sestini}}, \
  and\ \bibinfo {author} {\bibfnamefont {D.}~\bibnamefont {Trigiante}},\
  }\href@noop {} {\bibfield  {journal} {\bibinfo  {journal} {SIAM Journal on
  Numerical Analysis}\ }\textbf {\bibinfo {volume} {44}},\ \bibinfo {pages}
  {1954} (\bibinfo {year} {2006})}\BibitemShut {NoStop}%
\bibitem [{\citenamefont {Mazzia}(2022)}]{Mazzia2022555}%
  \BibitemOpen
  \bibfield  {author} {\bibinfo {author} {\bibfnamefont {F.}~\bibnamefont
  {Mazzia}},\ }\href@noop {} {\bibfield  {journal} {\bibinfo  {journal} {Annali
  dell'Universita di Ferrara}\ }\textbf {\bibinfo {volume} {68}},\ \bibinfo
  {pages} {555} (\bibinfo {year} {2022})}\BibitemShut {NoStop}%
\bibitem [{\citenamefont {Amodio}\ and\ \citenamefont
  {Settanni}(2009)}]{amse1}%
  \BibitemOpen
  \bibfield  {author} {\bibinfo {author} {\bibfnamefont {P.}~\bibnamefont
  {Amodio}}\ and\ \bibinfo {author} {\bibfnamefont {G.}~\bibnamefont
  {Settanni}},\ }\href@noop {} {\bibfield  {journal} {\bibinfo  {journal}
  {Journal of Numerical Analysis, Industrial and Applied Mathematics}\ }\textbf
  {\bibinfo {volume} {4}},\ \bibinfo {pages} {65} (\bibinfo {year}
  {2009})}\BibitemShut {NoStop}%
\bibitem [{\citenamefont {Amodio}\ and\ \citenamefont
  {Settanni}(2012)}]{amse2}%
  \BibitemOpen
  \bibfield  {author} {\bibinfo {author} {\bibfnamefont {P.}~\bibnamefont
  {Amodio}}\ and\ \bibinfo {author} {\bibfnamefont {G.}~\bibnamefont
  {Settanni}},\ }\href@noop {} {\bibfield  {journal} {\bibinfo  {journal}
  {Journal of Computational and Applied Mathematics}\ }\textbf {\bibinfo
  {volume} {236}},\ \bibinfo {pages} {3869} (\bibinfo {year}
  {2012})}\BibitemShut {NoStop}%
\bibitem [{Note9()}]{Note9}%
  \BibitemOpen
  \bibinfo {note} {All algorithms we used in our perfect-gas study \cite
  {dg2019ejmb}\ blew off and failed miserably when we attempted the same
  leap.}\BibitemShut {Stop}%
\bibitem [{Note10()}]{Note10}%
  \BibitemOpen
  \bibinfo {note} {Of course, in case of existence, one should still ponder
  about the physical meaningfulness or realisability of such
  values.}\BibitemShut {Stop}%
\bibitem [{Note11()}]{Note11}%
  \BibitemOpen
  \bibinfo {note} {There exists also an entropy minimum but, inexplicably, it
  is hardly mention in the literature. See \protect \mbox {Fig. 22}$_{\scalebox
  {0.65}{$\protect \mbox {\cite {dg2019ejmb}}$}}$.}\BibitemShut {Stop}%
\bibitem [{\citenamefont {Antonov}(1985)}]{va1985}%
  \BibitemOpen
  \bibfield  {author} {\bibinfo {author} {\bibfnamefont {V.~A.}\ \bibnamefont
  {Antonov}},\ }in\ \href@noop {} {\emph {\bibinfo {booktitle} {Dynamics of
  star clusters}}},\ \bibinfo {series and number} {Proceedings of the 113th
  Symposium held in Princeton, NJ, 29 May to 1 June 1984},\ \bibinfo {editor}
  {edited by\ \bibinfo {editor} {\bibfnamefont {J.}~\bibnamefont {Goodman}}\
  and\ \bibinfo {editor} {\bibfnamefont {P.}~\bibnamefont {Hut}}}\ (\bibinfo
  {publisher} {Reidel Publishing Company},\ \bibinfo {address} {Dordrecht, The
  Netherlands},\ \bibinfo {year} {1985})\ pp.\ \bibinfo {pages} {525--540},\
  \bibinfo {note} {{E}nglish translation of original article [Vest. Leningrad
  Univ., \textbf{7}, 135 (1962)]}\BibitemShut {NoStop}%
\bibitem [{\citenamefont {Lynden-{B}ell}\ and\ \citenamefont
  {Wood}(1968)}]{dl1968mnras}%
  \BibitemOpen
  \bibfield  {author} {\bibinfo {author} {\bibfnamefont {D.}~\bibnamefont
  {Lynden-{B}ell}}\ and\ \bibinfo {author} {\bibfnamefont {R.}~\bibnamefont
  {Wood}},\ }\href@noop {} {\bibfield  {journal} {\bibinfo  {journal} {Monthly
  Notices of the Royal Astronomical Society}\ }\textbf {\bibinfo {volume}
  {138}},\ \bibinfo {pages} {495} (\bibinfo {year} {1968})}\BibitemShut
  {NoStop}%
\bibitem [{\citenamefont {Lynden-{B}ell}\ and\ \citenamefont
  {Lynden-{B}ell}(1977)}]{dl1977mnras}%
  \BibitemOpen
  \bibfield  {author} {\bibinfo {author} {\bibfnamefont {D.}~\bibnamefont
  {Lynden-{B}ell}}\ and\ \bibinfo {author} {\bibfnamefont {R.~M.}\ \bibnamefont
  {Lynden-{B}ell}},\ }\href@noop {} {\bibfield  {journal} {\bibinfo  {journal}
  {Monthly Notices of the Royal Astronomical Society}\ }\textbf {\bibinfo
  {volume} {181}},\ \bibinfo {pages} {405} (\bibinfo {year}
  {1977})}\BibitemShut {NoStop}%
\bibitem [{\citenamefont {Lynden-{B}ell}(1999)}]{dl1999p}%
  \BibitemOpen
  \bibfield  {author} {\bibinfo {author} {\bibfnamefont {D.}~\bibnamefont
  {Lynden-{B}ell}},\ }\href@noop {} {\bibfield  {journal} {\bibinfo  {journal}
  {Physica A}\ }\textbf {\bibinfo {volume} {263}},\ \bibinfo {pages} {293}
  (\bibinfo {year} {1999})}\BibitemShut {NoStop}%
\bibitem [{\citenamefont {Thirring}(1970)}]{wt1970zfp}%
  \BibitemOpen
  \bibfield  {author} {\bibinfo {author} {\bibfnamefont {W.}~\bibnamefont
  {Thirring}},\ }\href@noop {} {\bibfield  {journal} {\bibinfo  {journal}
  {Zeitschrift f\"{u}r Physik A}\ }\textbf {\bibinfo {volume} {235}},\ \bibinfo
  {pages} {339} (\bibinfo {year} {1970})}\BibitemShut {NoStop}%
\bibitem [{\citenamefont {Hachisu}\ and\ \citenamefont
  {Sugimoto}(1978)}]{ih1978ptp}%
  \BibitemOpen
  \bibfield  {author} {\bibinfo {author} {\bibfnamefont {I.}~\bibnamefont
  {Hachisu}}\ and\ \bibinfo {author} {\bibfnamefont {D.}~\bibnamefont
  {Sugimoto}},\ }\href@noop {} {\bibfield  {journal} {\bibinfo  {journal}
  {Progress of Theoretical Physics}\ }\textbf {\bibinfo {volume} {60}},\
  \bibinfo {pages} {123} (\bibinfo {year} {1978})}\BibitemShut {NoStop}%
\bibitem [{\citenamefont {Thirring}\ \emph {et~al.}(2003)\citenamefont
  {Thirring}, \citenamefont {Narnhofer},\ and\ \citenamefont
  {Posch}}]{wt2003prl}%
  \BibitemOpen
  \bibfield  {author} {\bibinfo {author} {\bibfnamefont {W.}~\bibnamefont
  {Thirring}}, \bibinfo {author} {\bibfnamefont {H.}~\bibnamefont {Narnhofer}},
  \ and\ \bibinfo {author} {\bibfnamefont {H.~A.}\ \bibnamefont {Posch}},\
  }\href@noop {} {\bibfield  {journal} {\bibinfo  {journal} {Physical Review
  Letters}\ }\textbf {\bibinfo {volume} {91}},\ \bibinfo {pages} {13061: 1}
  (\bibinfo {year} {2003})}\BibitemShut {NoStop}%
\bibitem [{Note12()}]{Note12}%
  \BibitemOpen
  \bibinfo {note} {\protect \mbox {Equation (\ref {ient.vdw})} is obtained from
  \protect \mbox {Eq. (96)}$_{\scalebox {0.65}{$\protect \mbox {\cite
  {dg2019ejmb}}$}}$ by substituting into it the vdW-model specific entropy
  \protect \[ s_{g} = K + \protect \frac {3}{2}R\protect \,\ln T +R\protect
  \,\ln \left (v-B\right )\protect \] rather than the perfect-gas specific
  entropy given in \protect \mbox {Eq. (98)}$_{\scalebox {0.65}{$\protect \mbox
  {\cite {dg2019ejmb}}$}}$. It reduces to \protect \mbox {Eq.
  (102)}$_{\scalebox {0.65}{$\protect \mbox {\cite {dg2019ejmb}}$}}$ if
  \protect \mbox {$\protect \,\protect \betay \rightarrow 0$}.}\BibitemShut
  {Stop}%
\bibitem [{Note13()}]{Note13}%
  \BibitemOpen
  \bibinfo {note} {Stahl et al. use the symbol $\protect \etayf $ which would
  conflict with our nondimensional radial coordinate [\protect \mbox {Eq.~(\ref
  {ndv})}]; for this reason we reverted to the original symbol $y$ used by
  Carnahan and Starling.}\BibitemShut {Stop}%
\bibitem [{Note14()}]{Note14}%
  \BibitemOpen
  \bibinfo {note} {\protect \color {black} We made an interesting numerical
  experiment in the ``bvp'' variant of the P scheme by introducing a fictitious
  \protect \textit {differential} equation for $\protect \,\protect \etayf
  ^{\ast }$ that, after a suitable change of coordinate, we have solved
  together with the others on the interval $[0,1]$ with the boundary conditions
  defined by \protect \mbox {Eqs.~(\ref {bc.nd})}, \protect \mbox {Eq.~(\ref
  {gfc})}, \protect \mbox {Eqs.~(\ref {bc.i.P})}.}\BibitemShut {Stop}%
\bibitem [{Note15()}]{Note15}%
  \BibitemOpen
  \bibinfo {note} {By the way, in the circumstance of vanishing gravitational
  number, the configuration consisting of a liquid core surrounded by a gas
  layer is not the only possible one; the other configuration, gas core in
  \protect \mbox {$[0,\protect \etayf ^{\ast }]$} surrounded by liquid layer in
  \protect \mbox {$[\protect \etayf ^{\ast },1]$}, is also possible. The
  corresponding formulae are obtained from \protect \mbox {Eqs.~(\ref
  {pe.N=0.l})}, \protect \mbox {Eqs.~(\ref {pe.N=0.g})}, \protect \mbox
  {Eqs.~(\ref {gf.nd.N=0.ud.l})--(\ref {gf.nd.N=0.ud.g})} and \protect \mbox
  {Eq.~(\ref {nc.pe})} by swapping the subscripts $l,g$ and the accents
  $\protect \,\protect \check {}\protect \, , \protect \hat {}\protect \,$
  wherever they occur. \protect \mbox {Equation (\ref {peil})} becomes \protect
  \[\protect \etayf ^{\ast } = \left (\protect \frac {\protect \xiyf _{l} -
  1}{\protect \xiyf _{l} - \protect \xiyf _{g}}\right )^{1/3}\protect \] and
  the existence condition [\protect \mbox {Eq.~(\ref {pe.N=0.cond})}] remains
  unaltered. In the case of the numerical example of the text, the \protect
  \textit {gas-liquid} interface's location would be \protect \mbox {$\protect
  \etayf ^{\ast } \simeq 0.9999$}; thus, there would be an extremely thin
  liquid layer separating the gas core from the container.}\BibitemShut {Stop}%
\bibitem [{Note16()}]{Note16}%
  \BibitemOpen
  \bibinfo {note} {Chemical reaction rates, though, are affected by that tiny
  fraction of molecules and their determination requires more sophisticated
  approaches.}\BibitemShut {Stop}%
\bibitem [{\citenamefont {Jeans}(1926)}]{jj1926mnras}%
  \BibitemOpen
  \bibfield  {author} {\bibinfo {author} {\bibfnamefont {J.~H.}\ \bibnamefont
  {Jeans}},\ }\href@noop {} {\bibfield  {journal} {\bibinfo  {journal} {Monthly
  Notices of the Royal Astronomical Society}\ }\textbf {\bibinfo {volume}
  {87}},\ \bibinfo {pages} {36} (\bibinfo {year} {1926})}\BibitemShut {NoStop}%
\bibitem [{\citenamefont {Jeans}(1927{\natexlab{a}})}]{jj1927.03mnras}%
  \BibitemOpen
  \bibfield  {author} {\bibinfo {author} {\bibfnamefont {J.~H.}\ \bibnamefont
  {Jeans}},\ }\href@noop {} {\bibfield  {journal} {\bibinfo  {journal} {Monthly
  Notices of the Royal Astronomical Society}\ }\textbf {\bibinfo {volume}
  {87}},\ \bibinfo {pages} {400} (\bibinfo {year}
  {1927}{\natexlab{a}})}\BibitemShut {NoStop}%
\bibitem [{\citenamefont {Jeans}(1927{\natexlab{b}})}]{jj1927.07mnras}%
  \BibitemOpen
  \bibfield  {author} {\bibinfo {author} {\bibfnamefont {J.~H.}\ \bibnamefont
  {Jeans}},\ }\href@noop {} {\bibfield  {journal} {\bibinfo  {journal} {Monthly
  Notices of the Royal Astronomical Society}\ }\textbf {\bibinfo {volume}
  {87}},\ \bibinfo {pages} {720} (\bibinfo {year}
  {1927}{\natexlab{b}})}\BibitemShut {NoStop}%
\bibitem [{\citenamefont {Jeans}(1928{\natexlab{a}})}]{jj1928.02n}%
  \BibitemOpen
  \bibfield  {author} {\bibinfo {author} {\bibfnamefont {J.~H.}\ \bibnamefont
  {Jeans}},\ }\href@noop {} {\bibfield  {journal} {\bibinfo  {journal}
  {Nature}\ }\textbf {\bibinfo {volume} {121}},\ \bibinfo {pages} {173}
  (\bibinfo {year} {1928}{\natexlab{a}})}\BibitemShut {NoStop}%
\bibitem [{\citenamefont {Eddington}(1928{\natexlab{a}})}]{ae1928.02n}%
  \BibitemOpen
  \bibfield  {author} {\bibinfo {author} {\bibfnamefont {A.}~\bibnamefont
  {Eddington}},\ }\href@noop {} {\bibfield  {journal} {\bibinfo  {journal}
  {Nature}\ }\textbf {\bibinfo {volume} {121}},\ \bibinfo {pages} {278}
  (\bibinfo {year} {1928}{\natexlab{a}})}\BibitemShut {NoStop}%
\bibitem [{\citenamefont {Eddington}(1928{\natexlab{b}})}]{ae1928.03mnras}%
  \BibitemOpen
  \bibfield  {author} {\bibinfo {author} {\bibfnamefont {A.}~\bibnamefont
  {Eddington}},\ }\href@noop {} {\bibfield  {journal} {\bibinfo  {journal}
  {Monthly Notices of the Royal Astronomical Society}\ }\textbf {\bibinfo
  {volume} {88}},\ \bibinfo {pages} {352} (\bibinfo {year}
  {1928}{\natexlab{b}})}\BibitemShut {NoStop}%
\bibitem [{\citenamefont {Jeans}(1928{\natexlab{b}})}]{jj1928.03mnras}%
  \BibitemOpen
  \bibfield  {author} {\bibinfo {author} {\bibfnamefont {J.~H.}\ \bibnamefont
  {Jeans}},\ }\href@noop {} {\bibfield  {journal} {\bibinfo  {journal} {Monthly
  Notices of the Royal Astronomical Society}\ }\textbf {\bibinfo {volume}
  {88}},\ \bibinfo {pages} {393} (\bibinfo {year}
  {1928}{\natexlab{b}})}\BibitemShut {NoStop}%
\bibitem [{\citenamefont {Eddington}(1928{\natexlab{c}})}]{ae1928.03n}%
  \BibitemOpen
  \bibfield  {author} {\bibinfo {author} {\bibfnamefont {A.}~\bibnamefont
  {Eddington}},\ }\href@noop {} {\bibfield  {journal} {\bibinfo  {journal}
  {Nature}\ }\textbf {\bibinfo {volume} {121}},\ \bibinfo {pages} {496}
  (\bibinfo {year} {1928}{\natexlab{c}})}\BibitemShut {NoStop}%
\bibitem [{\citenamefont {Jeans}(1928{\natexlab{c}})}]{jj1928.03n}%
  \BibitemOpen
  \bibfield  {author} {\bibinfo {author} {\bibfnamefont {J.~H.}\ \bibnamefont
  {Jeans}},\ }\href@noop {} {\bibfield  {journal} {\bibinfo  {journal}
  {Nature}\ }\textbf {\bibinfo {volume} {121}},\ \bibinfo {pages} {496}
  (\bibinfo {year} {1928}{\natexlab{c}})}\BibitemShut {NoStop}%
\bibitem [{\citenamefont {Jeans}(1929)}]{jj1929}%
  \BibitemOpen
  \bibfield  {author} {\bibinfo {author} {\bibfnamefont {J.~H.}\ \bibnamefont
  {Jeans}},\ }\href@noop {} {\emph {\bibinfo {title} {Astronomy and
  Cosmogony}}},\ \bibinfo {edition} {2nd}\ ed.\ (\bibinfo  {publisher}
  {Cambridge University Press},\ \bibinfo {address} {Cambridge UK},\ \bibinfo
  {year} {1929})\BibitemShut {NoStop}%
\bibitem [{\citenamefont {Robitaille}(2011{\natexlab{a}})}]{pmr2011.Ipip}%
  \BibitemOpen
  \bibfield  {author} {\bibinfo {author} {\bibfnamefont {P.-M.}\ \bibnamefont
  {Robitaille}},\ }\href@noop {} {\bibfield  {journal} {\bibinfo  {journal}
  {Progress in Physics}\ }\textbf {\bibinfo {volume} {3}},\ \bibinfo {pages}
  {3} (\bibinfo {year} {2011}{\natexlab{a}})}\BibitemShut {NoStop}%
\bibitem [{\citenamefont {Robitaille}(2011{\natexlab{b}})}]{pmr2011.IIpip}%
  \BibitemOpen
  \bibfield  {author} {\bibinfo {author} {\bibfnamefont {P.-M.}\ \bibnamefont
  {Robitaille}},\ }\href@noop {} {\bibfield  {journal} {\bibinfo  {journal}
  {Progress in Physics}\ }\textbf {\bibinfo {volume} {3}},\ \bibinfo {pages}
  {41} (\bibinfo {year} {2011}{\natexlab{b}})}\BibitemShut {NoStop}%
\bibitem [{\citenamefont {Schwarzschild}(1916)}]{ks1916skpaw}%
  \BibitemOpen
  \bibfield  {author} {\bibinfo {author} {\bibfnamefont {K.}~\bibnamefont
  {Schwarzschild}},\ }\href@noop {} {\bibfield  {journal} {\bibinfo  {journal}
  {Sitzungberichte der K\"{o}niglich Preussischen Akademie der Wissenschaften}\
  }\textbf {\bibinfo {volume} {Erster Halbblad}},\ \bibinfo {pages} {424}
  (\bibinfo {year} {1916})}\BibitemShut {NoStop}%
\bibitem [{\citenamefont {Tolman}(1939)}]{rt1939pr}%
  \BibitemOpen
  \bibfield  {author} {\bibinfo {author} {\bibfnamefont {R.}~\bibnamefont
  {Tolman}},\ }\href@noop {} {\bibfield  {journal} {\bibinfo  {journal}
  {Physical Review}\ }\textbf {\bibinfo {volume} {55}},\ \bibinfo {pages} {364}
  (\bibinfo {year} {1939})}\BibitemShut {NoStop}%
\bibitem [{\citenamefont {Oppenheimer}\ and\ \citenamefont
  {Volkoff}(1939)}]{ro1939pr}%
  \BibitemOpen
  \bibfield  {author} {\bibinfo {author} {\bibfnamefont {J.~R.}\ \bibnamefont
  {Oppenheimer}}\ and\ \bibinfo {author} {\bibfnamefont {G.~M.}\ \bibnamefont
  {Volkoff}},\ }\href@noop {} {\bibfield  {journal} {\bibinfo  {journal}
  {Physical Review}\ }\textbf {\bibinfo {volume} {55}},\ \bibinfo {pages} {374}
  (\bibinfo {year} {1939})}\BibitemShut {NoStop}%
\bibitem [{\citenamefont {Misner}\ \emph {et~al.}(1970)\citenamefont {Misner},
  \citenamefont {Thorne},\ and\ \citenamefont {Wheeler}}]{cm1970}%
  \BibitemOpen
  \bibfield  {author} {\bibinfo {author} {\bibfnamefont {C.~W.}\ \bibnamefont
  {Misner}}, \bibinfo {author} {\bibfnamefont {K.~S.}\ \bibnamefont {Thorne}},
  \ and\ \bibinfo {author} {\bibfnamefont {J.~A.}\ \bibnamefont {Wheeler}},\
  }\href@noop {} {\emph {\bibinfo {title} {Gravitation}}}\ (\bibinfo
  {publisher} {Freeman and Company},\ \bibinfo {address} {San Francisco CA},\
  \bibinfo {year} {1970})\BibitemShut {NoStop}%
\bibitem [{\citenamefont {Weinberg}(1972)}]{sw1972}%
  \BibitemOpen
  \bibfield  {author} {\bibinfo {author} {\bibfnamefont {S.}~\bibnamefont
  {Weinberg}},\ }\href@noop {} {\emph {\bibinfo {title} {Gravitation and
  cosmology: principles and applications of the general theory of
  relativity}}}\ (\bibinfo  {publisher} {John Wiley \& Sons},\ \bibinfo
  {address} {New York NY},\ \bibinfo {year} {1972})\BibitemShut {NoStop}%
\bibitem [{\citenamefont {Wald}(1984)}]{rw1984}%
  \BibitemOpen
  \bibfield  {author} {\bibinfo {author} {\bibfnamefont {R.}~\bibnamefont
  {Wald}},\ }\href@noop {} {\emph {\bibinfo {title} {General Relativity}}}\
  (\bibinfo  {publisher} {The University of Chicago Press},\ \bibinfo {address}
  {Chicago IL},\ \bibinfo {year} {1984})\BibitemShut {NoStop}%
\bibitem [{\citenamefont {Schutz}(2009)}]{bs2009}%
  \BibitemOpen
  \bibfield  {author} {\bibinfo {author} {\bibfnamefont {B.}~\bibnamefont
  {Schutz}},\ }\href@noop {} {\emph {\bibinfo {title} {A First Course in
  General Relativity}}}\ (\bibinfo  {publisher} {Cambridge University Press},\
  \bibinfo {address} {Cambridge UK},\ \bibinfo {year} {2009})\BibitemShut
  {NoStop}%
\bibitem [{\citenamefont {Carroll}(2019)}]{sc2019}%
  \BibitemOpen
  \bibfield  {author} {\bibinfo {author} {\bibfnamefont {S.}~\bibnamefont
  {Carroll}},\ }\href@noop {} {\emph {\bibinfo {title} {Spacetime and
  Geometry}}}\ (\bibinfo  {publisher} {Cambridge University Press},\ \bibinfo
  {address} {Cambridge UK},\ \bibinfo {year} {2019})\BibitemShut {NoStop}%
\bibitem [{\citenamefont {Boltzmann}(1964)}]{lb1986}%
  \BibitemOpen
  \bibfield  {author} {\bibinfo {author} {\bibfnamefont {L.}~\bibnamefont
  {Boltzmann}},\ }\href@noop {} {\emph {\bibinfo {title} {Lectures on gas
  theory}}}\ (\bibinfo  {publisher} {Dover Publications},\ \bibinfo {address}
  {New York NY},\ \bibinfo {year} {1964})\BibitemShut {NoStop}%
\bibitem [{\citenamefont {Fridman}(1934)}]{af1934}%
  \BibitemOpen
  \bibfield  {author} {\bibinfo {author} {\bibfnamefont {A.~A.}\ \bibnamefont
  {Fridman}},\ }\href@noop {} {\emph {\bibinfo {title} {An essay on
  hydrodynamics of compressible fluid}}}\ (\bibinfo  {publisher} {State
  Technical-Theoretical Publishing House},\ \bibinfo {address} {Moscow},\
  \bibinfo {year} {1934})\ \bibinfo {note} {{R}ussian publication of his
  doctoral thesis: \textwncyr{A. A. Fridman, Opyt Gidromekhaniki Szhimaemoi0
  Zhidkosti. ONTI Gosudarstvennoe Tekhniko-Teoreticheskoe Izdatelp1stvo,
  Moskva, 1934}.}\BibitemShut {Stop}%
\bibitem [{\citenamefont {Crocco}(1937)}]{lc1937zamm}%
  \BibitemOpen
  \bibfield  {author} {\bibinfo {author} {\bibfnamefont {L.}~\bibnamefont
  {Crocco}},\ }\href@noop {} {\bibfield  {journal} {\bibinfo  {journal}
  {Zeitschrift f\"{u}r Angewandte Mathematik und Mechanik}\ }\textbf {\bibinfo
  {volume} {17}},\ \bibinfo {pages} {1} (\bibinfo {year} {1937})}\BibitemShut
  {NoStop}%
\bibitem [{\citenamefont {V\'{a}zsonyi}(1945)}]{av1945qam}%
  \BibitemOpen
  \bibfield  {author} {\bibinfo {author} {\bibfnamefont {A.}~\bibnamefont
  {V\'{a}zsonyi}},\ }\href@noop {} {\bibfield  {journal} {\bibinfo  {journal}
  {Quarterly of Applied Mathematics}\ }\textbf {\bibinfo {volume} {III}},\
  \bibinfo {pages} {29} (\bibinfo {year} {1945})}\BibitemShut {NoStop}%
\bibitem [{\citenamefont {V\'{a}n}(2023)}]{pv2023pof}%
  \BibitemOpen
  \bibfield  {author} {\bibinfo {author} {\bibfnamefont {P.}~\bibnamefont
  {V\'{a}n}},\ }\href@noop {} {\bibfield  {journal} {\bibinfo  {journal}
  {Physics of Fluids}\ }\textbf {\bibinfo {volume} {35}},\ \bibinfo {pages}
  {057105 1} (\bibinfo {year} {2023})}\BibitemShut {NoStop}%
\end{thebibliography}
%\bibliography{/Users/dg/Library/texmf/bibtex/bib/mybibreflibrary.bib}   % name your BibTeX data base

\end{document}